\documentclass[letter, 12pt]{article}

\usepackage{graphicx}
\usepackage{txfonts}

\begin{document}

\title{\textbf{QUANTUM DYNAMICS OF LOOP QUANTUM GRAVITY}}

\author{\\ \\ \\ \\ \\ \\ \\ \\ \normalsize A Thesis\\ \normalsize\\
\normalsize Submitted to the Graduate Faculty of the\\
\normalsize Louisiana State University and\\
\normalsize Agricultural and Mechanical College\\
\normalsize in partial fulfillment of the\\
\normalsize requirements for the degree of\\
\normalsize Master of Science\\ \normalsize\\
\normalsize in\\ \normalsize\\
\normalsize The Department of Physics and Astronomy\\ \\ \\ \\ \\ \\
\normalsize by\\
\normalsize Muxin Han\\
\normalsize B.S., Beijing Normal University, Beijing, 2005}
\date{\normalsize May 2007}

\maketitle

\thispagestyle{empty}

\pagenumbering{roman}

\newpage

%\begin{center}
\section*{Acknowledgements}
%\end{center}

First of all, I am grateful to Dr. Jorge Pullin for his advise and
many corrections of this thesis, and to Dr. Jonathan Dowling for all
his kind help in these two years. I also would like to thank Dr.
Hwang Lee for his kind support and being a member of my committee.

I would like to thank all the people who have discussed issues with
me concerning the subject in the thesis. They are: Dr. Abhay
Ashtekar, Dr. Lai-Him Chan, Dr. Weiming Huang, Dr. Jerzy
Lewandowski, Dr. Yongge Ma, Dr. Andrzej Okolow, Dr. Jorge Pullin,
Dr. Carlo Rovelli, Dr. Thomas Thiemann, Dr. Dmitry Uskov, Dr. Robert
M. Wald, and Dr. Hongbao Zhang.

I also want to thank my wife, Yan Li, for all her support and help, to push me finishing this work successfully.

This work is supported by the assistantship of LSU, the Horace
Hearne Institute for Theoretical Physics at LSU, the funding from
Advanced Research and Development Activity, and the support of NSFC 10205002 and NSFC
10675019.

\newpage

\tableofcontents

\addcontentsline {section}{Acknowledgements}{ii}

\addcontentsline {section}{Abstract}{iv}

\addcontentsline {section}{References}{81}

\addcontentsline {section}{Vita}{91}

\newpage

%\begin{center}
\section*{Abstract}
%\end{center}

In the last 20 years, loop quantum gravity, a background independent
approach to unify general relativity and quantum mechanics, has been
widely investigated. The aim of loop quantum gravity is to construct
a mathematically rigorous, background independent, nonperturbative
quantum theory for the Lorentzian gravitational field on a
four-dimensional manifold. In this approach, the principles of
quantum mechanics are combined with those of general relativity
naturally. Such a combination provides us a picture of "quantum
Riemannian geometry", which is discrete at a fundamental scale. In
the investigation of quantum dynamics, the classical expressions of
constraints are quantized as operators. The quantum evolution is
contained in the solutions of the quantum constraint equations. On
the other hand, the semi-classical analysis has to be carried out in
order to test the semiclassical limit of the quantum dynamics.

In this thesis, the structure of the dynamical theory in loop
quantum gravity is presented pedagogically. The outline is as
follows: first we review the classical formalism of general
relativity as a dynamical theory of connections. Then the
kinematical Ashtekar-Isham-Lewandowski representation is introduced
as a foundation of loop quantum gravity. We discuss the construction
of a Hamiltonian constraint operator and the master constraint
programme, for both the cases of pure gravity and matter field
coupling. Finally, some strategies are discussed concerning testing
the semiclassical limit of the quantum dynamics.

\newpage

\pagenumbering{arabic}

\section{Introduction}
\subsection{Motivation of Quantum Gravity}

The current view of physics is that there exist four fundamental
interactions: strong interaction, weak interaction, electromagnetic
interaction and gravitational interaction. The description for the
former three kinds of forces is quantized in the well-known standard
model. The interactions are transmitted via the exchange of
particles. However, the last kind of interaction, gravitational
interaction, is described by Einstein's theory of general
relativity, which is a classical theory which describes the
gravitational field as a smooth metric tensor field on a manifold,
i.e., a 4-dimensional spacetime geometry. There is no $\hbar$ and
hence no quantum structure of spacetime. Thus there is a fundamental
inconsistency in our current description of the whole physical
world. Physicists widely accept the assumption that our world is
quantized at fundamental level. So all interactions should be
brought into the framework of quantum mechanics fundamentally. As a
result, the gravitational field should also have "quantum
structure".

Throughout the last century, our understanding of nature has
considerably improved from macroscale to microscale, including the
phenomena at molecule, atom, sub-atom, and elementary particle
scale. The standard model of particle physics agrees with all
present experimental tests in laboratory (see e.g. \cite{weinberg}).
However, because unimaginably large amount of energy would be
needed, no experimental tests exist for processes that happen near
the Planck scale $\ell_p\equiv(G\hbar/{c^3})^{1/2}\sim10^{-33}cm$
and $t_p\equiv(G\hbar/{c^5})^{1/2}\sim10^{-43}s $, which are viewed
as the most fundamental scales. The Planck scale arises naturally in
attempts to formulate a quantum theory of gravity, since $\ell_p$
and $t_p$ are unique combinations of speed of light $c$, Planck
constant $\hbar$, and gravitational constant $G$, which have the
dimensions of length and time respectively. The dimensional
arguments suggest that at Planck scale the smooth structure of
spacetime should break down, and therefore the well-known quantum
field theory is invalid since it depends on a fixed smooth
background spacetime. Hence we believe that physicists should go
beyond the successful standard model to explore the new physics near
Planck scale, which is, perhaps, a quantum field theory without a
background spacetime, and this quantum field theory should include
the quantum theory of gravity. Moreover, current theoretical physics
is thirsting for a quantum theory of gravity to solve at least the
following fundamental difficulties.
\begin{itemize}
\item { Classical Gravity - Quantum Matter Inconsistency}

The equation relating matter and the gravitational field is the
famous Einstein field equation:
\begin{equation}
R_{\alpha\beta}[g]-\frac{1}{2}R[g]g_{\alpha\beta}=\kappa
T_{\alpha\beta}[g],\label{ein0}
\end{equation}
where the left hand side of the equation concerns spacetime geometry
which has classical smooth structure, while the right hand side
concerns also matter field which is fundamentally quantum mechanical
in standard model. In quantum field theory the energy-momentum
tensor of matter field should be an operator-valued tensor
$\hat{T}_{\alpha\beta}$. One possible way to keep classical geometry
consistent with quantum matter is to replace $T_{\alpha\beta}[g]$ by
the expectation value $<\hat{T}_{\alpha\beta}[g]>$ with respect to
some quantum state of the matter on a fixed spacetime. However, in
the solution of this equation the background $g_{\alpha\beta}$ has
to be changed due to the non-vanishing of
$<\hat{T}_{\alpha\beta}[g]>$. So one has to feed back the new metric
into the definition of the vacuum expectation value etc. The result
of the iterations does not converge in general \cite{FM}. On the
other hand, some other arguments show that such a semiclassical
treatment may violate the principle of superposition in quantum
mechanics \cite{carlip}. This inconsistency motivates us to quantize
the background geometry to arrive at an operator formula also on the
left hand side of Eq.(\ref{ein0}).

\item { Singularities in General Relativity}

Einstein's theory of General Relativity is considered as one of the
most elegant theories in the 20th century. Many experimental tests
confirm the theory in the classical domain \cite{will}. However,
Penrose and Hawking proved that singularities are inevitable in
general spacetimes with matter satisfying certain conditions in, by
now well known, singularity theorems (for a summary, see
\cite{hawking}\cite{wald}). Thus general relativity as a classical
theory breaks down in certain regions of spacetime in a generic way.
One naturally expects that, in extra strong gravitational field
domains near the singularities, the gravitational theory would
probably be replaced by an unknown quantum theory of gravity.

\item { Infinities in Quantum Field Theory}

It is well known that there are infinity problems in quantum field
theory in Minkowski spacetime. In curved spacetime, the problem of
divergences is even more complicated, since the renormalization
process in curved spacetime is ambiguous, the expectation value of
stress tensor can be fixed up to some local curvature terms, and it
also depends on a fundamental scale of spacetime. Although much
progress on the renormalization have been made
\cite{wald2}\cite{wald1}, a fundamentally satisfactory theory is
still far from reaching. So it is expected that some quantum gravity
theory, playing a fundamental role at Planck scale, could provide a
natural cut-off to cure the infinities in quantum field theory. The
situation of quantum field theory on a fixed spacetime looks just
like that of quantum mechanics for particles in electromagnetic
field before the establishing of quantum electrodynamics, where the
particle mechanics (actress) is quantized but the background
electromagnetic field (stage) is classical. The history suggests
that such a semi-classical situation is only an approximation which
should be replaced by a much more fundamental and satisfactory
theory.

\end{itemize}

\subsection{Purpose of Loop Quantum Gravity}

The research on quantum gravity is quite active. Many quantization
programmes for gravity are being carried out (for a summary see e.g.
\cite{thiemann2}). In these different kinds of approaches, Among
these different kinds of approaches, the idea of loop quantum
gravity finds its roots in researchers from the general relativity
community. It follows closely the motivations of general relativity,
and hence it is a quantum theory born with background independence.
Roughly speaking, loop quantum gravity is an attempt to construct a
mathematically rigorous, non-perturbative, background independent
quantum theory of four-dimensional, Lorentzian general relativity
plus all known matter in the continuum. The project of loop quantum
gravity inherits the basic idea of Einstein that gravity is
fundamentally spacetime geometry. Here one believes in that the
theory of quantum gravity is a quantum theory of spacetime geometry
with diffeomorphism invariance (this legacy is discussed
comprehensively in Rovelli's book \cite{rovelli}). To carry out the
quantization procedure, one first casts general relativity into the
Hamiltonian formalism as a diffeomorphism invariant Yang-Mills gauge
field theory with a compact internal gauge group. Thus the
construction of loop quantum gravity can also be applied to all
background independent gauge field theories. One can therefore claim
that the theory can also be called as a background independent
quantum gauge field theory.

All classical fields theories, other than the gravitational field,
are defined on a fixed spacetime, which provides a foundation to the
perturbative Fock space quantization. However general relativity is
only defined on a manifold and hence is a background independent
classical field theory, since gravity itself is the background. So
the situation for gravity is much different from other fields by
construction \cite{rovelli}, namely gravity is not only the
background stage, but also the dynamical actress. Such a double
character for gravity leads to many difficulties in the
understanding of general relativity and its quantization, since we
cannot follow the strategy in ordinary quantum theory of matter
fields. However, an amazing result in loop quantum gravity is that
the background independent programme can even help us to avoid the
difficulties in ordinary quantum field theory. In perturbative
quantum field theory in curved spacetime, the definition of some
basic physical quantities, such as the expectation value of
energy-momentum, is ambiguous and it is difficult to calculate the
back-reaction of quantum fields to the background spacetime
\cite{wald1}. One could speculate that the difficulty is related to
the fact that the present formulation of quantum field theories is
background dependent. For instance, the vacuum state of a quantum
field is closely related to spacetime structure, which plays an
essential role in the description of quantum field theory in curved
spacetime and their renormalization procedures. However, if the
quantization programme is by construction background independent and
non-perturbative, it may be possible to solve the problems
fundamentally. In loop quantum gravity there is no assumption of a
priori background metric at all and the gravitational field and
matter fields are coupled and fluctuating naturally with respect to
each other on a common manifold.

In the following sections, we will review pedagogically the basic
construction of a completely new, background independent quantum
field theory, which is completely different from the known quantum
field theory. For completeness and accuracy, we will use detailed
mathematical terminology. However, for simplicity, we will skip the
complicated proofs of many important statements. One may find the
missing details in the references cited. Thus our review will not be
comprehensive. We refer to Ref.\cite{thiemann2} and \cite{HM3} for a
more detailed exploration, Refs. \cite{AL} and \cite{lecture} for
more advanced topics. It turns out that in the framework of loop
quantum gravity all theoretical inconsistencies introduced in the
previous section are likely to be cured. More precisely, one will
see that there is no UV divergence in quantum fields of matter if
they are coupled with gravity in the background independent
approach. Also recent works show that the singularities in general
relativity can be smeared out in symmetry-reduced models
\cite{bojowald}\cite{singular}\cite{boj5}. The crucial point is that
gravity and matter are coupled and consistently quantized
non-perturbatively so that the problems of classical gravity and
quantum matter inconsistency disappear.

\newpage

\section{Classical Framework}
\subsection{Lagrangian Formalism}
In order to canonically quantize classical gravity, a Hamiltonian
analysis has to be performed to obtain a canonical formalism of the
classical theory suitable to be represented on a Hilbert space. A
well known canonical formalism of general relativity is the ADM
formalism (geometrodynamics) derived from the Einstein--Hilbert
action\cite{wald}\cite{Liang}, which has been problematic to cast
into a quantum theory rigorously. Another well-known action of
general relativity is the Palatini formalism, where the tetrad and
the connection are regarded as independent dynamical variables.
However, the dynamics of the Palatini action has similar
difficulties at the time of quantization as the dynamics derived
from the Einstein--Hilbert action \cite{Ash}\cite{han}. In 1986,
Ashtekar presented a formalism of true connection dynamics for
general relativity with a relatively simple Hamiltonian constraint,
and thus opened the door to apply quantization techniques from gauge
field theory \cite{Ash1}\cite{Ash2}\cite{RS}. However, a drawback of
that formalism is that the canonical variables are complex, which
need the implementation of complicated reality conditions if one is
to represent real general relativity. Moreover, the quantization
based on the complex connection could not be carried out rigorously,
since the internal gauge group is noncompact. In 1995, Barbero
modified the Ashtekar new variables to give a system of real
canonical variables for dynamical theory of connections
\cite{barbero}. Then Holst constructed a generalized Palatini action
to support Barbero's real connection dynamics \cite{holst}. Although
there is a free parameter (Barbero-Immirzi parameter $\beta$) in the
generalized Palatini action and the Hamiltonian constraint is more
complicated than the Ashtekar one, the generalized Palatini
Hamiltonian with the real connections is widely used by loop
theorists for the quantization programme. All the following analysis
is based on the generalized Palatini formalism.

Consider a 4-manifold, $M$, on which the basic dynamical variables
in the generalized Palatini framework are a tetrad $e_I^{\alpha}$
and an $so(1,3)$-valued connection $\omega_{\alpha}^{\ IJ}$(not
necessarily torsion-free), where the capital Latin indices $I,\
J,...$ refer to the internal $SO(1,3)$ group and the Greek indices
$\alpha,\ \beta,...$ denote spacetime indices. A tensor with both
spacetime indices and internal indices is named as a generalized
tensor. The internal space is equipped with a Minkowskian metric
$\eta_{IJ}$ (of signature $-, +, +, +$) fixed once for all, such
that the spacetime metric reads:
\begin{displaymath}
g_{\alpha\beta}=\eta_{IJ}e^{I}_{\alpha}e^{J}_{\beta}.
\end{displaymath}
The generalized Palatini action in which we are interested is
given by \cite{AL}:
\begin{eqnarray}
S_{p}[e_{K}^{\beta},\omega_{\alpha}^{\ IJ}]
=\frac{1}{2\kappa}\int_{M}d^4x(e)
e_{I}^{\alpha}e_{J}^{\beta}(\Omega_{\alpha\beta}^{\ \
IJ}+\frac{1}{2\beta}\epsilon^{IJ}_{\ \ KL}\Omega_{\alpha\beta}^{\ \
KL}) \label{action},
\end{eqnarray}
where $e$ is the square root of the determinant of the metric
$g_{\alpha\beta}$, $\epsilon^{IJ}_{\ \ KL}$ is the internal
Levi-Civita symbol, $\beta$ is the real Barbero-Immirzi parameter,
and the $so(1,3)$-valued curvature 2-form $\Omega_{\alpha\beta}^{\ \
IJ}$ of the connection $\omega_{\alpha}^{\ IJ}$ reads:
\begin{eqnarray}
\Omega_{\alpha\beta}^{\ \
IJ}:=2\mathcal{D}_{[\alpha}\omega_{\beta]}^{\ IJ} =
\partial_\alpha\omega^{\ IJ}_{\beta}-\partial_\beta\omega^{\ IJ}_{\alpha}+\omega_{\alpha}^{\
IK}\wedge\omega_{\beta K}^{\ \ \ J},\nonumber
\end{eqnarray}
here $\mathcal{D}_\alpha$ denote the $so(1,3)$ generalized covariant
derivative with respect to $\omega_{\alpha}^{\ IJ}$ acting on both
spacetime and internal indices. Note that the generalized Palatini
action returns to the Palatini action when $\frac{1}{\beta}=0$ and
gives the (anti)self-dual Ashtekar formalism when one sets
$\frac{1}{\beta}=\pm i$. Moreover, besides spacetime diffeomorphism
transformations, the action is also invariant under internal
$SO(1,3)$ rotations:
\begin{eqnarray}
(e,\omega)\mapsto({e'},{\omega'})=(b^{-1}{e},b^{-1}{\omega}b+b^{-1}{d}b),
\label{gauge}\nonumber
\end{eqnarray}
for any $SO(1,3)$ valued function $b$ on $M$. The gravitational
field equations are obtained by varying this action with respect
to $e_{I}^{\alpha}$ and $\omega_{\alpha}^{\ IJ}$. We first study
the variation with respect to the connection $\omega_{\alpha}^{\
IJ}$. One has
\begin{eqnarray}
\delta\Omega_{\alpha\beta}^{\ \ IJ}=(d\ \delta\omega^{\
IJ})_{\alpha\beta}+\delta\omega_{\alpha}^{\ IK}\wedge\omega_{\beta
K}^{\ \ \ J}+\omega_{\alpha}^{\ IK}\wedge\delta\omega_{\beta K}^{\
\ \ J}=2\mathcal{D}_{[\alpha}\delta\omega_{\beta]}^{\ IJ}\nonumber
\end{eqnarray}
by the definition of covariant generalized derivative
$\mathcal{D}_\alpha$. Note that $\delta\omega_\alpha^{\ IJ}$ is a
Lorentz covariant generalized tensor field since it is the
difference between two Lorentz connections
\cite{peldan}\cite{nakahara}. One thus obtains
\begin{eqnarray}
\delta S_{p}&=&\frac{1}{2\kappa}\int_{M}d^4x(e)
e_{I}^{\alpha}e_{J}^{\beta}(\delta\Omega_{\alpha\beta}^{\ \
IJ}+\frac{1}{2\beta}\epsilon^{IJ}_{\ \
KL}\delta\Omega_{\alpha\beta}^{\ \ KL})\nonumber\\
&=&-\frac{1}{\kappa}\int_M(\delta\omega_\beta^{\ \
IJ}+\frac{1}{2\beta}\epsilon^{IJ}_{\ \ KL}\delta\omega_{\beta}^{\ \
KL})\mathcal{D}_\alpha[(e)e_{I}^{\alpha}e_{J}^{\beta}],\nonumber
\end{eqnarray}
where we have used the fact that $\mathcal{D}_\alpha
\widetilde{\lambda}^\alpha=\partial_\alpha\widetilde{\lambda}^\alpha$
for all vector density $\widetilde{\lambda}^\alpha$ of weight $+1$
and neglected the surface term. Then it gives the equation of
motion:
\begin{eqnarray}
\mathcal{D}_\alpha[(e)e_{I}^{\alpha}e_{J}^{\beta}]
=-\frac{1}{4}\mathcal{D}_\alpha[\widetilde{\eta}^{\alpha\beta\gamma\delta}\epsilon_{IJKL}e^K_\gamma
e^L_\delta]=0,\nonumber
\end{eqnarray}
where $\widetilde{\eta}^{\alpha\beta\gamma\delta}$ is the
spacetime Levi-Civita symbol. This equation leads to the
torsion-free Cartan's first equation:
\begin{eqnarray}
\mathcal{D}_{[\alpha} e^I_{\beta]}=0,\nonumber
\end{eqnarray}
which means that the connection $\omega_\alpha^{\ IJ}$ is the
unique torsion-free Levi-Civita spin connection compatible with
the tetrad $e^\alpha_I$. As a result, the second term in the
action (\ref{action}) can be calculated as:
\begin{eqnarray}
(e)e_{I}^{\alpha}e_{J}^{\beta}\epsilon^{IJKL}\Omega_{\alpha\beta
KL}=\eta^{\alpha\beta\gamma\delta}R_{\alpha\beta\gamma\delta},\nonumber
\end{eqnarray}
which is exactly vanishing, because of the symmetric properties of
Riemann tensor. So the generalized Palatini action returns to the
Palatini action, which will certainly give the Einstein field
equation.
\subsection{Hamiltonian Formalism}
To carry out the Hamiltonian analysis of action (\ref{action}),
suppose the spacetime $M$ is topologically
$\Sigma\times\mathbf{R}$ for some 3-dimensional compact manifold
$\Sigma$ without boundary. We introduce a foliation parameterized
by a smooth function $t$ and a time-evolution vector field
$t^\alpha$ such that $t^\alpha (dt)_\alpha=1$ in $M$, where
$t^\alpha$ can be decomposed with respect to the unit normal
vector $n^\alpha$ of $\Sigma$ as:
\begin{equation}
t^\alpha=Nn^\alpha+N^\alpha, \nonumber
\end{equation}
here $N$ is called the lapse function and $N^\alpha$ the shift
vector \cite{wald}\cite{Liang}. The internal normal vector is
defined as $n_I\equiv n_\alpha e^\alpha_I$. It is convenient to
carry out a partial gauge fixing, i.e., fix a internal constant
vector field $n^I$ with $\eta_{IJ}n^In^J=-1$. Note that the gauge
fixing puts no restriction on the real dynamics\footnote{However,
there are some arguments that such a gauge fixing is a non-natural
way to break the internal Lorentz symmetry (see e.g.
\cite{samuel}).}. Then the internal vector space $V$ is 3+1
decomposed with a 3-dimensional subspace $W$ orthogonal to $n^I$,
which will be the internal space on $\Sigma$. With respect to the
internal normal $n^I$ and spacetime normal $n^\alpha$, the internal
and spacetime projection maps are denoted by $q_i^I$ and
$q_a^\alpha$ respectively, where we use $i,j,k,...$ to denote the
3-dimensional internal space indices and $a,b,c,...$ to denote the
indices of space $\Sigma$. Then an internal reduced metric
$\delta_{ij}$ and a reduced spatial metric on $\Sigma$, $q_{ab}$,
are obtained by these two projection maps. The two metrics are
related by:
\begin{equation}
q_{ab}=\delta_{ij}e^i_ae^j_b, \nonumber
\end{equation}
where the orthonormal co-triad on $\Sigma$ is defined by
$e^i_a:=e^I_\alpha q^i_Iq^\alpha_a$. Now the internal gauge group
$SO(1,3)$ is reduced to its subgroup $SO(3)$ which leaves $n^I$
invariant. Finally, two Levi-Civita symbols are obtained
respectively as
\begin{eqnarray}
\epsilon_{ijk}&:=&q_i^Iq_j^Jq_k^Kn^L\epsilon_{LIJK},\nonumber\\
\underline{\eta}_{abc}&:=&q_a^\alpha q_b^\beta q_c^\gamma
t^\mu\underline{\eta}_{\mu\alpha\beta\gamma}, \nonumber
\end{eqnarray}
where the internal Levi-Civita symbol $\epsilon_{ijk}$ is an
isomorphism of Lie algebra $so(3)$. Using the connection 1-form
$\omega_\alpha^{\ IJ}$, one can defined two $so(3)$-valued 1-form
on $\Sigma$:
\begin{eqnarray}
\Gamma_a^i&:=&\frac{1}{2}q^\alpha_aq^i_I\epsilon^{IJ}_{\ \ KL}n_J\omega_\alpha^{\ KL},\nonumber\\
K_a^i&:=&q^i_Iq^\alpha_a\omega_\alpha^{\ IJ}n_J, \nonumber
\end{eqnarray}
where $\Gamma$ is a spin connection on $\Sigma$ and $K$ will be
related to the extrinsic curvature of $\Sigma$ on shell. After the
3+1 decomposition and the Legendre transformation, action
(\ref{action}) can be expressed as \cite{holst}:
\begin{eqnarray}
S_{p}=\int_\mathbf{R}dt\int_{\Sigma}d^3x[\widetilde{P}^a_i\mathcal{L}_tA^i_a-\mathcal{H}_{tot}
(A^i_a,\widetilde{P}^b_j,\Lambda^i,N,N^c)] \label{action2},
\end{eqnarray}
from which the symplectic structure on the classical phase space
is obtained as
\begin{eqnarray}
\{A^i_a(x),\widetilde{P}^b_j(y)\}:=\delta^i_j\delta^a_b\delta^3(x,y),
\label{symplectic}
\end{eqnarray}
where the configuration and conjugate momentum are defined
respectively by:
\begin{eqnarray}
A_a^i&:=&\Gamma^i_a+\beta K^i_a,\nonumber\\
\widetilde{P}^a_i&:=&\frac{1}{2\kappa\beta}\widetilde{\eta}^{abc}\epsilon_{ijk}e^j_be^k_c
\ =\ \frac{1}{\kappa\beta}\sqrt{|\det q|}e^a_i,\nonumber
\end{eqnarray}
here $\det q$ is the determinant of the 3-metric $q_{ab}$ on
$\Sigma$ and hence $\det q=(\kappa\beta)^3 \det P$. In the
definition of the configuration variable $A^i_a$, we should
emphasize that $\Gamma_a^i$ is restricted to be the unique torsion
free $so(3)$-valued spin connection compatible with the triad
$e^a_i$. This conclusion is obtained by solving a second class
constraint in the Hamiltonian analysis \cite{holst}. In the
Hamiltonian formalism, one starts with the fields
$(A_a^i,\widetilde{P}^a_i)$. Then neither the basic dynamical
variables nor their Poisson brackets depend on the Barbero-Immirzi
parameter $\beta$. Hence, for the case of pure gravitational field,
the dynamical theories with different $\beta$ are related by a
canonical transformation. However, as we will see, the spectrum of
geometric operators are modified by different value of $\beta$, and
the non-perturbative calculation of black hole entropy is compatible
with Bekenstein-Hawking's formula only for a specific value of
$\beta$ \cite{lewandowski1}. In addition, it is argued that the
Barbero-Immerzi parameter $\beta$ may lead to observable effects in
principle when the gravitational field is coupled with Fermions
\cite{rovelli7}. In the decomposed action (\ref{action2}), the
Hamiltonian density $\mathcal{H}_{tot}$ is a linear combination of
constraints:
\begin{eqnarray}
\mathcal{H}_{tot}=\Lambda^iG_i+N^aC_a+NC,\nonumber
\end{eqnarray}
where $\Lambda^i\equiv-\frac{1}{2}\epsilon^i_{\ jk}\omega^{\ jk}_t$,
$N^a$ and $N$ are Lagrange multipliers. The three kinds of
constraints in the Hamiltonian are expressed as \cite{AL}:
\begin{eqnarray}
G_i&=&D_a\widetilde{P}^a_i\ :=\
\partial_a\widetilde{P}^a_i+\epsilon_{ij}^{\
\ k}A_a^j\widetilde{P}^a_k,\nonumber\\
C_a&=&\widetilde{P}^b_iF_{ab}^i-\frac{1+\beta^2}{\beta}K^i_aG_i,\nonumber\\
C&=&\frac{\kappa\beta^2}{2\sqrt{|\det
q|}}\widetilde{P}^a_i\widetilde{P}^b_j[\epsilon^{ij}_{\ \
k}F^k_{ab}-2(1+\beta^2)K^i_{[a}K^j_{b]}]\nonumber\\
&+&\kappa(1+\beta^2)\partial_a\big(\frac{\widetilde{P}^a_i}{\sqrt{|\det
q|}}\big)G^i,\label{constraint}
\end{eqnarray}
where the configuration variable $A_a^i$ performs as a
$so(3)$-valued connection on $\Sigma$ and $F_{ab}^i$ is the
$so(3)$-valued curvature 2-form of $A_a^i$ with the well-known
expression:
\begin{eqnarray}
F_{ab}^i:=2D_{[a}A^i_{b]}=\partial_aA^i_b-\partial_bA^i_a+\epsilon^i_{\
jk}A^j_aA^k_b.\nonumber
\end{eqnarray}
In any dynamical system with constraints, the constraint analysis is
essentially important because they reflect the gauge invariance of
the system. From the above three kinds of constraints of general
relativity, one can know the gauge invariance of the theory. The
Gauss constraint $G_i=0$ has crucial importance in formulating the
general relativity into a dynamical theory of connections. The
corresponding smeared constraint function,
$\mathcal{G}(\Lambda):=\int_\Sigma d^3x\Lambda^i(x)G_i(x)$,
generates a transformation on the phase space as:
\begin{eqnarray}
\{A^i_a(x),\ \mathcal{G}(\Lambda)\}&=&-D_a\Lambda^i(x)\nonumber\\
\{\widetilde{P}^a_i(x), \
\mathcal{G}(\Lambda)\}&=&\epsilon_{ij}^{\ \
k}\Lambda^j(x)\widetilde{P}^a_k(x),\nonumber
\end{eqnarray}
which are just the infinitesimal versions of the following gauge
transformation for the $so(3)$-valued connection 1-form
$\textbf{A}$ and internal rotation for the $so(3)$-valued
densitized vector field $\widetilde{\textbf{P}}$ respectively:
\begin{eqnarray}
(\textbf{A}_a,\
\widetilde{\textbf{P}}^b)\mapsto(g^{-1}\textbf{A}_ag+g^{-1}(d
g)_a,\ g^{-1}\widetilde{\textbf{P}}^bg).\nonumber
\end{eqnarray}
To display the meaning of the vector constraint $C_a=0$, one may
consider the smeared constraint function:
\begin{eqnarray}
\mathcal{V}(\vec{N}):=\int_\Sigma
d^3x(N^a\widetilde{P}^b_iF^i_{ab}-(N^aA^i_a)G_i).\nonumber
\end{eqnarray}
It generates the infinitesimal spatial diffeomorphism by the
vector field $N^a$ on $\Sigma$ as:
\begin{eqnarray}
\{A^i_a(x),\ \mathcal{V}(\vec{N})\}=\mathcal{L}_{\vec{N}}A^i_a(x),\nonumber\\
\{\widetilde{P}^a_i(x),\
\mathcal{V}(\vec{N})\}=\mathcal{L}_{\vec{N}}\widetilde{P}^a_i(x).\nonumber
\end{eqnarray}
The smeared scalar constraint is weakly equivalent to the
following function, which is re-expressed for quantization purpose
as
\begin{eqnarray}
\mathcal{S}(N)&:=&\int_\Sigma d^3x
N(x)\widetilde{C}(x)\nonumber\\
&=&\frac{\kappa\beta^2}{2}\int_\Sigma
d^3xN\frac{\widetilde{P}^a_i\widetilde{P}^b_j}{\sqrt{|\det
q|}}[\epsilon^{ij}_{\ \
k}F^k_{ab}-2(1+\beta^2)K^i_{[a}K^j_{b]}].\label{scalar}
\end{eqnarray}
It generates the infinitesimal time evolution off $\Sigma$. The
constraints algebra, i.e., the Poisson brackets between these
constraints, play a crucial role in the quantization programme. It
can be shown that the constraints algebra of (\ref{constraint})
has the following form:
\begin{eqnarray}
\{\mathcal{G}(\Lambda),\ \mathcal{G}(\Lambda')\}&=&\mathcal{G}([\Lambda,\ \Lambda']),\nonumber\\
\{\mathcal{G}(\Lambda),\ \mathcal{V}(\vec{N})\}&=&-\mathcal{G}(\mathcal{L}_{\vec{N}}\Lambda),\nonumber\\
\{\mathcal{G}(\Lambda),\ \mathcal{S}(N)\}&=&0,\nonumber\\
\{\mathcal{V}(\vec{N}),\ \mathcal{V}(\vec{N}')\}&=&\mathcal{V}([\vec{N},\ \vec{N}']),\nonumber\\
\{\mathcal{V}(\vec{N}),\ \mathcal{S}(M)\}&=&-\mathcal{S}(\mathcal{L}_{\vec{N}}M),\nonumber\\
\{\mathcal{S}(N),\ \mathcal{S}(M)\}&=&-\mathcal{V}((N\partial_bM-M\partial_bN)q^{ab})\nonumber\\
&&-\mathcal{G}((N\partial_bM-M\partial_bN)q^{ab}A_a))\nonumber\\
&&-(1+\beta^2)\mathcal{G}(\frac{[\widetilde{P}^a\partial_aN,\widetilde{P}^b\partial_bM]}{|\det
q|}),\label{constraint algebra}
\end{eqnarray}
where $|\det
q|q^{ab}=\kappa^2\beta^2\widetilde{P}^a_i\widetilde{P}^b_j\delta^{ij}$.
Hence the constraints algebra is closed under the Poisson brackets,
i.e., the constraints are all first class. Note that the evolution
of constraints is consistent since the Hamiltonian $H=\int_\Sigma
d^3x \mathcal{H}_{tot}$ is a linear combination of the constraints
functions. The evolution equations of the basic canonical pair read
\begin{eqnarray}
\mathcal{L}_tA^i_a=\{A^i_a,\ H\},\ \ \ \ \ \
\mathcal{L}_t\widetilde{P}^a_i=\{\widetilde{P}^a_i,\ H\}.\nonumber
\end{eqnarray}
Together with the three constraint equations, they are completely
equivalent to the Einstein field equations. Thus general relativity
is cast as a dynamical theory of connections with a compact
structure group. Before finishing the discussion of this section,
several issues should be emphasized.
\begin{itemize}
\item {{Canonical Transformation Viewpoint}}

The above construction can be reformulated in the language of
canonical transformations, since the phase space of connection
dynamics is the same as that of triad geometrodynamics. In the triad
formalism the basic conjugate pair consists of densitized triad
$\widetilde{E}^a_i=\beta\widetilde{P}^a_i$ and "extrinsic curvature"
$K^i_a$. The Hamiltonian and constraints read
\begin{eqnarray}
\mathcal{H}_{tot}&=&\Lambda^iG'_{i}+N^aC_a+NC\nonumber\\
G'_{i}&=&\epsilon_{ij}^{\ \ k}K_a^j\widetilde{E}^a_{k},\label{g}\\
C_a&=&\widetilde{E}^b_j\nabla_{[a}K_{b]}^j,\\
C&=&\frac{1}{\sqrt{|\det q|}}[\frac{1}{2}|\det
q|R+\widetilde{E}_i^{[a}\widetilde{E}_j^{b]}K^i_aK^j_b],
\end{eqnarray}
where $\nabla_a$ is the $SO(3)$ generalized derivative operator
compatible with triad $e^a_i$ and $R$ is the scalar curvature with
respect to it. Since $\widetilde{E}^a_i$ is a vector density of
weight one, we have
\begin{eqnarray}
\nabla_a\widetilde{E}^a_i=\partial_a\widetilde{E}^a_i+\epsilon_{ij}^{\
\ k}\Gamma_a^j\widetilde{E}_{k}^{a}=0.\nonumber
\end{eqnarray}
One can construct the desired Gauss law by
\begin{eqnarray}
G_i&:=&\frac{1}{\beta}\nabla_a\widetilde{E}^a_i+
G'_i,\nonumber\\
&=&\partial_a\widetilde{P}^a_i+\epsilon_{ij}^{\ \
k}(\Gamma_a^j+\beta K_a^j)\widetilde{P}_{k}^{a},\nonumber
\end{eqnarray}
which is weakly zero by construction. This motivates us to define
the connection $A^a_i=\Gamma_a^i+\beta K_a^i$. Moreover, the
transformation from the pair $(\widetilde{E}^a_i,K_b^j)$ to
$(\widetilde{P}^a_i,A^j_b)$ can be proved to be a canonical
transformation \cite{barbero}\cite{thiemann2}, i.e., the Poisson
algebra of the basic dynamical variables is preserved under the
transformation:
\begin{eqnarray}
\widetilde{E}^a_i&\mapsto&\widetilde{P}^a_i=\widetilde{E}^a_i/\beta\nonumber\\
K_b^j&\mapsto& A^j_b=\Gamma_b^j+\beta K_b^j,\nonumber
\end{eqnarray}
as
\begin{eqnarray}
\{\widetilde{P}^a_i(x),\ A^j_b(y)\}\ =&\{\widetilde{E}^a_i(x),K^j_b(y)\}\ =
&\delta^a_b\delta^j_i\delta(x,y),\nonumber\\
\{A^i_a(x),\ A^j_b(y)\}\ =&\{K^i_a(x),K^j_b(y)\}\ =&0,\nonumber\\
\{\widetilde{P}^a_i(x),\ \widetilde{P}^b_j(y)\}\
=&\{\widetilde{E}^a_i(x),\widetilde{E}^b_j(y)\}\ =&0.\nonumber
\end{eqnarray}

\item {{The Preparation for Quantization}}

The advantage of a dynamical theory of connections is that it is
convenient to be quantized in a background independent fashion. In
the following procedure of quantization, the quantum algebra of the
elementary observables will be generated by \textit{holonomy}, i.e.,
connection smeared on a curve, and \textit{electric flux}, i.e., a
densitized triad smeared on a 2-surface. So no information about a
background geometry is needed to define the quantum algebra. In the
remainder of the thesis, in order to incorporate also spinors, we
will enlarge the internal gauge group to be $SU(2)$. This does not
damage the prior constructions because the Lie algebra of $SU(2)$ is
the same as that of $SO(3)$. Due to the well-known nice properties
of compact Lie group $SU(2)$, such as the Haar measure and
Peter-Weyl theorem, one can obtain the background independent
representation of the quantum algebra and the spin-network
decomposition of the kinematic Hilbert space.

\item {{Analysis on Constraint Algebra}}

The classical constraint algebra (\ref{constraint algebra}) is an
infinite dimensional Poisson algebra. Unfortunately, it is not a Lie
algebra since the Poisson bracket between two scalar constraints has
structure function depending on dynamical variables. This causes
problems when solving the constraints quantum mechanically. On the
other hand, one can see from Eq.(\ref{constraint algebra}) that the
algebra generated by Gauss constraints forms not only a subalgebra
but also a 2-side ideal in the full constraint algebra. Thus one can
first solve the Gauss constraints independently. It is convenient to
find the quotient algebra with respect to the Gauss constraint
subalgebra as
\begin{eqnarray}
\{\mathcal{V}(\vec{N}),\ \mathcal{V}(\vec{N}')\}&=&\mathcal{V}([\vec{N},\vec{N}']),\nonumber\\
\{\mathcal{V}(\vec{N}),\ \mathcal{S}(M)\}&=&-\mathcal{S}(\mathcal{L}_{\vec{N}}M),\nonumber\\
\{\mathcal{S}(N),\
\mathcal{S}(M)\}&=&-\mathcal{V}((N\partial_bM-M\partial_bN)q^{ab}),\nonumber
\end{eqnarray}
which plays a crucial role in solving the constraints quantum
mechanically. But the subalgebra generated by the diffeomorphism
constraints can not form an ideal. Hence the procedures of solving
the diffeomorphism constraints and solving Hamiltonian constraints
are entangled with each other. This leads to certain ambiguity in
the construction of a Hamiltonian constraint operator
\cite{thiemann1}\cite{thiemann3}. Fortunately, the \textbf{Master
Constraint Project} addresses the above two problems as a whole by
introducing a new classical constraint algebra \cite{thiemann3}. The
new algebra is a Lie algebra where the diffeomorphism constraints
form a 2-side ideal. We will come back to this point in the
discussion on quantum dynamics of loop quantum gravity.

\end{itemize}

\newpage

\section{Foundations of Loop Quantum Gravity}

In this chapter, we will begin to quantize the above classical
dynamics of connections as a background independent quantum field
theory. The main purpose of the chapter is to construct a suitable
kinematical Hilbert space $\mathcal{H}_{kin}$ for the representation
of quantum observables. In the following discussion, we formulate
the construction in the language of algebraic quantum field theory
\cite{haag}. It should be emphasized that the following
constructions can be generalized to all background independent
non-perturbative gauge field theories with compact gauge groups.

\subsection{General Programme for Algebraic Quantization}

In the strategy of loop quantum gravity, a canonical programme is
performed to quantize general relativity, which has been cast into a
diffeomorphism invariant gauge field theory, or more generally, a
dynamical system with constraints. The following is a summary for a
general procedure to quantize a dynamical system with first class
constraints \cite{thiemannlecture}\cite{Ash}.
\begin{itemize}
\item { Algebra of Classical Elementary Observables}

One starts with the classical phase space $(\mathcal{M}, \{,\})$ and
$R$ ($R$ can be countable infinity\footnote{This includes the case
of field theory with infinite many degree of freedom, since one can
introduce the expression
$C_{n,\mu}=\int_{\Sigma}d^3x\phi_n(x)C_\mu(x)$, where
$\{\phi_n(x)\}^\infty_{n=1}$ forms a system of basis in
$L^2(\Sigma,d^3x)$.} ) first-class constraints $C_r(r=1...R)$ such
that $\{C_r , C_s\}=\Sigma_{t=1}^R f_{rs}^{\ \ t}C_t$, where
$f_{rs}^{\ \ t}$ is generally a function on phase space, namely, the
structure function of Poisson algebra. The algebra of classical
elementary observables $\mathfrak{P}$ is
defined as:\\ \\
\textbf{Definition 3.1.1}: \textit{The algebra of classical
elementary observables $\mathfrak{P}$ is a collection of functions
$f(m),m\in\mathcal{M}$ on the
phase space such that \\
$(1)$ $f(m)\in\mathfrak{P}$ should separate the points of
$\mathcal{M}$, i.e., for any $m\neq m'$, there exists
$f(m)\in\mathfrak{P}$, such that $f(m)\neq f(m')$;
$($analogy to the $p$ and $q$ in $\mathcal{M}=\mathrm{T}^*\mathbf{R}$.$)$\\
$(2)$ $f(m), f'(m)\in\mathfrak{P}\ \Rightarrow\ \{f(m),
f'(m)\}\in\mathfrak{P}$
$($closed under Poisson bracket$)$;\\
$(3)$ $f(m)\in\mathfrak{P}\ \Rightarrow\ \bar{f}(m)\in\mathfrak{P}$
$($closed under complex conjugation$)$. }\\ \\
So $\mathfrak{P}$ forms a sub $*$-Poisson algebra of
$C^\infty(\mathcal{M})$. In the case of
$\mathcal{M}=\mathrm{T}^*\mathbf{R}$, $\mathfrak{P}$ is generated by
the conjugate pair $(q, p)$ with $\{q,p\}=1$.

\item { Quantum Algebra of Elementary Observables}

Given the algebra of classical elementary observables
$\mathfrak{P}$, a quantum algebra of elementary observables can be
constructed as follows. Consider the formal finite sequences of
classical observable $(f_1...f_n)$ with $f_k\in\mathfrak{P}$. Then
the operations of multiplication and involution are defined as
\begin{eqnarray}
(f_1,...,f_n)\cdot(f'_1,...,f_m')&:=&(f_1,...,f_n,f_1',...,f_m'),\nonumber\\
(f_1,..,f_n)^*&:=&(\bar{f}_n,...,\bar{f}_1).\nonumber
\end{eqnarray}
One can define the formal sum of different sequences with different
number of elements. Then the general element of the newly
constructed free $*$-algebra $F(\mathfrak{P})$ of $\mathfrak{P}$, is
formally expressed as $\sum^N_{k=1}(f^{(k)}_1,...f^{(k)}_{n_k})$,
where $f^{(i)}_{n_i}\in\mathfrak{P}$. Consider the elements of the
form (sequences consisting of only one element)
\begin{eqnarray}
(f+f')-(f)-(f'),\ \ (zf)-z(f),\ \
[(f),(f')]-i\hbar(\{f,f'\}),\nonumber
\end{eqnarray}
where $z\in \mathbf{C}$ is a complex number, and the canonical
commutation bracket is defined as
\begin{eqnarray}
[(f),(f')]:=(f)\cdot(f')-(f')\cdot(f).\nonumber
\end{eqnarray}
A 2-side ideal $\mathfrak{I}$ of $F(\mathfrak{P})$ can be
constructed from these element, and is preserved by the action of
involution $*$. One thus obtains\\ \\
\textbf{Definition 3.1.2}: \textit{The quantum algebra
$\mathfrak{A}$ of elementary observables is defined to be the
quotient $*$-algebra $F(\mathfrak{P})/\mathfrak{I}$.}\\ \\
Note that the motivation to construct a quantum algebra of
elementary observables is to avoid the problem of operators ordering
in quantum theory so that the quantum algebra $\mathfrak{A}$ can be
represented on a Hilbert space without ordering ambiguities.

\item { Representation of Quantum Algebra}

In order to obtain a quantum theory, we need to quantize the
classical observables in the dynamical system. The, so called,
quantization is nothing but a $*$-representation map\footnote{A map
$\pi$: $\mathfrak{A}\rightarrow\mathcal{L}(\mathcal{H})$ is a
*-representation if and only if (1) there exists a dense subspace
$\mathcal{D}$ of $\mathcal{H}$ contained in
$\cap_{a\in\mathfrak{A}}[D(\pi(a))\cap D(\pi(a^*))]$ where
$D(\pi(a))$ is the domain of the operator $\pi(a)$ and (2) for every
$a,b\in\mathfrak{A}$ and $\lambda\in\mathbf{C}$ the following
conditions are satisfied in $\mathcal{D}$,
\begin{eqnarray}
\pi(a+b)=\pi(a)+\pi(b),&&\ \ \pi(\lambda a)=\lambda\pi(a),\nonumber\\
\pi(a\cdot b)=\pi(a)\pi(b),&&\ \ \pi(a^*)=\pi(a)^\dagger.\nonumber
\end{eqnarray}
Note that $\mathcal{L}(\mathcal{H})$ fails to be an algebra because
the domains of unbounded operators cannot be the whole Hilbert
space. However, the collection of bounded operators on any Hilbert
space is really a $*$-algebra.} $\pi$ from the quantum algebra of
elementary observable $\mathfrak{A}$ to the collection of linear
operators $\mathcal{L}(\mathcal{H})$ on a Hilbert Space
$\mathcal{H}$. At the level of quantum mechanics, the well-known
Stone-Von Neumann Theorem concludes that in quantum mechanics, there
is only one strongly continuous, irreducible, unitary representation
of the Weyl algebra, up to unitary equivalence (see, for example,
Ref.\cite{simon}). However, the conclusion of Stone-Von Neumann
cannot be generalized to the quantum field theory because the latter
has infinite many degrees of freedom (for detail, see, for example
\cite{wald1}). In quantum field theory, a representation can be
constructed by GNS(Gel'fand-Naimark-Segal)-construction for a
quantum algebra of elementary observable $\mathfrak{A}$, which is a
unital $*$-algebra actually. The GNS-construction for the
representation of quantum algebra $\mathfrak{A}$ is briefly
summarized as follows.\\ \\
\textbf{Definition 3.1.3}: \textit{Given a positive linear
functional $($a state$)$ $\omega$ on $\mathfrak{A}$, the null space
$\mathfrak{N}_{\omega}\in\mathfrak{A}$ with respect to $\omega$ is
defined as
$\mathfrak{N}_{\omega}:=\{a\in\mathfrak{A}|\omega(a^*\cdot a)=0\}$,
which is a left ideal in $\mathfrak{A}$. Then a quotient map can be
defined as $[.]$:
$\mathfrak{A}\rightarrow\mathfrak{A}/\mathfrak{N}_{\omega}$;
$a\mapsto[a]:=\{a+b|b\in\mathfrak{N}_{\omega}\}$. The
GNS-representation for $\mathfrak{A}$ with respect to $\omega$ is a
$*$-representation map: $\pi_{\omega}$:
$\mathfrak{A}\rightarrow\mathcal{L}(\mathcal{H}_{\omega})$, where
$\mathcal{H}_{\omega}:=\langle\mathfrak{A}/\mathfrak{N}_{\omega}\rangle$
and $\langle.\rangle$ denotes the completion with respect to the
naturally equipped well-defined inner product
\[<[a]|[b]>_{\mathcal{H}_{\omega}}:=\omega(a^*\cdot b)\] on
$\mathcal{H}_{\omega}$. This representation map is defined by
\[\pi_\omega(a)[b]:=[a\cdot b],\ \forall\ a\in\mathfrak{A}\
\mathrm{and}\ [b]\in\mathcal{H}_{\omega},\] where $\pi_\omega(a)$ is
an unbounded operator in general. Moreover, GNS-representation is a
cyclic representation, i.e., $\exists\
\Omega_\omega\in\mathcal{H}_{\omega}$, such that
$\langle\{\pi(a)\Omega_\omega|a\in\mathfrak{A}\}\rangle=\mathcal{H}_{\omega}$
and $\Omega_\omega$ is called a cyclic vector in the representation
space. In fact $\Omega_\omega:=[1]$ is a cyclic vector in
$\mathcal{H}_{\omega}$ and
$\langle\{\pi_\omega(a)\Omega_\omega|a\in\mathfrak{A}\}\rangle=\mathcal{H}_{\omega}$.
As a result, the positive linear functional with which we begin can
be expressed as \[\omega(a)=<\Omega_\omega|
\pi_\omega(a)\Omega_\omega>_{\mathcal{H}_{\omega}}.\] Thus a
positive linear functional on $\mathfrak{A}$ is equivalent to a
cyclic representation of $\mathfrak{A}$, which is a triple
$(\mathcal{H}_{\omega},\pi_\omega, \Omega_\omega)$. Moreover, every
non-degenerate representation is an orthogonal direct sum of
cyclic representations $($ for proof, see $\cite{conway}$ $)$ .}\\

So the kinematical Hilbert space
$\mathcal{H}_{kin}=\mathcal{H}_\omega$ for the system with
constrains can be obtained by GNS-construction. In the case that
there are gauge symmetries in our dynamical system, supposing that
there is a gauge group $G$ acting on $\mathfrak{A}$ by automorphisms
$\alpha_g: \mathfrak{A}\rightarrow\mathfrak{A},\ \forall\ g\in G$,
it is preferred to construct a gauge invariant representation of
$\mathfrak{A}$. So we require the positive linear functional
$\omega$ on $\mathfrak{A}$ to be gauge invariant, i.e.,
$\omega\circ\alpha_g=\omega$. Then the group $G$ is represented on
the Hilbert space $\mathcal{H}_\omega$ as:
\begin{eqnarray}
U(g)\pi_\omega(a)\Omega_\omega\:=\pi_\omega(\alpha_g(a))\Omega_\omega,\nonumber
\end{eqnarray}
and such a representation is a unitary representation of $G$. In
loop quantum gravity, it is crucial to construct an internal gauge
invariant and diffeomorphism invariant representation for the
quantum algebra of elementary observables.

\item {Implementation of the Constraints}

In the Dirac quantization programme for a system with constraints,
the constraints should be quantized as some operators in a
kinematical Hilbert space $\mathcal{H}_{kin}$. One then solves them
at quantum level to get a physical Hilbert space
$\mathcal{H}_{phys}$, that is, to find a quantum analogy $\hat{C}_r$
of the classical constraint formula $C_r$ and to solve the general
solution of the equation $\hat{C}_r\Psi=0$. However, there are
several problems in the construction of the constraint operator
$\hat{C}_r$.
\begin{itemize}
\item[(i)] $C_r$ is in general not in $\mathfrak{P}$, so there is a
factor ordering ambiguity in quantizing $C_r$ to be an operator
$\hat{C}_r$.

\item[(ii)] In quantum field theory, there are
ultraviolet(UV) divergence problems in constructing operators.
However, the UV divergence can be avoided in the background
independent approach.

\item[(iii)] Sometimes, quantum anomalies
may appear when there are structure functions in the Poisson
algebra. Although classically we have $\{C_r , C_s\}=\Sigma_{t=1}^R
f_{rs}^{\ \ t}C_t,\ r,s,t=1,...,R$, where $f_{rs}^{\ \ t}$ is a
function on phase space, quantum mechanically it is possible that
$[\hat{C}_r , \hat{C}_s]\neq i\hbar\Sigma_{t=1}^R \hat{f}_{rs}^{\ \
t}\hat{C}_t$ due to the ordering ambiguity between $\hat{f}_{rs}^{\
\ t}$ and $\hat{C}_t$. If one sets $[\hat{C}_r ,
\hat{C}_s]=\frac{i\hbar}{2}\Sigma_{t=1}^R (\hat{f}_{rs}^{\ \
t}\hat{C}_t+\hat{C}_t\hat{f}_{rs}^{\ \ t})$, for $\Psi$ satisfying
$\hat{C}_r\Psi=0$, we have
\begin{eqnarray}
[\hat{C}_r , \hat{C}_s]\Psi=\frac{i\hbar}{2}\sum_{t=1}^R
\hat{C}_t\hat{f}_{rs}^{\ \
t}\Psi=\frac{i\hbar}{2}\sum_{t=1}^R[\hat{C}_t, \hat{f}_{rs}^{\ \
t}]\Psi.\label{anomaly}
\end{eqnarray}
However, $[\hat{C}_t, \hat{f}_{rs}^{\ \ t}]\Psi$ are not necessary
to equal to zero for all $r,s,t=1...R$. If not, the problem of
quantum anomaly appears and the new quantum constraints $[\hat{C}_t,
\hat{f}_{rs}^{\ \ t}]\Psi=0$ have to be imposed on physical quantum
states, since the classical Poisson brackets $\{C_r , C_s\}$ are
weakly equal to zero on the constraint surface
$\overline{\mathcal{M}}\subset\mathcal{M}$. Thus too many
constraints are imposed and the physical Hilbert space
$\mathcal{H}_{phys}$ would be too small. Therefore this is not a
satisfactory solution and one needs to find a way to avoid the
quantum anomalies.
\end{itemize}

\item { Solving the Constraints and Physical Hilbert Space}

In general the original Dirac quantization approach can not be
carried out directly, since there is usually no nontrivial
$\Psi\in\mathcal{H}_{kin}$ such that $\hat{C}_r\Psi=0$. This happens
when the constraint operator $\hat{C}_r$ has "generalized
eigenfunctions" rather than eigenfunctions. One then develops the
so-called Refined Algebraic Quantization Programme, where the
solutions of the quantum constraint can be found in the algebraic
dual space of a dense subset in $\mathcal{H}_{kin}$ (see e.g.
\cite{GM}). The quantum diffeomorphism constraint in loop quantum
gravity is solved in this way. Another interesting way to solve the
quantum constraints is the \textbf{Master Constraint Approach}
proposed by Thiemann recently \cite{thiemann3}, which seems
especially suited to deal with the particular feature of the
constraint algebra of general relativity. A master constraint is
defined as
$\textbf{M}:=\frac{1}{2}\Sigma_{r,s=1}^R\textbf{K}_{rs}C_s\bar{C}_r$
for some real positive matrix $\textbf{K}_{rs}$. Classically one has
$\textbf{M}=0$ if and only if $C_r=0$ for all $r=1...R$. So quantum
mechanically one may consider solving the \textbf{Master Equation}:
$\hat{\textbf{M}}\Psi=0$ to obtain the physical Hilbert space
$\mathcal{H}_{phys}$ instead of solving $\hat{C}_r\Psi=0,\ \forall\
r=1...R$. Because the master constraint $\textbf{M}$ is classically
positive, one has opportunities to implement it as a self-adjoint
operator on $\mathcal{H}_{kin}$. If it is indeed the case and
$\mathcal{H}_{kin}$ is separable, one can use the direct integral
representation of $\mathcal{H}_{kin}$ associated with the
self-adjoint operator $\hat{\textbf{M}}$ to obtain
$\mathcal{H}_{phys}$:
\begin{eqnarray}
\mathcal{H}_{kin}&\sim&\int_\mathbf{R}^\oplus
d\mu(\lambda)\mathcal{H}^\oplus_{\lambda}\nonumber,\\
<\Phi|\Psi>_{kin}&=&\int_\mathbf{R}d\mu(\lambda)<\Phi|\Psi>_{\mathcal{H}^\oplus_{\lambda}},\label{did}
\end{eqnarray}
where $\mu$ is a so-called spectral measure and
$\mathcal{H}^\oplus_{\lambda}$ is the (generalized) eigenspace of
$\hat{\textbf{M}}$ with the eigenvalue $\lambda$. The physical
Hilbert space is then formally obtained as
$\mathcal{H}_{phys}=\mathcal{H}^\oplus_{\lambda=0}$ with the induced
physical inner product $<\ |\
>_{\mathcal{H}^\oplus_{\lambda=0}}\ $\footnote{One need to be careful for such a formal prescription,
see the later discussion of master constraint or \cite{thiemann9}.
}. Now the issue of quantum anomaly is represented in terms of the
size of $\mathcal{H}_{phys}$ and the existence of sufficient numbers
of semi-classical states.

\item {Physical Observables}

We denote $\mathcal{M}$ as the original unconstrained phase space,
$\overline{\mathcal{M}}$ as the constraint surface, i.e.,
$\overline{\mathcal{M}}:=\{m\in\mathcal{M}|C_r(m)=0,\ \forall\
r=1...R\}$, and $\hat{\mathcal{M}}$ as the reduced phase space, i.e.
the space of orbits for gauge transformations generated by all
$C_r$. The concept of Dirac observable is defined as the
follows.\\ \\
\textbf{Definition 3.1.4}: \\
\textit{$(1)$ A function $\mathcal{O}$ on $\mathcal{M}$ is called a
weak Dirac observable if and only if the function depends only on
points of $\hat{\mathcal{M}}$, i.e.,
$\{\mathcal{O},C_r\}|_{\overline{\mathcal{M}}}=0$ for all $r=1...R$.
For the quantum version, a self-adjoint operator $\hat{\mathcal{O}}$
is a weak Dirac observable
if and only if the operator can be well defined on the physical Hilbert space.\\
$(2)$ A function $\mathcal{O}$ on $\mathcal{M}$ is called a strong
Dirac observable if and only if
$\{\mathcal{O},C_r\}|_{\mathcal{M}}=0$ for all $r=1...R$. For the
quantum version, a self-adjoint operator $\hat{\mathcal{O}}$ is a
strong Dirac observable if and only if the operator can be defined
on the kinematic Hilbert space $\mathcal{H}_{kin}$ and
$[\hat{\mathcal{O}},\hat{C}_r]=0$ in $\mathcal{H}_{kin}$ for all
$r=1...R$.}\\ \\
A physical observable is at least a weak Dirac observable. While
Dirac observables have been found explicitly in symmetry reduced
models, some even with an infinite number of degrees of freedom, it
seems extremely difficult to find explicit expressions for them in
full general relativity. Moreover the Hamiltonian is a linear
combination of first-class constraints. So there is no dynamics in
the reduced phase space, and the meaning of time evolution of the
Dirac observables becomes subtle. However, using the concepts of
partial and complete observables
\cite{rovelli1}\cite{rovelli2}\cite{rovelli}, a systematic method to
get Dirac observables can be developed, and the problem of time in
such system with a Hamiltonian $H=\Sigma_{r=1}^R\beta_rC_r$ may also
be solved.

Classically, let $f(m)$ and $\{T_r(m)\}_{r=1}^R$ be gauge
non-invariant functions (partial observables) on phase space
$\mathcal{M}$, such that $A_{sr}\equiv\{C_s,T_r\}$ is a
non-degenerate matrix on $\mathcal{M}$. A system of classical weak
Dirac observables (complete observables) $F_{f,T}^\tau$ labelled by
a collection of real parameters $\tau\equiv\{\tau_r\}_{r=1}^R$ can
be constructed as
\begin{eqnarray}
F_{f,T}^\tau:=\sum^\infty_{\{n_1\cdot\cdot\cdot
n_R\}}\frac{(\tau_1-T_1)^{n_1}\cdot\cdot\cdot(\tau_R-T_R)^{n_R}}{n_1!\cdot\cdot\cdot
n_R!}
\widetilde{X}_1^{n_1}\circ\cdot\cdot\cdot\circ\widetilde{X}_R^{n_R}(f)\nonumber,
\end{eqnarray}

where $\ \widetilde{X}_r(f):=\{\Sigma_{s=1}^R A^{-1}_{rs}C_s,\
f\}\equiv\{\widetilde{C}_r,\ f\}\ $. It can be verified that $\
[\widetilde{X}_r,\ \widetilde{X}_s]|_{\overline{\mathcal{M}}}=0$ and
$\{F_{f,T}^\tau,C_r\}|_{\overline{\mathcal{M}}}=0$, for all
$r=1...R$ (for details see \cite{dittrich} and \cite{dittrich2}).

The partial observables $\{T_r(m)\}_{r=1}^R$ may be regarded as
clock variables, and $\tau_r$ is the time parameter for $T_r$. The
gauge is fixed by giving a system of functions $\{T_r(m)\}_{r=1}^R$
and corresponding parameters $\{\tau_r\}_{r=1}^R$, namely, a section
in $\overline{\mathcal{M}}$ is selected by $T_r(m)=\tau_r$ for each
$r$, and $F_{f,T}^\tau$ is the value of $f$ on the section. To solve
the problem of dynamics, one assumes another set of canonical
coordinates $(P_1,\cdot\cdot\cdot,
P_{N-R},\Pi_1,\cdot\cdot\cdot,\Pi_{R};Q_1,\cdot\cdot\cdot,Q_{N-R},
T_1,\cdot\cdot\cdot,T_R)$ by canonical transformations in the phase
space $(\mathcal{M},\{\ ,\ \})$, where $P_s$ and $\Pi_r$ are
conjugate to $Q_s$ and $T_r$ respectively. After solving the
complete system of constraints
$\{C_r(P_i,Q_j,\Pi_s,T_t)=0\}_{r=1}^R$, the Hamiltonian $H_r$ with
respect to the time $T_r$ is obtained as $H_r:=\Pi_r(P_i,Q_j,T_t)$.
Given a system of constants $\{(\tau_0)_r\}_{r=1}^R$, for an
observable $f(P_i,Q_j)$ depending only on $P_i$ and $Q_j$, the
physical dynamics is given by \cite{dittrich}\cite{thiemann14}:
\begin{eqnarray}
(\frac{\partial}{\partial\tau_r})_{\tau=\tau_0}F_{f,T}^\tau|_{\overline{\mathcal{M}}}=F_{\{H_r,f\},T}^{\tau_0}
|_{\overline{\mathcal{M}}}
=\{F_{H_r,T}^{\tau_0},F_{f,T}^{\tau_0}\}|_{\overline{\mathcal{M}}},
\nonumber
\end{eqnarray}
where $F_{H_r,T}^{\tau_0}$ is the physical Hamiltonian function
generating the evolution with respect to $\tau_r$. Thus one has
addressed the problem of time and dynamics as a result.

\item {Semi-classical Analysis}

An important issue in the quantization is to check whether the
quantum constraint operators have correct classical limits. This has
to be done by using the kinematical semiclassical states in
$\mathcal{H}_{kin}$. Moreover, the physical Hilbert space
$\mathcal{H}_{phys}$ must contain enough semi-classical states to
guarantee that the quantum theory one obtains can return to the
classical theory when $\hbar\rightarrow0$. The semi-classical states
in a Hilbert space $\mathcal{H}$ should have the following
properties.\\ \\
\textbf{Definition 3.1.5}:  \textit{Given a class of observables
$\mathcal{S}$ which comprises a subalgebra in the space
$\mathcal{L}(\mathcal{H})$ of linear operators on the Hilbert space,
a family of (pure) states $\{\omega_m\}_{m\in\mathcal{M}}$ are said
to be
semi-classical with respect to $\mathcal{S}$ if and only if:\\
$(1)$ The observables in $\mathcal{S}$ should have correct
semi-classical limit on semi-classical states and the fluctuations
should be small, i.e.,
\begin{eqnarray}
\lim_{\hbar\rightarrow0}|\frac{\omega_m(\hat{a})-a(m)}{a(m)}|=0,\nonumber\\
\lim_{\hbar\rightarrow0}|\frac{\omega_m(\hat{a}^2)-\omega_m(\hat{a})^2}{\omega_m(\hat{a})^2}|=0,\nonumber
\end{eqnarray}
for all $\hat{a}\in\mathcal{S}$.\\
$(2)$ The set of cyclic vectors $\Omega_m$ related to $\omega_m$ via
the $GNS$-representation
$(\pi_\omega,\mathcal{H}_\omega,\Omega_\omega)$ is dense in
$\mathcal{H}$.}\\

Seeking for semiclassical states are one of open issues of current
research in loop quantum gravity. Recent original works focus on the
construction of coherent states of loop quantum gravity in analogy
with the coherent states for harmonic oscillator system
\cite{thiemann10}\cite{thiemann11}\\
\cite{thiemann12}\cite{thiemann13}\cite{AL2}\cite{shadow}.

\end{itemize}
The above is the general programme to quantize a system with
constraints. In the following subsection, we will apply the
programme to the theory of general relativity and restrict our view
to the representation with the properties of background independence
and spatial diffeomorphism invariance.

\subsection{Quantum Configuration Space}

In quantum mechanics, the kinematical Hilbert space is
$L^2(\mathbf{R}^3,d^3x)$, where the simple $\mathbf{R}^3$ is the
classical configuration space of free particle which has finite
degrees of freedom, and $d^3x$ is the Lebesgue measure on
$\mathbf{R}^3$. In quantum field theory, it is expected that the
kinematical Hilbert space is also the $L^2$ space on the
configuration space of the field, which is infinite dimensional,
with respect to some Borel measure naturally defined. However, it is
often hard to define a concrete Borel measure on the classical
configuration space, since the integral theory on infinite
dimensional space is involved \cite{dewitt}. Thus the intuitive
expectation should be modified, and the concept of quantum
configuration space should be introduced as a suitable enlargement
of the classical configuration space so that an infinite dimensional
measure, often called cylindrical measure, can be well defined on
it. The example of a scalar field can be found in the references
\cite{AL}\cite{Ash3}. For quantum gravity, it should be emphasized
that the construction for quantum configuration space must be
background independent. Fortunately, general relativity has been
reformulated as a dynamical theory of $SU(2)$ connections, which
would be great helpful for our further development.

The classical configuration space for gravitational field, which
is denoted by $\mathcal{A}$, is a collection of the $su(2)$-valued
connection 1-form field smoothly distributed on $\Sigma$. The idea
of the construction for quantum configuration
is due to the concept of holonomy.\\ \\
\textbf{Definition 3.2.1}:  \textit{Given a smooth $SU(2)$
connection field $A_a^i$ and an analytic curve $c$ with the
parameter $t\in[0,1]$ supported on a compact subset $($compact
support $)$ of $\Sigma$, the corresponding holonomy is defined by
the solution of the parallel transport equation \cite{nakahara}
\begin{eqnarray}
\frac{d}{dt}A(c,t)=-[A_a^i\dot{c}^a\tau_i]A(c,t),\label{transport}
\end{eqnarray}
with the initial value $A(c,0)=1$, where $\dot{c}^a$ is the tangent
vector of the curve and $\tau_i\in su(2)$ constitute an orthonormal
basis with respect to the Killing-Cartan metric
$\eta(\xi,\zeta):=-2\mathrm{Tr}(\xi\zeta)$, which satisfy
$[\tau_i,\tau_j]=\epsilon^k_{\ ij}\tau_k$ and are fixed once for
all. Thus the holonomy is an element in $SU(2)$, which can be
expressed as
\begin{eqnarray}
A(c)=\mathcal{P}\exp\big(-\int_0^1[A_a^i\dot{c}^a\tau_i]\ dt
\big),\label{holonomy}
\end{eqnarray}
where $A(c)\in SU(2)$ and $\mathcal{P}$ is a path-ordering
operator along the curve $c$ (see the footnote at p382 in
\cite{nakahara}).}\\ \\
The definition can be well extended to the case of piecewise
analytic curves via the relation:
\begin{eqnarray}
A(c_1\circ c_2)=A(c_1)A(c_2),\label{1}
\end{eqnarray}
where $\circ$ stands for the composition of two curves. It is easy
to see that a holonomy is invariant under the re-parametrization
and is covariant under changing the orientation, i.e.,
\begin{eqnarray}
A(c^{-1})=A(c)^{-1}.\label{2}
\end{eqnarray}
So one can formulate the properties of holonomy in terms of the
concept of the equivalent classes of curves.\\ \\
\textbf{Definition 3.2.2}:  \textit{Two analytic curves $c$ and $c'$
are said to be equivalent if and only if they have the same source
$s(c)$ $($beginning point $)$ and the same target $t(c)$ $($end
point $)$, and the holonomies of the two curves are equal to each
other, i.e., $A(c)=A(c')$ $\forall A\in\mathcal{A}$. A equivalent
class of analytic curves is defined to be an edge, and
a piecewise analytic path is an composition of edges.}\\ \\
To summarize, the holonomy is actually defined on the set
$\mathcal{P}$ of piecewise analytic paths with compact supports.
The two properties (\ref{1}) and (\ref{2}) mean that each
connection in $\mathcal{A}$ is a homomorphism from $\mathcal{P}$,
which is so-called a groupoid by definition \cite{velh}, to our
compact gauge group $SU(2)$. Note that the internal gauge
transformation and spatial diffeomorphism act covariantly on a
holonomy as
\begin{eqnarray}
A(e)\mapsto g(t(e))^{-1}A(e)g(s(e))\ \ \ \mathrm{and} \ \ \
A(e)\mapsto A(\varphi\circ e),\label{trans}
\end{eqnarray}
for any $SU(2)$-valued function $g(x)$ on $\Sigma$ and spatial
diffeomorphism $\varphi$. All above discussion is for classical
smooth connections in $\mathcal{A}$. The quantum configuration
space for loop quantum gravity can be constructed by extending the
concept of holonomy, since its definition does not depend on an
extra background. One thus obtains the quantum configuration space
$\overline{\mathcal{A}}$ of loop quantum gravity as the following.\\ \\
\textbf{Definition 3.2.3}:  \textit{The quantum configuration space
$\overline{\mathcal{A}}$ is a collection of all quantum connections
$A$, which are algebraic homomorphism maps without any continuity
assumption from the collection of piecewise analytic paths with
compact supports, $\mathcal{P}$, on $\Sigma$ to the gauge group
$SU(2)$, i.e.,
$\overline{\mathcal{A}}:=\mathrm{Hom}(\mathcal{P},SU(2))$\footnote{It
is easy to see that the definition of $\overline{\mathcal{A}}$ does
not depend on the choice of local section in $SU(2)$-bundle, since
the internal gauge transformations leave $\overline{\mathcal{A}}$
invariant.}. Thus for any $A\in \overline{\mathcal{A}}$ and edge $e$
in $\mathcal{P}$,
\begin{eqnarray}
A(e_1\circ e_2)=A(e_1)A(e_2)\ \ \ \mathrm{and}\ \ \
A(e^{-1})=A(e)^{-1}.\nonumber
\end{eqnarray}
The transformations of quantum connections under internal gauge
transformations and diffeomorphisms are defined by Eq.(\ref{trans}).}\\ \\
The above discussion on the smooth connections shows that the
classical configuration space $\mathcal{A}$ can be understood as a
subset in the quantum configuration space $\overline{\mathcal{A}}$.
Moreover, the Giles theorem \cite{giles} shows precisely that a
smooth connection can be recovered from its holonomies by varying
the length and location of the paths.

On the other hand, it was shown in \cite{velh}\cite{thiemann2} that
the quantum configuration space $\overline{\mathcal{A}}$ can be
constructed via a projective limit technique and admits a natural
defined topology. To make the discussion precise, we begin with a
few definitions:\\ \\
\textbf{Definition 3.2.4}:
\textit{\begin{enumerate}
\item A finite set $\{e_1,...,e_N\}$ of edges is
said to be independent if the edges $e_i$ can only intersect each
other at their sources $s(e_i)$ or targets $t(e_i)$.
\item A finite graph is a collection of a finite set $\{e_1,...,e_N\}$ of independent edges and their
vertices, i.e. their sources $s(e_i)$ and targets $t(e_i)$. We
denote by $E(\gamma)$ and $V(\gamma)$ respectively as the sets of
independent edges and vertices of a given finite graph $\gamma$. And
$N_\gamma$ is the number of elements in $E(\gamma)$.
\item A subgroupoid $\alpha(\gamma)\subset\mathcal{P}$ can be generated
from $\gamma$ by identifying $V(\gamma)$ as the set of objects and
all $e\in E(\gamma)$ together with their inverses and finite
compositions as the set of homomorphisms. This kind of subgoupoid in
$\mathcal{P}$ is called tame subgroupoid. $\alpha(\gamma)$ is
independent of the orientation of $\gamma$, so the graph $\gamma$
can be recovered from tame subgroupoid $\alpha$ up to the
orientations on the edges. We will also denote by $N_\alpha$ the
number of elements in $E(\gamma)$ where $\gamma$ is recovered by the
tame subgroupoid $\alpha$.
\item $\mathcal{L}$ denotes the set of all tame subgroupoids in
$\mathcal{P}$.
\end{enumerate}}

One can equip a partial order relation $\prec$ on $\mathcal{L}$
\footnote{A partial order on $\mathcal{L}$ is a relation, which is
reflective ($\alpha\prec\alpha$), symmetric ($\alpha\prec\alpha',\
\alpha'\prec\alpha\Rightarrow\alpha'=\alpha$) and transitive
($\alpha\prec\alpha',\
\alpha'\prec\alpha''\Rightarrow\alpha'\prec\alpha''$). Note that not
all pairs in $\mathcal{L}$ need to have a relation.}, defined by
$\alpha\prec\alpha'$ if and only if $\alpha$ is a subgroupoid in
$\alpha'$. Obviously, for any two tame subgroupoids
$\alpha\equiv\alpha(\gamma)$ and $\alpha'\equiv\alpha(\gamma')$ in
$\mathcal{L}$, there exists
$\alpha''\equiv\alpha(\gamma'')\in\mathcal{L}$ such that
$\alpha,\alpha'\prec\alpha''$, where
$\gamma''\equiv\gamma\cup\gamma'$. Define $X_\alpha\equiv
Hom(\alpha, SU(2))$ as the set of all homomorphisms from the
subgroupoid $\alpha(\gamma)$ to the group $SU(2)$. Note that an
element $A_\alpha\in X_{\alpha}$ ($\alpha=\alpha(\gamma)$) is
completely determined by the $SU(2)$ group elements $A(e)$ where
$e\in E(\gamma)$, so that one has a bijection $\lambda:
X_{\alpha(\gamma)} \rightarrow SU(2)^{N_\gamma}$, which induces a
topology on $X_{\alpha(\gamma)}$ such that $\lambda$ is a
topological homomorphism. For any pair $\alpha\prec\alpha'$, one can
define a surjective projection map $P_{\alpha'\alpha}$ from
$X_{\alpha'}$ to $X_\alpha$ by restricting the domain of the map
$A_{\alpha'}$ from $\alpha'$ to the subgroupoid $\alpha$, and these
projections satisfy the consistency condition
$P_{\alpha'\alpha}\circ P_{\alpha''\alpha'}=P_{\alpha''\alpha}$.
Thus a projective family
$\{X_{\alpha},P_{\alpha'\alpha}\}_{\alpha\prec\alpha'}$ is obtained
by above constructions. Then the projective limit
$\lim_{\alpha}(X_{\alpha})$ is naturally
obtained.\\ \\
\textbf{Definition 3.2.5}:  \textit{The projective limit
$\lim_{\alpha}(X_{\alpha})$ of the projective family
$\{X_{\alpha},P_{\alpha'\alpha}\}_{\alpha\prec\alpha'}$ is a subset
of the direct product space
$X_{\infty}:=\prod_{\alpha\in\mathcal{L}}X_{\alpha}$ defined by
\begin{eqnarray}
\lim_{\alpha}(X_{\alpha}):=
\{\{A_{\alpha}\}_{\alpha\in\mathcal{L}}|P_{\alpha'\alpha}A_{\alpha'}=A_{\alpha},\
\forall\ \alpha\prec\alpha'\}.\nonumber
\end{eqnarray}}\\
Note that the projection $P_{\alpha'\alpha}$ is surjective and
continuous with respect to the topology of $X_{\alpha}$. One can
equip the direct product space $X_{\infty}$ with the so-called
Tychonov topology. Since any $X_{\alpha}$ is a compact Hausdorff
space, by Tychonov theorem $X_{\infty}$ is also a compact Hausdorff
space. One then can prove that the projective limit,
$\lim_{\alpha}(X_{\alpha})$, is a closed subset in $X_{\infty}$ and
hence a compact Hausdorff space with respect to the topology induced
from $X_{\infty}$. At last, one can find the relation between the
projective limit and the prior constructed quantum configuration
space $\overline{\mathcal{A}}$. As one might expect, there is a
bijection $\Phi$ between $\overline{\mathcal{A}}$ and
$\lim_{\alpha}(X_{\alpha})$ \cite{thiemann2}:
\begin{eqnarray}
\Phi:\ \
\overline{\mathcal{A}}&\rightarrow&\lim_{\alpha}(X_{\alpha});\nonumber\\
A&\mapsto&\{A|_{\alpha}\}_{\alpha\in\mathcal{L}},\nonumber\nonumber
\end{eqnarray}
where $A|_{\alpha}$ means the restriction of the domain of the map
$A\in\overline{\mathcal{A}}=\mathrm{Hom}(\mathcal{P},SU(2))$. As a
result, the quantum configuration space is identified with the
projective limit space and hence can be equipped with the topology.
In conclusion, the quantum configuration space
$\overline{\mathcal{A}}$ is constructed to be a compact Hausdorff
topological space.

\subsection{Cylindrical Functions on Quantum Configuration Space}

Given the projective family
$\{X_{\alpha},P_{\alpha'\alpha}\}_{\alpha\prec\alpha'}$, the
cylindrical function on its projective limit
$\overline{\mathcal{A}}$ is well defined as follows.\\ \\
\textbf{Definition 3.3.1}:  \textit{Let $C(X_{\alpha})$ be the set
of all continuous complex functions on $X_{\alpha}$, two functions
$f_{\alpha}\in C(X_{\alpha})$ and $f_{\alpha'}\in C(X_{\alpha'})$
are said to be equivalent or cylindrically consistent, denoted by
$f_{\alpha}\sim f_{\alpha'}$, if and only if
$P^*_{\alpha''\alpha}f_\alpha=P^*_{\alpha''\alpha'}f_{\alpha'}$,
$\forall\alpha''\succ\alpha,\alpha'$, where $P^*_{\alpha''\alpha}$
denotes the pullback map induced from $P_{\alpha''\alpha}$. Then the
space $Cyl(\overline{\mathcal{A}})$ of cylindrical functions on the
projective limit $\overline{\mathcal{A}}$ is defined to be the space
of equivalent classes $[f]$, i.e.,
\begin{eqnarray}
Cyl(\overline{\mathcal{A}}):=\big[\cup_\alpha
C(X_{\alpha})\big]/\sim.\nonumber
\end{eqnarray}}\\
One then can easily prove the following proposition by
definition.\\ \\
\textbf{Proposition 3.3.1}: \\
\textit{All continuous functions $f_{\alpha}$ on $X_{\alpha}$ are
automatically cylindrical since each of them can generate a
equivalent class $[f_{\alpha}]$ via the pullback map
$P^*_{\alpha'\alpha}$ for all $\alpha'\succ\alpha$, and the
dependence of $P^*_{\alpha'\alpha}f_{\alpha}$ on the groups
associated to the edges in $\alpha'$ but not in $\alpha$ is trivial,
i.e., by the definition of the pull back map,
\begin{eqnarray}
(P^*_{\alpha'\alpha}f_{\alpha})(A(e_1),...,A(e_{N_\alpha}),...,A(e_{N_{\alpha'}}))=
f_{\alpha}(A(e_1),...,A(e_{N_\alpha})),\label{func}
\end{eqnarray}
where $N_\alpha$ denotes the number of independent edges in the
graph recovered from the groupoid $\alpha$. On the other hand, by
definition, given a cylindrical function $f\in
Cyl(\overline{\mathcal{A}})$ there exists a suitable groupoid
$\alpha$ such that $f=[f_{\alpha}]$, so one can identify $f$ with
$f_\alpha$. Moreover, given two cylindrical functions $f,\ f'\in
Cyl(\overline{\mathcal{A}})$, by definition of cylindrical functions
and the property of projection map, there exists a common groupoid
$\alpha$ and $f_{\alpha},\ f'_{\alpha}\in C(X_{\alpha})$ such that
$f=[f_{\alpha}]$ and
$f'=[f'_{\alpha}]$.}\\ \\

Let $f$, $f'\in Cyl(\overline{\mathcal{A}})$, there exists graph
$\alpha$ such that $f=[f_{\alpha}]$, and $f'=[f'_{\alpha}]$, then
the following operations are well defined
\begin{eqnarray}
f+f':=[f_\alpha+f'_\alpha],\ ff':=[f_\alpha f'_\alpha],\
zf:=[zf_\alpha],\ \bar{f}:=[\bar{f}_\alpha],\nonumber
\end{eqnarray}
where $z\in\mathbf{C}$ and $\bar{f}$ denotes complex conjugate. So
we construct $Cyl(\overline{\mathcal{A}})$ as an Abelian
$*$-algebra. In addition, there is a unital element in the algebra
because $Cyl(\overline{\mathcal{A}})$ contains constant functions.
Moreover, we can well define the sup-norm for $f=[f_{\alpha}]$ by
\begin{eqnarray}
\|f\|:=\sup_{A_\alpha\in X_\alpha}|f_\alpha(A_\alpha)|,\label{norm}
\end{eqnarray}
which satisfies the $C^*$ property $\|f\bar{f}\|=\|f\|^2$. Then
$\overline{Cyl(\overline{\mathcal{A}})}$ is a unital Abelian
$C^*$-algebra, after the completion with respect to the norm.

From the theory of $C^*$-algebra, it is known that a unital Abelian
$C^*$-algebra is identical to the space of continuous functions on
its spectrum space via an isometric isomorphism, the so-called
Gel'fand transformation (see e.g. \cite{thiemann2}). So we have
the following theorem \cite{AL1}\cite{AL3}, which finishes this section.\\
\\
\textbf{Theorem 3.3.1}:  \\
\textit{$(1)$ The space $Cyl(\overline{\mathcal{A}})$ has a structure of a unital Abelian $C^*$-algebra after completion with respect to the sup-norm.\\
$(2)$ Quantum configuration space $\overline{\mathcal{A}}$ is the
spectrum space of completed $\overline{Cyl(\overline{\mathcal{A}})}$
such that $\overline{Cyl(\overline{\mathcal{A}})}$ is identical to
the space $C(\overline{\mathcal{A}})$ of continuous functions on
$\overline{\mathcal{A}}$. }

\subsection{Loop Quantum Kinematics}

In analogy with the quantization procedure of section 3.1, in this
subsection we would like to perform the background-independent
construction of algebraic quantum field theory for general
relativity. First we construct the algebra of classical observables.
Taking account of the future quantum analogs, we define the algebra
of classical observables $\mathfrak{P}$ as the Poission
$*$-subalgebra generated by the functions of holonomies (cylindrical
functions) and the fluxes of triad fields smeared on some 2-surface.
Namely, one can define the classical algebra in analogy with
geometric quantization in finite dimensional phase space case by the
so-called classical Ashtekar-Corichi-Zapata holonomy-flux
$*$-algebra as the following \cite{unique}.\\ \\
\textbf{Definition 3.4.1}\\
\textit{The classical Ashtekar-Corichi-Zapata holonomy-flux
$*$-algebra is defined to be a vector space
$\mathfrak{P}_{ACZ}:=Cyl(\overline{\mathcal{A}})\times\mathcal{V}^{\mathbf{C}}(\overline{\mathcal{A}})$,
where $\mathcal{V}^{\mathbf{C}}(\overline{\mathcal{A}})$ is the
vector space of algebraic vector fields spanned by the vector fields
$\psi Y_f(S)$ $\psi\in Cyl(\overline{\mathcal{A}})$, and their
commutators, here the smeared flux vector field $Y_f(S)$ is defined
by acting on any cylindrical function:
\begin{eqnarray}
Y_f(S)\psi:=\{\int_S\underline{\eta}_{abc}\widetilde{P}^c_if^i, \
\psi\},\nonumber
\end{eqnarray}
for any $su(2)$-valued function $f^i$ with compact supports on $S$
and $\psi$ are cylindrical functions on $\overline{\mathcal{A}}$. We
equip $\mathfrak{P}_{ACZ}$
with the structure of an $*$-Lie algebra by:\\
(1) Lie bracket $\{\ ,\ \}:\
\mathfrak{P}_{ACZ}\times\mathfrak{P}_{ACZ}\rightarrow\
\mathfrak{P}_{ACZ}$ is defined by
\begin{eqnarray}
\{(\psi,Y),\ (\psi',Y')\}:=(Y\circ\psi'-Y'\circ\psi,[Y,
Y']),\nonumber
\end{eqnarray}
for all $(\psi, Y),\ (\psi', Y')\in\mathfrak{P}_{ACZ}$ with $\psi,
\psi'\in Cyl(\overline{\mathcal{A}})$ and $Y,
Y'\in\mathcal{V}^{\mathbf{C}}(\overline{\mathcal{A}})$.\\ \\
(2) Involution: $p\mapsto\bar{p}\ \forall\ p\in\mathfrak{P}_{ACZ}$
is defined by complex conjugate of cylindrical functions and vector
fields, i.e., $\bar{p}:=(\overline{\psi}, \overline{Y})\ \forall\
p=(\psi, Y)\in\mathfrak{P}_{ACZ}$, where
$\overline{Y}\psi:=\overline{Y\overline{\psi}}$.\\ \\
(3) $\mathfrak{P}_{ACZ}$ admits a natural action of
$Cyl(\overline{\mathcal{A}})$ by
\begin{eqnarray}
\psi'\circ(\psi,Y):=(\psi'\psi,\psi'Y),\nonumber
\end{eqnarray}
which gives $\mathfrak{P}_{ACZ}$ a module
structure.}\\ \\
Note that the action of flux vector field $Y_f(S)$ on can be
expressed explicitly on any cylindrical function $\psi_\gamma\in
C^1(X_{\alpha(\gamma)})$ via a suitable
regularization\cite{thiemann2}:
\begin{eqnarray}
Y_f(S)\psi_\gamma&=&\{\int_S\underline{\eta}_{abc}\widetilde{P}^c_if^i,
\
\psi_\gamma\},\nonumber\\
&=&\sum_{e\in
E(\gamma)}\{\int_S\underline{\eta}_{abc}\widetilde{P}^c_if^i,\
A(e)_{mn}\}\frac{\partial}{\partial
A(e)_{mn}}\psi_\gamma\nonumber\\
&=&\sum_{e\in E(\gamma)}\frac{\kappa(S,\ e)}{2}f^i(S\cap
e)[\delta_{S\cap e,s(e)}(A(e)\tau_i)_{mn}-\delta_{S\cap
e,t(e)}(\tau_iA(e))_{mn}]
\frac{\partial}{\partial A(e)_{mn}}\psi_\gamma\nonumber\\
&=&\sum_{v\in V(\gamma)\cap S}\sum_{e\ at\ v}\frac{\kappa(S,\
e)}{2}f^i(v)X^{(e,v)}_i\psi_\gamma,\nonumber
\end{eqnarray}
where $A(e)_{mn}$ is the $SU(2)$ matrix element of the holonomy
along the edge $e$, $X^{(e,v)}_i$ is the left(right) invariant
vector field $L^{(\tau_i)}(R^{(\tau_i)})$ of the group associated
with the edge $e$ if $v$ is the source(target) of edge $e$ by
definition:
\begin{eqnarray}
L^{(\tau_i)}\psi\big(A(e)\big)&:=&\frac{d}{dt}|_{t=0}\psi\big(A(e)\exp(t\tau_i)\big),\nonumber\\
R^{(\tau_i)}\psi\big(A(e)\big)&:=&\frac{d}{dt}|_{t=0}\psi\big(\exp(-t\tau_i)A(e)\big),\nonumber
\end{eqnarray}
and
\begin{eqnarray}
\kappa(S,\ e)=\left\{%
\begin{array}{ll}
    0, & \hbox{if $e\cap S=\emptyset$, or $e$ lies in $S$;} \\
    1, & \hbox{if $e$ lies above $S$ and $e\cap S=p$;} \\
    -1, & \hbox{if $e$ lies below $S$ and $e\cap S=p$.} \\
\end{array}%
\right.\nonumber
\end{eqnarray}
Since the surface $S$ is oriented with normal $n_a$, "above" means
$n_a\dot{e}^a|_p>0$, and "below" means $n_a\dot{e}^a|_p<0$, where
$\dot{e}^a|_p$ is the tangent vector of $e$ at $p$. And one should
consider $e\cap S$ contained in the set $V(\gamma)$ and some edges
are written as the union of elementary edges which either lie in
$S$, or intersect $S$ at their source or target. On the other hand,
from the commutation relations for the left(right) invariant vector
fields, one can see that the commutators between flux vector fields
do not necessarily vanish when $S\cap S'\neq\emptyset$. This unusual
property is the classical origin of the non-commutativity of quantum
Riemannian structures \cite{noncommut}.

The classical Ashtekar-Corichi-Zapata holonomy-flux $*$-algebra
serves as a classical algebra of elementary observables in our
dynamical system of gauge fields. Then one can construct the quantum
algebra of elementary observables from $\mathfrak{P}_{ACZ}$ in
analogy with Definition 3.1.2.\\ \\
\textbf{Definition 3.4.2}\cite{unique} \\
\textit{The abstract free algebra $F(\mathfrak{P}_{ACZ})$ of the
classical $*$-algebra is defined by the formal direct sum of finite
sequences of classical observables $(p_1,...,p_n)$ with
$p_k\in\mathfrak{P}_{ACZ}$, where the operations of multiplication
and involution are defined as
\begin{eqnarray}
(p_1,...,p_n)\cdot(p'_1,...,p'_m)&:=&(p_1,...,p_n,p_1',...,p_m'),\nonumber\\
(p_1,..,p_n)^*&:=&(\bar{p}_n,...,\bar{p}_1).\nonumber
\end{eqnarray}
A $2$-sided ideal $\mathfrak{I}$ can be generated by the following
elements,
\begin{eqnarray}
(p+p')-(p)-(p'),\ \ (zp)-z(p),\nonumber\\
\ [(p),(p')]-i\hbar(\{p,p'\}),\ \ \ \ \ \ \ \nonumber\\
((\psi,0), p)-(\psi\circ p),\ \ \ \ \ \ \ \ \ \nonumber
\end{eqnarray}
where the canonical commutation bracket is defined by
\begin{eqnarray}
[(p),(p')]:=(p)\cdot(p')-(p')\cdot(p).\nonumber
\end{eqnarray}
Note that the ideal $\mathfrak{I}$ is preserved by the involution $*$,
and the last set of generators in the ideal $\mathfrak{I}$ cancels
the overcompleteness generated from the module structure of $\mathfrak{P}_{ACZ}$ \cite{Ash}. \\
The quantum holonomy-flux $*$-algebra is defined by the quotient
$*$-algebra $\mathfrak{A}=F(\mathfrak{P}_{ACZ})/\mathfrak{I}$, which
contains the unital element $1:=((1,0))$. Note that a sup-norm has
been defined by Eq.$(\ref{norm})$ for the Abelian sub-$*$-algebra
$Cyl(\overline{\mathcal{A}})$ in
$\mathfrak{A}$.}\\ \\
For simplicity, we denote the one element sequences (equivalence
classes) $\widehat{((\psi,0))}$ and $\widehat{((0,Y))}$ $\forall\
\psi\in Cyl(\overline{\mathcal{A}}),\ Y\in
\mathcal{V}^{\mathbf{C}}(\overline{\mathcal{A}})$ in $\mathfrak{A}$
by $\hat{\psi}$ and $\hat{Y}$ respectively, where the "hat" denotes
the equivalence class with respect to the quotient. In particular,
for all cylindrical functions $\hat{\psi}$ and flux vector fields
$\hat{Y}_f(S)$,
\begin{eqnarray}
\hat{\psi}^*=\hat{\bar{\psi}}\ \ \ \mathrm{and}\ \ \
\hat{Y}_f(S)^*=\hat{Y}_f(S).\nonumber
\end{eqnarray}
It can be seen that the free algebra $F(\mathfrak{P}_{ACZ})$ is
simplified a great deal after the quotient, and every element of the
quantum algebra $\mathfrak{A}$ can be written as a finite linear
combination of elements of the form
\begin{eqnarray}
&&\hat{\psi},\nonumber\\
&&\hat{\psi}_1\cdot\hat{Y}_{f_{11}}(S_{11}),\nonumber\\
&&\hat{\psi}_2\cdot\hat{Y}_{f_{21}}(S_{21})\cdot\hat{Y}_{f_{22}}(S_{22}),\nonumber\\
&&...\nonumber\\
&&\hat{\psi}_k\cdot\hat{Y}_{f_{k1}}(S_{k1})\cdot\hat{Y}_{f_{k2}}(S_{k2})\cdot...\cdot\hat{Y}_{f_{kk}}(S_{kk}),\nonumber\\
&&...\nonumber
\end{eqnarray}
Moreover, given a cylindrical function $\psi$ and a flux vector
field $Y_f(S)$, one has the relation from the commutation relation:
\begin{eqnarray}
\hat{Y}_f(S)\cdot\hat{\psi}=i\hbar
\widehat{Y_f(S)\psi}+\hat{\psi}\cdot\hat{Y}_f(S).\label{Y}
\end{eqnarray}
Then the kinematical Hilbert space $\mathcal{H}_{kin}$ can be
obtained properly via the GNS-construction for unital $*$-algebra
$\mathfrak{A}$ in the same way as in Definition 3.1.3. By the
GNS-construction, a positive linear functional, i.e. a state
$\omega_{kin}$, on $\mathfrak{A}$ defines a cyclic representation
$(\mathcal{H}_{kin},\pi_{kin},\Omega_{kin})$ for $\mathfrak{A}$. In
our case of quantum holonomy-flux $*$-algebra, the state with both
Yang-Mills gauge invariance and diffeomorphism invariance is defined
for any $\psi_\gamma\in Cyl(\overline{\mathcal{A}})$ and
non-vanishing flux vector field $Y_f(S)\in
\mathcal{V}^{\mathbf{C}}(\overline{\mathcal{A}})$ as \cite{unique}:
\begin{eqnarray}
&&\omega_{kin}(\hat{\psi_\gamma}):=\int_{SU(2)^{N_\gamma}}
\prod_{e\in E(\gamma)}d\mu_H(A(e))\psi_\gamma(\{A(e)\}_{e\in E(\gamma)}),\nonumber\\
&&\omega_{kin}(\hat{a}\cdot\hat{Y}_f(S)):=0,\ \ \
\forall\hat{a}\in\mathfrak{A},\nonumber
\end{eqnarray}
where $d\mu_H$ is the Haar measure on the compact group $SU(2)$ and
$N_\gamma$ is the number of elements in $E(\gamma)$. This
$\omega_{kin}$ is called Ashtekar-Isham-Lewandowski state. The null
space $\mathfrak{N}_{kin}\in\mathfrak{A}$ with respect to
$\omega_{kin}$ is defined as
$\mathfrak{N}_{kin}:=\{\hat{a}\in\mathfrak{A}|\omega_{kin}(\hat{a}^*\cdot
\hat{a})=0\}$, which is a left ideal. Then a quotient map can be
defined as:
\begin{eqnarray}
[.]:\ \mathfrak{A}&\rightarrow&\mathfrak{A}/\mathfrak{N}_{kin};\nonumber\\
\hat{a}&\mapsto&[\hat{a}]:=\{\hat{a}+\hat{b}|\hat{b}
\in\mathfrak{N}_{kin}\}.\nonumber
\end{eqnarray}
The GNS-representation for $\mathfrak{A}$ with respect to
$\omega_{kin}$ is a representation map: $\pi_{kin}$:
$\mathfrak{A}\rightarrow\mathcal{L}(\mathcal{H}_{kin})$ such that
$\pi_{kin}(\hat{a}\cdot\hat{b})=\pi_{kin}(\hat{a})\pi_{kin}(\hat{b})$,
where
$\mathcal{H}_{kin}:=\langle\mathfrak{A}/\mathfrak{N}_{kin}\rangle=\langle
Cyl(\overline{\mathcal{A}})\rangle$ by straightforward verification
and the $\langle\cdot\rangle$ denotes the completion with respect to
the natural equipped inner product on $\mathcal{H}_{kin}$,
\begin{eqnarray}
<[\hat{a}]|[\hat{b}]>_{kin}:=\omega_{kin}(\hat{a}^*\cdot\hat{b}).\nonumber
\end{eqnarray}
To show how this inner product works, given any two cylindrical
functions $\psi=[\psi_\alpha], \psi'=[\psi'_{\alpha'}]\in
Cyl(\overline{\mathcal{A}})$, the inner product between them is
expressed as
\begin{eqnarray}
<[\hat{\psi}]|[\hat{\psi'}]>_{kin}:=\int_{X_{\alpha''}}(P^*_{\alpha''\alpha}\overline{\psi}_\alpha)
(P^*_{\alpha''\alpha'}\psi'_{\alpha'}) d\mu_{\alpha''},
\label{inner}
\end{eqnarray}
for any groupoid $\alpha''$ containing both $\alpha$ and $\alpha'$.
The measure $d\mu_\alpha$ on $X_\alpha$ is defined by the pull back
of the product Haar measure $d\mu^{N_\alpha}_H$ on the product group
$SU(2)^{N_\alpha}$ via the identification bijection between
$X_\alpha$ and $SU(2)^{N_\alpha}$, where $N_\alpha$ is number of
maximal analytic edges generating $\alpha$. In addition, a nice
result shows that given such a family of measures
$\{\mu_\alpha\}_{\alpha\in\mathcal{L}}$, a probability measure $\mu$
is uniquely well-defined on the quantum configuration space
$\overline{\mathcal{A}}$ \cite{AL1}, such that the kinematical
Hilbert space $\mathcal{H}_{kin}$ coincides with the collection of
the square-integrable functions with respect to the measure $\mu$ on
the quantum configuration space, i.e.
$\mathcal{H}_{kin}=L^2(\overline{\mathcal{A}},\ d\mu)$, just as we
expected at the beginning of our construction.

The representation map $\pi_{kin}$ is defined by
\begin{eqnarray}
\pi_{kin}(\hat{a})[\hat{b}]:=[\hat{a}\cdot\hat{b}],\ \ \forall\
\hat{a}\in \mathfrak{A}, \ \mathrm{and}\
[\hat{b}]\in\mathcal{H}_{kin}.\nonumber
\end{eqnarray}
Note that $\pi_{kin}(\hat{a})$ is an unbounded operator in general.
It is easy to verify that
\begin{eqnarray}
\pi_{kin}(\hat{Y}_f(S))[\hat{\psi}]=i\hbar[\widehat{Y_f(S)\psi}]\nonumber
\end{eqnarray}
via Eq.(\ref{Y}), which gives the canonical momentum operator. In
the following context, we denote the operator
$\pi_{kin}(\hat{Y}_f(S))$ by $\hat{P}_f(S)$ on $\mathcal{H}_{kin}$,
and just denote the elements $[\hat{\psi}]$ in $\mathcal{H}_{kin}$
by $\psi$ for simplicity.

Moreover, since $\Omega_{kin}:=1$ is a cyclic vector in
$\mathcal{H}_{kin}$, the positive linear functional which we begin
with can be expressed as
\begin{eqnarray}
\omega_{kin}(\hat{a})=<\Omega_{kin}|
\pi_{kin}(\hat{a})\Omega_{kin}>_{kin}.\nonumber
\end{eqnarray}
Thus the Ashtekar-Isham-Lewandowski state $\omega_{kin}$ on
$\mathfrak{A}$ is equivalent to a cyclic representation
$(\mathcal{H}_{kin},\pi_{kin}, \Omega_{kin})$ for $\mathfrak{A}$,
which is the Ashtekar-Isham-Lewandowski representation for quantum
holonomy-flux $*$-algebra of background independent gauge field
theory. One thus obtains the kinematical representation of loop
quantum gravity via the construction of algebraic quantum field
theory. It is important to note that the Ashtekar-Isham-Lewandowski
state is the unique state on the quantum holonomy-flux $*$-algebra
$\mathfrak{A}$ invariant under internal gauge transformations and
spatial diffeomorphisms\footnote{The proof of this conclusion
depends on the compact support property of the smear functions $f^i$
(see \cite{unique} for detail).}, which are both automorphisms
$\alpha_g$ and $\alpha_\varphi$ on $\mathfrak{A}$ and can be
verified that $\omega_{kin}\circ\alpha_g=\omega_{kin}$ and
$\omega_{kin}\circ\alpha_\varphi=\omega_{kin}$. So these gauge
transformations are represented as unitary transformations on
$\mathcal{H}_{kin}$, while the cyclic vector $\Omega_{kin}$,
representing "no geometry vacuum" state, is the unique state in
$\mathcal{H}_{kin}$ invariant under internal gauge transformations
and spatial diffeomorphisms. This is a very crucial uniqueness
theorem for canonical quantization of gauge field theory
\cite{unique}:\\ \\
\textbf{Theorem 3.4.1}: \textit{There exists exactly one Yang-Mills
gauge invariant and spatial diffeomorphism invariant state
$($positive linear functional$)$ on the quantum holonomy-flux
$*$-algebra. In other words, there exists a unique Yang-Mills gauge
invariant and spatial diffeomorphism invariant cyclic representation
for the quantum holonomy-flux $*$-algebra, which is called
Ashtekar-Isham-Lewandowski representation. Moreover, this
representation is irreducible with respect to an exponential version
of the quantum holonomy-flux algebra (defined in
\cite{thiemann4}), which is analogous to the Weyl algebra.} \\
\\
Hence we have finished the construction of kinematical Hilbert space
for background independent gauge field theory and represented the
quantum holonomy-flux algebra on it. Then following the general
programme presented in the last subsection, we should impose the
constraints as operators on the kinematical Hilbert space since we
are dealing with a gauge system.

\subsection{Spin-network Decomposition of Kinematical Hilbert Space}

The kinematical Hilbert space $\mathcal{H}_{kin}$ for loop quantum
gravity has been well defined. In this subsection, it will be shown
that $\mathcal{H}_{kin}$ can be decomposed into the orthogonal
direct sum of 1-dimensional subspaces and find a basis, called
spin-network basis, in the Hilbert space, which consists of
uncountably infinite elements. So the kinematic Hilbert space is
non-separable. In the following, we will show the decomposition in
three steps.

\begin{itemize}

\item {{Spin-network Decomposition on a Single Edge}}

Given a graph consisting of only one edge $e$, which naturally
associates with a group $SU(2)=X_{\alpha(e)}$, the elements of
$X_{\alpha(e)}$ are the quantum connections only taking nontrivial
values on $e$. Then we consider the decomposition of the Hilbert
space $\mathcal{H}_{\alpha(e)}=L^2(X_{\alpha(e)},
d\mu_{\alpha(e)})\simeq L^2(SU(2), d\mu_H)$, which is nothing but
the space of square integrable functions on the compact group
$SU(2)$ with the natural $L^2$ inner product. It is natural to
define several operators on $\mathcal{H}_{\alpha(e)}$. First, the
so-called configuration operator $\hat{f}\big(A(e)\big)$ whose
operation on any $\psi$ in a dense domain of $L^2(SU(2), d\mu_H)$ is
nothing but multiplication by the function $f\big(A(e)\big)$, i.e.,
\begin{eqnarray}
\hat{f}\big(A(e)\big)\psi\big(A(e)\big):=f\big(A(e)\big)\psi\big(A(e)\big),\nonumber
\end{eqnarray}
where $A(e)\in SU(2)$. Second, given any vector $\xi\in su(2)$, it
generates left invariant vector field $L^{(\xi)}$ and right
invariant vector field $R^{(\xi)}$ on $SU(2)$ by
\begin{eqnarray}
L^{(\xi)}\psi\big(A(e)\big):=\frac{d}{dt}|_{t=0}\psi\big(A(e)\exp(t\xi)\big),\nonumber\\
R^{(\xi)}\psi\big(A(e)\big):=\frac{d}{dt}|_{t=0}\psi\big(\exp(-t\xi)A(e)\big),\nonumber
\end{eqnarray}
for any function $\psi\in C^1(SU(2))$. Then one can define the
so-called momentum operators on the single edge by
\begin{eqnarray}
\hat{J}_i^{(L)}=iL^{(\tau_i)}\ \ \ \mathrm{and}\ \ \
\hat{J}_i^{(R)}=iR^{(\tau_i)},\nonumber
\end{eqnarray}
where the generators $\tau_i\in su(2)$ constitute an orthonormal
basis with respect to the Killing-Cartan metric. The momentum
operators have the well-known commutation relation of the angular
momentum operators in quantum mechanics:
\begin{eqnarray}
[\hat{J}^{(L)}_i,\hat{J}^{(L)}_j]=i\epsilon^k_{\
ij}\hat{J}^{(L)}_k,\
[\hat{J}^{(R)}_i,\hat{J}^{(R)}_j]=i\epsilon^k_{\
ij}\hat{J}^{(R)}_k,\ [\hat{J}^{(L)}_i,\hat{J}^{(R)}_j]=0.\nonumber
\end{eqnarray}
Third, the Casimir operator on $\mathcal{H}_e$ can be expressed as
\begin{eqnarray}
\hat{J}^2:=\delta^{ij}\hat{J}^{(L)}_i\hat{J}^{(L)}_j=\delta^{ij}\hat{J}^{(R)}_i\hat{J}^{(R)}_j.\label{casimir}
\end{eqnarray}

The decomposition of $\mathcal{H}_e=L^2(SU(2),d\mu_H)$ is provided
by
the following Peter-Weyl Theorem.\\ \\
\textbf{Theorem 3.5.1} \cite{dieck}:  \\
\textit{Given a compact group $G$, the function space
$L^2(G,d\mu_H)$ can be decomposed as an orthogonal direct sum of
finite dimensional Hilbert spaces, and the matrix elements of the
equivalence classes of finite dimensional irreducible
representations of $G$ form an orthogonal basis in
$L^2(G,d\mu_H)$.}\\

Note that a finite dimensional irreducible representation of $G$
can be regarded as a matrix-valued function on $G$, so the matrix
elements are functions on $G$. Using this theorem, one can find
the decomposition of the Hilbert space:
\begin{eqnarray}
L^2(SU(2),d\mu_H)=\oplus_j[\mathcal{H}_j\otimes
\mathcal{H}^*_j],\nonumber
\end{eqnarray}
where $j$, labelling irreducible representations of $SU(2)$, are
the half integers, $\mathcal{H}_j$ denotes the carrier space of
the $j$-representation of dimension $2j+1$, and $ \mathcal{H}^*_j$
is its dual space. The basis $\{\mathbf{e}^j_m\otimes
\mathbf{e}_n^{j*}\}$ in $\mathcal{H}_j\otimes \mathcal{H}^{*}_j$
maps a group element $g\in SU(2)$ to a matrix $\{\pi^j_{mn}(g)\}$,
where $m,n=-j,...,j$. Thus the space $\mathcal{H}_j\otimes
\mathcal{H}^*_j$ is spanned by the matrix element functions
$\pi^j_{mn}$ of equivalent $j$-representations. Moreover, the
spin-network basis can be defined.\\

\textbf{Proposition 3.5.1} \cite{carmeli}\\
\textit{The system of spin-network functions on
$\mathcal{H}_{\alpha(e)}$, consisting of matrix elements
$\{\pi^j_{mn}\}$ in finite dimensional irreducible representations
labelled by half-integers $\{j\}$, satisfies
\begin{eqnarray}
\hat{J}^2\pi^j_{mn}=j(j+1)\pi^j_{mn},\
\hat{J}^{(L)}_3\pi^j_{mn}=m\pi^j_{mn},\
\hat{J}^{(R)}_3\pi^j_{mn}=n\pi^j_{mn}\nonumber,
\end{eqnarray}
where $j$ is called angular momentum quantum number and
$m,n=-j,...,j$ magnetic quantum number. The normalized functions
$\{\sqrt{2j+1}\pi^j_{mn}\}$ form an orthonormal basis in
$\mathcal{H}_{\alpha(e)}$ by the above theorem and
\begin{eqnarray}
\int_{\overline{\mathcal{A}}_e}\overline{\pi^{j'}_{m'n'}}\pi^j_{mn}d\mu_e
=\frac{1}{2j+1}\delta^{j'j}\delta_{m'm}\delta_{n'n},\nonumber
\end{eqnarray}
which is called the spin-network basis on $\mathcal{H}_{\alpha(e)}$.
So the Hilbert space on a single edge
has been decomposed into one dimensional subspaces. }\\

Note that the system of operators
$\{\hat{J}^2,\hat{J}^{(R)}_3,\hat{J}^{(L)}_3\}$ forms a complete set
of commutable operators in $\mathcal{H}_{\alpha(e)}$. There is a
cyclic "vacuum state" in the Hilbert space, which is the
$(j=0)$-representation $\Omega_{\alpha(e)}=\pi^{j=0}=1$,
representing that there is no geometry on the edge.

\item {{Spin-network Decomposition on Finite Graph}}

Given a groupoid $\alpha$ generated by a graph $\gamma$ with $N$
oriented edges $e_i$ and $M$ vertices, one can define the
configuration operators on the corresponding Hilbert space
$\mathcal{H}_\alpha=L^2(X_\alpha, d\mu_\alpha)\simeq L^2(SU(2)^N,
d\mu_H^N)$ by
\begin{eqnarray}
\hat{f}\big(A(e_i)\big)\psi\big(A(e_1),...,A(e_{N})\big):=f\big(A(e_i)\big)\psi
\big(A(e_1),...,A(e_{N})\big).\nonumber
\end{eqnarray}
The momentum operators $\hat{J_i}^{(e,v)}$ associated with a edge
$e$ connecting a vertex $v$ are defined as
\begin{eqnarray}
\hat{J_i}^{(e,v)}:=(1\otimes...\otimes
\hat{J}_i\otimes...\otimes1),\nonumber
\end{eqnarray}
where we set $\hat{J}_i=\hat{J}_i^{(L)}$ if $v=s(e)$ and
$\hat{J}_i=\hat{J}_i^{(R)}$ if $v=t(e)$, so
$\hat{J_i}^{(e,v)}=iX_i^{(e,v)}$. Note that $\hat{J_i}^{(e,v)}$ only
acts nontrivially on the Hilbert space associated with the edge $e$.
Then one can define a vertex operator associated with vertex $v$ in
analogy with the total angular momentum operator via
\begin{eqnarray}
[\hat{J}^v]^2:=\delta^{ij}\hat{J}_i^v\hat{J}_j^v,\nonumber
\end{eqnarray}
where
\begin{eqnarray}
\hat{J}_i^v:=\sum_{e\ at\ v}\hat{J}^{(e,v)}_i.\nonumber
\end{eqnarray}
Obviously, $\mathcal{H}_\alpha$ can be firstly decomposed by the
representations on each edge $e$ of $\alpha$ as:
\begin{eqnarray}
\mathcal{H}_\alpha&=&\otimes_{e}\mathcal{H}_{\alpha(e)}=\otimes_{e}[\oplus_j(\mathcal{H}^e_j\otimes
\mathcal{H}^{e*}_j)]=\oplus_{\mathbf{j}}[\otimes_e(\mathcal{H}^e_j\otimes
\mathcal{H}^{e*}_j)]\nonumber\\
&=&\oplus_{\mathbf{j}}[\otimes_v(\mathcal{H}^{v=s(e)}_{\mathbf{j}(s)}\otimes
\mathcal{H}^{v=t(e)}_{\mathbf{j}(t)})],\nonumber
\end{eqnarray}
where $\mathbf{j}:=(j_1,...,j_N)$ assigns to each edge an
irreducible representation of $SU(2)$, in the fourth step the
Hilbert spaces associated with the edges are allocated to the
vertexes where these edges meet so that for each vertex $v$,
\begin{eqnarray}
\mathcal{H}^{v=s(e)}_{\mathbf{j}(s)}\equiv\otimes_{s(e)=v}\mathcal{H}^e_j
\ \ \ \mathrm{and}\ \ \
\mathcal{H}^{v=t(e)}_{\mathbf{j}(t)}\equiv\otimes_{t(e)=v}\mathcal{H}^{e*}_j.\nonumber
\end{eqnarray}
The group of gauge transformations $g(v)\in SU(2)$ at each vertex
is reducibly represented on the Hilbert space
$\mathcal{H}^{v=s(e)}_{\mathbf{j}(s)}\otimes
\mathcal{H}^{v=t(e)}_{\mathbf{j}(t)}$ in a natural way. So this
Hilbert space can be decomposed as a direct sum of irreducible
representation spaces via Clebsch-Gordon decomposition:
\begin{eqnarray}
\mathcal{H}^{v=s(e)}_{\mathbf{j}(s)}\otimes
\mathcal{H}^{v=t(e)}_{\mathbf{j}(t)}=\oplus_l\mathcal{H}^v_{\mathbf{j}(v),l}\
.\nonumber
\end{eqnarray}
As a result, $\mathcal{H}_\alpha$ can be further decomposed as:
\begin{eqnarray}
\mathcal{H}_\alpha=\oplus_{\mathbf{j}}[\otimes_v(\oplus_l\mathcal{H}^v_{\mathbf{j}(v),l})]
=\oplus_{\mathbf{j}}[\oplus_{\mathbf{l}}(\otimes_v\mathcal{H}^v_{\mathbf{j}(v),l})]\equiv
\oplus_{\mathbf{j}}[\oplus_{\mathbf{l}}\mathcal{H}_{\alpha,\mathbf{j},\mathbf{l}}].\label{decomposition}
\end{eqnarray}
It can also be viewed as the eigenvector space decomposition of the
commuting operators $[\hat{J}^v]^2$ (with eigenvalues $l(l+1)$) and
$[\hat{J}^e]^2\equiv \delta^{ij}\hat{J}^e_i\hat{J}^e_j$. Note that
$\mathbf{l}:=(l_1,...,l_M)$ assigns to each vertex(objective) of
$\alpha$ an irreducible representation of $SU(2)$. One may also
enlarge the set of commuting operators to further refine the
decomposition of the Hilbert space. Note that the subspace of
$\mathcal{H}_\alpha$ with $\mathbf{l}=0$ is Yang-Mills gauge
invariant, since the representation of gauge transformations is
trivial.

\item {{Spin-network Decomposition of $\mathcal{H}_{kin}$}}

Since $\mathcal{H}_{kin}$ has the structure
$\mathcal{H}_{kin}=\langle\ \cup_{\alpha\in
\mathcal{L}}\mathcal{H}_\alpha\ \rangle$, one may consider to
construct it as a direct sum of $\mathcal{H}_\alpha$ by canceling
some overlapping components. The
construction is precisely described as a theorem below.\\ \\
\textbf{Theorem 3.5.2}:  \\
\textit{Consider assignments $\mathbf{j}=(j_1,...,j_N)$ to the edges
of any groupoid $\alpha\in\mathcal{L}$ and assignments
$\mathbf{l}=(l_1,...,l_M)$ to the vertices. The edge representation
$j$ is non-trivial on each edge, and the vertex representation $l$
is non-trivial at each spurious\footnote{A vertex $v$ is spurious if
it is bivalent and $e\circ e'$ is itself analytic edge with $e,\ e'$
meeting at $v$.} vertex, unless it is the base point of a close
analytic loop. Let $\mathcal{H}'_{\alpha}$ be the Hilbert space
composed by the subspaces
$\mathcal{H}_{\alpha,\mathbf{j},\mathbf{l}}$ (assigned the above
conditions) according to Eq.(\ref{decomposition}). Then
$\mathcal{H}_{kin}$ can be decomposed as the direct sum of the
Hilbert spaces $\mathcal{H}'_{\alpha}$, i.e.,
\begin{eqnarray}
\mathcal{H}_{kin}=\oplus_{\alpha\in
\mathcal{L}}\mathcal{H}'_{\alpha}\oplus\mathbf{C}.\nonumber
\end{eqnarray}  }\\
\textbf{Proof}:  \\
Since the representation on each edge is non-trivial, by definition
of the inner product, it is easy to see that $\mathcal{H}'_\alpha$
and $\mathcal{H}'_{\alpha'}$ are mutual orthogonal if one of the
groupoids $\alpha$ and $\alpha'$ has at leat an edge $e$ more than
the other due to
\begin{eqnarray}
\int_{\overline{\mathcal{A}}_e}\pi^j_{mn}d\mu_e=\int_{\overline{\mathcal{A}}_e}1\cdot\pi^j_{mn}d\mu_e=0\nonumber
\end{eqnarray}
for any $j\neq0$. Now consider the case of the spurious vertex. An
edge $e$ with $j$-representation in a graph is assigned the
Hilbert space $\mathcal{H}^e_j\otimes\mathcal{H}^{e*}_j$.
Inserting a vertex $v$ into the edge, one obtains two edges $e_1$
and $e_2$ split by $v$ both with $j$-representations, which belong
to a different graph. By the decomposition of the corresponding
Hilbert space,
\begin{eqnarray}
\mathcal{H}^{e_1}_j\otimes\mathcal{H}^{e_1*}_j\otimes\mathcal{H}^{e_2}_j\otimes\mathcal{H}^{e_2*}_j
=\mathcal{H}^{e_1}_j\otimes(\oplus_{l=0...2j}\mathcal{H}^v_l)\otimes\mathcal{H}^{e_2*}_j,\nonumber
\end{eqnarray}
the subspace for all $l\neq0$ are orthogonal to the space
$\mathcal{H}^e_j\otimes\mathcal{H}^{e*}_j$, while the subspace for
$l=0$ coincides with $\mathcal{H}^e_j\otimes\mathcal{H}^{e*}_j$
since $\mathcal{H}^v_{l=0}=\mathbf{C}$ and $A(e)=A(e_1)A(e_2)$.
This completes the proof.\\ \\
Since there are uncountably many graphs on $\Sigma$, the kinematical
Hilbert $\mathcal{H}_{kin}$ is non-separable. We denote the
spin-network basis in $\mathcal{H}_{kin}$ by $\Pi_s,\
s=(\gamma(s),\mathbf{j}_s,\mathbf{m}_s,\mathbf{n}_s)$ and vacuum
$\Omega_{kin}\equiv\Pi_0=1$, where
\begin{eqnarray}
\Pi_s:=\prod_{e\in E(\gamma(s))}\sqrt{2j_e+1}\pi^{j_e}_{m_e n_e}\ \
\ \ \ (j_e\neq0),\nonumber
\end{eqnarray}
which form a orthonormal basis with the relation
$<\Pi_s|\Pi_{s'}>_{kin}=\delta_{ss'}$. And
$Cyl_\gamma(\overline{\mathcal{A}})\subset
Cyl(\overline{\mathcal{A}})$ denotes the linear span of the spin
network functions $\Pi_s$ for $\gamma(s)=\gamma$.
\end{itemize}

The spin-network basis can be used to construct the so-called spin
network representation of loop quantum gravity.\\ \\
\textbf{Definition 3.5.1}:  \textit{The spin-network representation
is a vector space $\widetilde{\mathcal{H}}$ of complex valued
functions
\begin{eqnarray}
\widetilde{\Psi}:\ S\ \rightarrow\ \mathbf{C}; \ \ s\ \mapsto\
\widetilde{\Psi}(s),\nonumber
\end{eqnarray}
where $S$ is the set of the labels $s$ for the spin network
states. $\widetilde{\mathcal{H}}$ is equipped with the scalar
product
\begin{eqnarray}
<\widetilde{\Psi},\ \widetilde{\Psi}'>:=\sum_{s\in
S}\overline{\widetilde{\Psi}(s)}\widetilde{\Psi}(s)'\nonumber
\end{eqnarray}
between square summable functions.}\\ \\
The relation between the Hilbert spaces $\widetilde{\mathcal{H}}$
and
$\mathcal{H}_{kin}$ is clarified by the following proposition \cite{thiemann2}.\\
\\
\textbf{Proposition 3.5.2:} \\
\textit{The spin-network transformation
\begin{eqnarray}
T:\ \mathcal{H}_{kin}\ \rightarrow\ \widetilde{\mathcal{H}}; \ \
\Psi\ \mapsto\ \widetilde{\Psi}(s):=<\Pi_s,\ \Psi>_{kin}\nonumber
\end{eqnarray}
is a unitary transformation with inverse
\begin{eqnarray}
T^{-1}\Psi=\sum_{s\in S}\widetilde{\Psi}(s)\Pi_s.\nonumber
\end{eqnarray} }\\
Thus the connection representation and the spin-network
representation are ''Fourier transforms'' of each other, where the
role of the kernel of the integral is played by the spin-network
basis. Note that, in the gauge invariant Hilbert space of loop
quantum gravity which we will define later, the Fourier transform
with respect to the gauge invariant spin-network basis is the
so-called loop transform, which leads to the unitary equivalent loop
representation of the theory \cite{rov2}\cite{gp1}\cite{rovelli}.

To conclude this subsection, we show the explicit representation of
elementary observables on the kinematical Hilbert space
$\mathcal{H}_{kin}$. The action of canonical momentum operator
$\hat{P}_f(S)$ on differentiable cylindrical functions
$\psi_\gamma\in Cyl_\gamma(\overline{\mathcal{A}})$ can be
expressed as
\begin{eqnarray}
\hat{P}_f(S)\psi_\gamma\big(\{A(e)\}_{e\in
E(\gamma)}\big)&=&\frac{\hbar}{2}\sum_{v\in V(\gamma)\cap
S}f^i(v)\big[\sum_{e\ at\ v}\kappa(S,\
e)\hat{J}_i^{(e,v)}\big]\psi_\gamma\big(\{A(e)\}_{e\in
E(\gamma)}\big)\nonumber\\
&=&\frac{\hbar}{2}\sum_{v\in V(\gamma)\cap
S}f^i(v)\big[\hat{J}^{(S,v)}_{i(u)}-\hat{J}^{(S,v)}_{i(d)}\big]\psi_\gamma\big(\{A(e)\}_{e\in
E(\gamma)}\big),\label{momentum}
\end{eqnarray}
where
\begin{eqnarray}
\hat{J}^{(S,v)}_{i(u)}&\equiv&\hat{J}^{(e_1,v)}_{i}+...+\hat{J}^{(e_u,v)}_{i},\nonumber\\
\hat{J}^{(S,v)}_{i(d)}&\equiv&\hat{J}^{(e_{u+1},v)}_{i}+...+\hat{J}^{(e_{u+d},v)}_{i},\label{updown}
\end{eqnarray}
for the edges $e_1,...,e_u$ lying above $S$ and
$e_{u+1},...,e_{u+d}$ lying below $S$. And it was proved that the
operator $\hat{P}_f(S)$ is essentially self-adjoint on
$\mathcal{H}_{kin}$ \cite{thiemann2}. On the other hand, it is
obvious to construct configuration operators by spin-network
functions:
\begin{eqnarray}
\hat{\Pi}_s\psi_\gamma\big(\{A(e)\}_{e\in
E(\gamma)}\big):=\Pi_s(\{A(e)\}_{e\in
E(\gamma(s))})\psi_\gamma(\{A(e)\}_{e\in E(\gamma)}).\nonumber
\end{eqnarray}
Note that $\hat{\Pi}_s$ may change the graph, i.e., $\hat{\Pi}_s$:
$Cyl_\gamma(\overline{\mathcal{A}})\rightarrow
Cyl_{\gamma\cup\gamma(s)}(\overline{\mathcal{A}})$. So far, the
elementary operators of quantum kinematics have been well defined on
$\mathcal{H}_{kin}$.

\subsection{Quantum Riemannian Geometry}

The well-established quantum kinematics of loop quantum gravity is
now in the same status as Riemannian geometry before the appearance
of general relativity and Einstein's equation, giving general
relativity mathematical foundation and offering living place to the
Einstein equation. Instead of classical geometric quantities, such
as length, area, volume etc., the quantities in quantum geometry are
operators on the kinematical Hilbert space $\mathcal{H}_{kin}$, and
their spectrum serve as the possible values of the quantities in
measurements. So far, the kinematical quantum geometric operators
constructed properly in loop quantum gravity include length operator
\cite{length},  area operator \cite{rovelli8}\cite{area}, two
different volume operators \cite{AL3}\cite{rovelli8}\cite{volume},
$\hat{Q}$ operator \cite{ma2}, etc.. Recently, a consistency check
was proposed for the different regularizations of the volume
operator \cite{thiemann17}\cite{thiemann18}. We thus will only
introduce the volume operator defined by Ashtekar and Lewandowski
\cite{volume}, which is shown to be correct in the consistency
check.

First, we define the area operator with respect to a 2-surface $S$
by the elementary operators. Given a closed 2-surface or a surface
$S$ with boundary, we can divide it into a large number $N$ of
small area cells $S_I$. Taking account of the classical expression
of an area, we set the area of the 2-surface to be the limit of
the Riemannian sum
\begin{eqnarray}
A_S:=\lim_{N\rightarrow\infty}[A_S]_N=\lim_{N\rightarrow\infty}\kappa\beta\sum_{I=1}^N\sqrt{P_i(S_I)P_j(S_I)
\delta^{ij}}.\nonumber
\end{eqnarray}
Then one can unambiguously obtain a quantum area operator from the
canonical momentum operators $\hat{P}_i(S)$ smeared by constant
functions. Given a cylindrical function $\psi_\gamma\in
Cyl_\gamma(\overline{\mathcal{A}})$ which has second order
derivatives, the action of the area operator on $\psi_\gamma$ is
defined in the limit by requiring that each area cell contains at
most only one intersecting point $v$ of the graph $\gamma$ and $S$
as
\begin{eqnarray}
\hat{A}_S\psi_\gamma:=\lim_{N\rightarrow\infty}[\hat{A}_S]_N\psi_\gamma
=\lim_{N\rightarrow\infty}\kappa\beta\sum_{I=1}^N
\sqrt{\hat{P}_i(S_I)\hat{P}_j(S_I)\delta^{ij}}\
\psi_\gamma.\nonumber
\end{eqnarray}
The regulator $N$ is easy to remove, since the result of the
operation of the operator $\hat{P}_i(S_I)$ does not change when
$S_I$ shrinks to a point. Since the refinement of the partition does
not affect the result of action of $[\hat{A}_S]_N$ on $\psi_\gamma$,
the limit area operator $\hat{A}_S$, which is shown to be
self-adjoint \cite{area}, is well defined on $\mathcal{H}_{kin}$ and
takes the explicit expression as:
\begin{eqnarray}
\hat{A}_S\psi_\gamma=4\pi\beta \ell_p^2\sum_{v\in V(\gamma\cap
S)}\sqrt{(\hat{J}^{(S,v)}_{i(u)}-\hat{J}^{(S,v)}_{i(d)})
(\hat{J}^{(S,v)}_{j(u)}-\hat{J}^{(S,v)}_{j(d)})\delta^{ij}}\
\psi_\gamma,\nonumber
\end{eqnarray}
where $\hat{J}^{(S,v)}_{i(u)}$ and $\hat{J}^{(S,v)}_{i(d)}$ have
been defined in Eq.(\ref{updown}). It turns out that the finite
linear combinations of spin-network basis in $\mathcal{H}_{kin}$
diagonalizes $\hat{A}_S$ with eigenvalues given by finite sums,
\begin{eqnarray}
a_S=4\pi\beta \ell_p^2 \sum_v \sqrt{2j_v^{(u)}(j_v^{(u)}+1) +
2j_v^{(d)}(j_v^{(d)}+1) -
j_v^{(u+d)}(j_v^{(u+d)}+1)},\label{spectrum}
\end{eqnarray}
where $j^{(u)}, j^{(d)}$ and $j^{(u+d)}$ are arbitrary
half-integers subject to the standard condition
\begin{eqnarray}
j^{(u+d)}\ \in \{|j^{(u)}- j^{(d)}|, |j^{(u)}- j^{(d)}|+1, ... ,
j^{(u)}+j^{(d)}\}. \label{jconstraint}
\end{eqnarray}
Hence the spectrum of the area operator is fundamentally pure
discrete, while its continuum approximation becomes excellent
exponentially rapidly for large eigenvalues. However, in fundamental
level, the area is discrete and so is the quantum geometry. One can
see that the eigenvalue of $\hat{A}_S$ does not vanish even in the
case where only one edge intersects the surface at a single point,
whence the quantum geometry is distributional.

The form of Ashtekar and Lewandowski's volume operator was
introduced for the first time in \cite{AL3}, and its detailed
properties are discussed in \cite{volume}. Given a region $R$ with a
fixed coordinate system $\{x^a\}_{a=1,2,3}$ in it, one can introduce
a partition of $R$ in the following way. Divide $R$ into small
volume cells $C$ such that, each cell $C$ is a cube with coordinate
volume less than $\epsilon$ and two different cells only share the
points on their boundaries. In each cell $C$, we introduce three
2-surfaces $s=(S^1,S^2,S^3)$ such that $x^a$ is constant on the
surface $S^a$. We denote this partition $(C,s)$ as
$\mathcal{P}_\epsilon$. Then the volume of the region $R$ can be
expressed classically as
\begin{eqnarray}
V_R^{s}&=&\lim_{\epsilon\rightarrow0}\sum_C\sqrt{|q_{C,s}|},\nonumber
\end{eqnarray}
where
\begin{eqnarray}
q_{C,s}&=&\frac{(\kappa\beta)^3}{3!}\epsilon^{ijk}\underline{\eta}_{abc}P_i(S^a)P_j(S^b)P_k(S^c).\nonumber
\end{eqnarray}
This motivates us to define the volume operator by naively
changing $P_i(S^a)$ to $\hat{P}_i(S^a)$:
\begin{eqnarray}
\hat{V}_R^{s}&=&\lim_{\epsilon\rightarrow0}\sum_C\sqrt{|\hat{q}_{C,s}|},\nonumber\\
 \hat{q}_{C,s}&=&\frac{(\kappa\beta)^3}{3!}\epsilon^{ijk}\underline{\eta}_{abc}
\hat{P}_i(S^a)\hat{P}_j(S^b)\hat{P}_k(S^c).\nonumber
\end{eqnarray}
Note that, given any cylindrical function $\psi_\gamma\in
Cyl_\gamma(\overline{\mathcal{A}})$, we require the vertexes of the
graph $\gamma$ to be at the intersecting points of the triples of
2-surfaces $s=(S^1,S^2,S^3)$ in corresponding cells. Thus the limit
operator will trivially exist due to the same reason in the case of
the area operator. However, the volume operator defined here depends
on the choice of orientations for the triples of surfaces
$s=(S^1,S^2,S^3)$, or essentially, the choice of coordinate systems.
So it is not uniquely defined. Since, for all choice of
$s=(S^1,S^2,S^3)$, the resulting operators have correct
semi-classical limit, one settles up the problem by averaging
different operators labelled by different $s$ \cite{volume}. The
process of averaging removes the freedom in defining the volume
operator up to an overall constant $\kappa_0$. The resulting
self-adjoint operator acts on any cylindrical function
$\psi_\gamma\in Cyl_\gamma(\overline{\mathcal{A}})$ as
\begin{eqnarray}
\hat{V}_R\ \psi_\gamma&=&\kappa_0\sum_{v\in
V(\alpha)}\sqrt{|\hat{q}_{v,\gamma}|}\ \psi_\gamma, \nonumber
\end{eqnarray}
where
\begin{eqnarray}
\hat{q}_{v,\gamma}&=&(8\pi\beta\ell_p^2)^3
\frac{1}{48}\sum_{e,e',e''\ at\ v}\epsilon^{ijk}\epsilon(e,e',e'')
\hat{J}^{(e,v)}_i\hat{J}^{(e',v)}_j\hat{J}^{(e'',v)}_k, \nonumber
\end{eqnarray}
here
$\epsilon(e,e',e'')\equiv\mathrm{sgn}(\epsilon_{abc}\dot{e}^a\dot{e}'^b\dot{e}''^c)|_v$
with $\dot{e}^a$ as the tangent vector of edge $e$ and
$\epsilon_{abc}$ as the orientation of $\Sigma$. The only
unsatisfactory point in the present volume operator is the choice
ambiguity of $\kappa_0$. However, fortunately, the most recent
discussion shows that the overall undetermined constant $\kappa_0$
can be fixed to be $\sqrt{6}$ by the consistency check between the
volume operator and the triad operator
\cite{thiemann17}\cite{thiemann18}.

\newpage
\section{Implementation of Quantum Constraints}

After constructing the kinematical Hilbert space $\mathcal{H}_{kin}$
of loop quantum gravity, one should implement the constraints on it
to obtain the physical Hilbert space which encodes the complete
information of quantum dynamics of general relativity, since the
Hamiltonian of general relativity is a linear combination of the
constraints. Recalling the constraints (\ref{constraint}) in the
Hamiltonian formalism and the Poission algebra (\ref{constraint
algebra}) among them, the subalgebra generated by the Gauss
constraints $\mathcal{G}(\Lambda)$ forms a Lie algebra and a 2-sided
ideal in the constraints algebra. So in this section, we first solve
the Gaussian constraints independently of the other two kinds of
constraints and find the solution space $\mathcal{H}^G$, which is
constituted by internal gauge invariant quantum states. Then,
although the subalgebra generated by the diffeomorphism constraints
is not an ideal in the constraint algebra, we still would like to
solve them independently of the scalar constraints for technical
convenience. After that, the quantum operator corresponding to the
Hamiltonian constraint(scalar constraint) is defined on the
kinematical Hilbert space, and we will also discuss an alterative
for the implementation of the scalar constraint, which is called the
master constraint programme by modifying the classical constraint
algebra.

\subsection{Solutions of Quantum Gaussian Constraint}

Recall that the classical expression of Gauss constraints reads
\begin{eqnarray}
\mathcal{G}(\Lambda)=\int_{\Sigma}d^3x\Lambda^iD_a\widetilde{P}^a_i
=-\int_{\Sigma}d^3x\widetilde{P}^a_iD_a\Lambda^i\equiv-P(D\Lambda),\nonumber
\end{eqnarray}
where $D_a\Lambda^i=\partial_a\Lambda^i+\epsilon^i_{\
jk}A^j_a\Lambda^k$. As the situation of triad flux, the Gauss
constraints can be defined as cylindrically consistent vector fields
$Y_{D\Lambda}$ on $\overline{\mathcal{A}}$, which act on any
cylindrical function $f_\gamma\in
Cyl_\gamma(\overline{\mathcal{A}})$ by
\begin{eqnarray}
Y_{D\Lambda}\circ f_\gamma(\{A(e)\}_{e\in
E(\gamma)}):=\{-P(D\Lambda),f_\gamma(\{A(e)\}_{e\in
E(\gamma)})\}.\nonumber
\end{eqnarray}
Then the Gauss constraint operator can be defined in analogy with
the momentum operator, which acts on $f_\gamma$ as:
\begin{eqnarray}
\hat{\mathcal{G}}(\Lambda)f_\gamma\big(\{A(e)\}_{e\in
E(\gamma)}\big)&:=&i\hbar Y_{D\Lambda}f_\gamma\big(\{A(e)\}_{e\in
E(\gamma)}\big)\nonumber\\&=&\hbar\sum_{v\in
V(\gamma)}[\Lambda^i(v)\hat{J}^{v}_i]f\big(\{A(e)\}_{e\in
E(\gamma)}\big),\nonumber
\end{eqnarray}
which is the generator of internal gauge transformations on
$Cyl_\gamma(\overline{\mathcal{A}})$. The kernel of the operator is
easily obtained in terms of the spin-network decomposition, which is
the internal gauge invariant Hilbert space:
\begin{eqnarray}
\mathcal{H}^G=\oplus_{\alpha,\mathbf{j}}\mathcal{H}'_{\alpha,\mathbf{j},\mathbf{l}=0}\oplus\mathbf{C}.\nonumber
\end{eqnarray}
One then naturally gets the gauge invariant spin-network basis
$T_s,\ s=(\gamma(s),\mathbf{j}_s,\mathbf{i}_s)$ in $\mathcal{H}^G$
via a group averaging technique at each
vertex\cite{rovelli9}\cite{ALM}\cite{baez1}(we will call $T_s$
spin-network state in the following context):
\begin{eqnarray}
T_{s=(\gamma,\mathbf{j},\mathbf{i})}=\bigotimes_{v\in
V(\gamma)}i_v\bigotimes_{e\in E(\gamma)}\pi^{j_e}(A(e)),\ \
(j_e\neq0)\nonumber
\end{eqnarray}
assigning a non-trivial spin representation $j$ on each edge and a
invariant tensor $i$ (intertwiner) on each vertex. We denote the
vector space of finite linear combinations of vacuum state and gauge
invariant spin-network states $Cyl(\overline{\mathcal{A/G}})$, which
is dense in $\mathcal{H}^G$. And
$Cyl_\gamma(\overline{\mathcal{A/G}})\subset
Cyl(\overline{\mathcal{A/G}})$ denotes the linear span of the gauge
invariant spin network functions $T_s$ for $\gamma(s)=\gamma$. All
Yang-Mills gauge invariant operators are well defined on
$\mathcal{H}^G$. However, the condition of acting on gauge invariant
states often changes the structure of the spectrum of quantum
geometric operators. For the area operator, the spectrum depends on
certain global properties of the surface $S$ (see
\cite{AL}\cite{area} for details). For the volume operators, a
non-zero spectrum arises from at least 4-valent vertices.

\subsection{Solutions of Quantum Diffeomorphism Constraint}

Unlike the strategy in solving Gaussian constraint, one cannot
define an operator for the quantum diffeomorphism constraint as the
infinitesimal generator of finite diffeomorphism transformations
(unitary operators since the measure is diffeomorphism invariant)
represented on $\mathcal{H}_{kin}$. The representation of finite
diffeomorphisms is a family of unitary operators $\hat{U}_\varphi$
acting on cylindrical functions $\psi_\gamma$ by
\begin{eqnarray}
\hat{U}_\varphi \psi_\gamma:=\psi_{\varphi\circ\gamma},
\end{eqnarray}
for any spatial diffeomorphism $\varphi$ on $\Sigma$. An 1-parameter
subgroup $\varphi_t$ in the group of spatial diffeomorphisms is then
represented as an 1-parameter unitary group $\hat{U}_{\varphi_t}$ on
$\mathcal{H}_{kin}$. However, $\hat{U}_{\varphi_t}$ is not weakly
continuous, since the subspaces $\mathcal{H}'_{\alpha(\gamma)}$ and
$\mathcal{H}'_{\alpha(\varphi_t\circ\gamma)}$ are orthogonal to each
other no matter how small the parameter $t$ is. So one always has
\begin{eqnarray}
|<\psi_\gamma|\hat{U}_{\varphi_t}|\psi_\gamma>_{kin}-<\psi_\gamma|\psi_\gamma>_{kin}|=<\psi_\gamma|\psi_\gamma>_{kin}
\neq 0, \label{weak continuous}
\end{eqnarray}
even in the limit when $t$ goes to zero. Therefore, the
infinitesimal generator of $\hat{U}_{\varphi_t}$ does not exist. In
the strategy to solve the diffeomorphism constraint, due to the Lie
algebra structure of diffeomorphism constraints subalgebra, the
so-called group averaging technique is employed. We now outline the
procedure. First, given a colored graph (a graph $\gamma$ and a
cylindrical function living on it), one can define the group of
graph symmetries $GS_\gamma$ by
\begin{eqnarray}
GS_\gamma:=Diff_\gamma/TDiff_\gamma,\nonumber
\end{eqnarray}
where $Diff_\gamma$ is the group of all diffeomorphisms preserving
the colored $\gamma$, and $TDiff_\gamma$ is the group of
diffeomorphisms which trivially acts on $\gamma$. We define a
projection map by averaging with respect to $GS_\gamma$ to obtain
the subspace in $Cyl_\gamma$ which is invariant under the
transformation of $GS_\gamma$:
\begin{eqnarray}
\hat{P}_{Diff,\gamma}\psi_\gamma:=\frac{1}{n_\gamma}\sum_{\varphi\in
GS_\gamma}\hat{U}_{\varphi}\psi_\gamma,\nonumber
\end{eqnarray}
for all cylindrical functions
$\psi_\gamma\in\mathcal{H}'_{\alpha(\gamma)}$, where $n_\gamma$ is
the number of the finite elements of $GS_\gamma$. Second, we average
with respect to all remaining diffeomorphisms which move the graph
$\gamma$. For each cylindrical function $\psi_\gamma\in
Cyl_\gamma(\overline{\mathcal{A/G}})$, there is an element
$\eta(\psi_\gamma)$ associated to it in the algebraic dual space
$Cyl^\star$ of $Cyl(\overline{\mathcal{A/G}})$, which acts on any
cylindrical function $\phi_{\gamma'}\in
Cyl_\gamma(\overline{\mathcal{A/G}})$ as:
\begin{eqnarray}
\eta(\psi_\gamma)[\phi_{\gamma'}]:=\sum_{\varphi\in
Diff(\Sigma)/Diff_\gamma}<\hat{U}_{\varphi}\hat{P}_{Diff,\gamma}\psi_\gamma|\phi_{\gamma'}>_{kin}.\nonumber
\end{eqnarray}
It is well defined since, for any given graph $\gamma'$, only finite
terms are non-zero in the summation. It is easy to verify that
$\eta(\psi_\gamma)$ is invariant under the group action of
$Diff(\Sigma)$, since
\begin{eqnarray}
\eta(\psi_\gamma)[\hat{U}_{\varphi}\phi_{\gamma'}]=\eta(\psi_\gamma)[\phi_{\gamma'}].\nonumber
\end{eqnarray}
Thus we have defined a rigging map $\eta:\
Cyl(\overline{\mathcal{A/G}})\rightarrow Cyl^\star_{Diff}$, which
maps every cylindrical function to a diffeomorphism invariant one,
where $Cyl^\star_{Diff}$ is spanned by vacuum state $T_0=1$ and
rigged spin-network functions $T_{[s]}\equiv\{\eta(T_{s})\},\
{[s]=([\gamma],\mathbf{j},\mathbf{i})}$ associated with
diffeomorphism classes $[\gamma]$ of graphs $\gamma$. Moreover a
Hermitian inner product can be defined on $Cyl^\star_{Diff}$ by the
natural action of the algebraic functional:
\begin{eqnarray}
<\eta(\psi_\gamma)|\eta(\phi_{\gamma'})>_{Diff}:=\eta(\psi_\gamma)[\phi_{\gamma'}].\nonumber
\end{eqnarray}
The diffeomorphism invariant Hilbert space $\mathcal{H}_{Diff}$ is
defined by the completion of $Cyl^\star_{Diff}$ with respect to
the above inner product $<\ |\
>_{Diff}$. The diffeomorphism invariant spin-network functions
$T_{[s]}$ form an orthonormal basis in $\mathcal{H}_{Diff}$.
Finally, we have obtained the general solutions invariant under both
Yang-Mills gauge transformations and spatial diffeomorphisms.

In general relativity, the problem of observables is a subtle
issue due to the diffeomorphism invariance
\cite{rovelli6}\cite{rovelli3}\cite{rovelli4}. Now we discuss the
operators as diffeomorphism invariant observables on
$\mathcal{H}_{Diff}$. We call an operator
$\hat{\mathcal{O}}\in\mathcal{L}(\mathcal{H}_{kin})$ a strong
observable if and only if
$\hat{U}^{-1}_{\varphi}\hat{\mathcal{O}}\hat{U}_{\varphi}=\hat{\mathcal{O}},
\ \forall \ \varphi\in Diff(\Sigma)$. We call it a weak observable
if and only if $\hat{\mathcal{O}}$ leaves $\mathcal{H}_{Diff}$
invariant. Then it is easy to see that a strong observable
$\hat{\mathcal{O}}$  must be a weak one. One notices that a strong
observable $\hat{\mathcal{O}}$ can first be defined on
$\mathcal{H}_{Diff}$ by its dual operator
$\hat{\mathcal{O}}^{\star}$ as
\begin{eqnarray}
(\hat{\mathcal{O}}^{\star}\Phi_{Diff})[\psi]:=\Phi_{Diff}[\hat{\mathcal{O}}\psi],\nonumber
\end{eqnarray}
then one gets
\begin{eqnarray}
(\hat{\mathcal{O}}^{\star}\Phi_{Diff})[\hat{U}_{\varphi}\psi]=\Phi_{Diff}[\hat{\mathcal{O}}\hat{U}_{\varphi}\psi]
=\Phi_{Diff}[\hat{U}^{-1}_{\varphi}\hat{\mathcal{O}}\hat{U}_{\varphi}\psi]=(\hat{\mathcal{O}}^{\star}\Phi_{Diff})
[\psi],\nonumber
\end{eqnarray}
for any $\Phi_{Diff}\in\mathcal{H}_{Diff}$ and
$\psi\in\mathcal{H}_{kin}$. Hence
$\hat{\mathcal{O}}^{\star}\Phi_{Diff}$ is also diffeomorphism
invariant. In addition, a strong observable also has the property of
$\hat{\mathcal{O}}^{\star}\eta(\psi_\gamma)=\eta(\hat{\mathcal{O}}^\dagger\psi_\gamma)$
since, $\forall \ \phi_{\gamma'}, \psi_\gamma\in\mathcal{H}_{kin}$,
\begin{eqnarray}
&&<\hat{\mathcal{O}}^{\star}\eta(\psi_\gamma)|\eta(\phi_{\gamma'})>_{Diff}
=(\hat{\mathcal{O}}^{\star}\eta(\psi_\gamma))[\phi_{\gamma'}]=\eta(\psi_\gamma)[\hat{\mathcal{O}}\phi_{\gamma'}]\nonumber\\
&=&\sum_{\varphi\in
Diff(\Sigma)/Diff_\gamma}<\hat{U}_{\varphi}\hat{P}_{Diff,\gamma}\psi_\gamma|\hat{\mathcal{O}}
\phi_{\gamma'}>_{kin}\nonumber\\
&=&\frac{1}{n_\gamma}\sum_{\varphi\in
Diff(\Sigma)/Diff_\gamma}\sum_{\varphi'\in
GS_\gamma}<\hat{U}_{\varphi}
\hat{U}_{\varphi'}\psi_\gamma|\hat{\mathcal{O}}\phi_{\gamma'}>_{kin}\nonumber\\
&=&\frac{1}{n_\gamma}\sum_{\varphi\in
Diff(\Sigma)/Diff_\gamma}\sum_{\varphi'\in
GS_\gamma}<\hat{U}_{\varphi}
\hat{U}_{\varphi'}\hat{\mathcal{O}}^\dagger\psi_\gamma|\phi_{\gamma'}>_{kin}\nonumber\\
&=&<\eta(\hat{\mathcal{O}}^{\dagger}\psi_\gamma)|\eta(\phi_{\gamma'})>_{Diff}.\nonumber
\end{eqnarray}
Note that the Hilbert space $\mathcal{H}_{Diff}$ is still
non-separable if one considers the $C^n$ diffeomorphisms with $n>0$.
However, if one extends the diffeomorphisms to be semi-analytic
diffeomotphisms, i.e. homomorphisms that are analytic
diffeomorphisms up to finite isolated points (which can be viewed as
an extension of the classical concept to the quantum case), the
Hilbert space $\mathcal{H}_{Diff}$ would be separable
\cite{FR}\cite{AL}.

\subsection{Hamiltonian Constraint Operator}

In the following, we consider the issue of scalar constraint in loop
quantum gravity. One may first construct a Hamiltonian constraint
(scalar constraint) operator in $\mathcal{H}_{kin}$ or
$\mathcal{H}_{Diff}$, then attempt to find the physical Hilbert
space $\mathcal{H}_{phys}$ by solving the quantum Hamiltonian
constraint. However, difficulties arise here due to the special role
played by the scalar constraints in the constraint algebra
(\ref{constraint algebra}).  First, the scalar constraints do not
form a Lie subalgebra. Hence the strategy of group averaging cannot
be used directly on $\mathcal{H}_{kin}$ for them. Second, modulo the
Gaussian constraint, there is still a structure function in the
Poisson bracket between two scalar constraints:
\begin{eqnarray}
\{\mathcal{S}(N),\mathcal{S}(M)\}=-\mathcal{V}((N\partial_bM-M\partial_bN)q^{ab}),\label{s-algebra}
\end{eqnarray}
which raises the danger of quantum anomalyies in quantization.
Moreover, the diffeomorphism constraints do not form an ideal in the
quotient constraint algebra modulo the Gaussian constraints. This
fact results in that the scalar constraint operator cannot be well
defined on $\mathcal{H}_{Diff}$, as it does not commute with the
diffeomorphism transformations $\hat{U}_\varphi$. Thus the previous
construction of $\mathcal{H}_{Diff}$ does not appear very useful for
the final construction of $\mathcal{H}_{phys}$, which is our final
goal. However, one may still first try to construct a Hamiltonian
constraint operator in $\mathcal{H}_{kin}$ for technical
convenience.

We recall the classical expression of Hamiltonian constraint:
\begin{eqnarray}
\mathcal{S}(N)&:=&\frac{\kappa\beta^2}{2}\int_\Sigma
d^3xN\frac{\widetilde{P}^a_i\widetilde{P}^b_j}{\sqrt{|\det
q|}}[\epsilon^{ij}_{\ \ k}F^k_{ab}-2(1+\beta^2)K^i_{[a}K^j_{b]}]\nonumber\\
&=&\mathcal{S}_{E}(N)-2(1+\beta^2)\mathcal{T}(N).\label{classicalH}
\end{eqnarray}
The main idea of the construction is to first express
$\mathcal{S}(N)$ in terms of the combination of Poisson brackets
between the variables which have been represented as operators on
$\mathcal{H}_{kin}$, then replace the Poisson brackets by canonical
commutators between the operators. We will use the volume functional
for a region $R\subset\Sigma$ and the extrinsic curvature functional
defined by:
\begin{eqnarray}
K&:=&\kappa\beta\int_{\Sigma}d^3x\widetilde{P}^a_iK^i_a.\nonumber
\end{eqnarray}
A key trick here is to consider the following classical identity
of the co-triad $e^i_a(x)$ \cite{thiemann1}:
\begin{eqnarray}
e^i_a(x)=\frac{(\kappa\beta)^2}{2}\underline{\eta}_{abc}\epsilon^{ijk}\frac{\widetilde{P}^b_j\widetilde{P}^c_k}
{\sqrt{\det q}}(x)=\frac{2}{\kappa\beta}\{A^i_a(x),V_R\},\nonumber
\end{eqnarray}
where $V_R$ is the volume functional for a neighborhood $R$
containing $x$. And the expression of the extrinsic curvature 1-form
$K^i_a(x)$:
\begin{eqnarray}
K^i_a(x)=\frac{1}{\kappa\beta}\{A^i_a(x),K\}.\nonumber
\end{eqnarray}
Note that $K$ can be expressed by a Poisson bracket between the
constant-smeared Euclidean Hamiltonian constraint and the total
volume of the space $\Sigma$:
\begin{eqnarray}
K=\beta^{-2}\{\mathcal{S}_E(1),V_\Sigma\}.
\end{eqnarray}
Thus one can obtain the equivalent classical expressions of
$\mathcal{S}_{E}(N)$ and $\mathcal{T}(N)$ as:
\begin{eqnarray}
\mathcal{S}_{E}(N)&=&\frac{\kappa\beta^2}{2}\int_\Sigma
d^3xN\frac{\widetilde{P}^a_i\widetilde{P}^b_j}{\sqrt{|\det
q|}}\epsilon^{ij}_{\ \ k}F^k_{ab} \nonumber\\
&=&-\frac{2}{\kappa^2\beta}\int_\Sigma
d^3xN(x)\widetilde{\eta}^{abc}\mathrm{Tr}\big(\mathbf{F}_{ab}(x)\{\mathbf{A}_c(x),V_{R_x}\}\big),\nonumber\\
\mathcal{T}(N)&=&\frac{\kappa\beta^2}{2}\int_\Sigma
d^3xN\frac{\widetilde{P}^a_i\widetilde{P}^b_j}{\sqrt{|\det
q|}}K^i_{[a}K^j_{b]}\nonumber\\
&=&-\frac{2}{\kappa^4\beta^3}\int_\Sigma
d^3xN(x)\widetilde{\eta}^{abc}\mathrm{Tr}\big(\{\mathbf{A}_a(x),K\}\{\mathbf{A}_b(x),K\}
\{\mathbf{A}_c(x),V_{R_x}\}\big),\nonumber
\end{eqnarray}
where $\mathbf{A}_a=A_a^i\tau_i$, $\mathbf{F}_{ab}=F_{ab}^i\tau_i$,
$\mathrm{Tr}$ represents the trace of the Lie algebra matrix, and
$R_x\subset\Sigma$ denotes an arbitrary neighborhood of
$x\in\Sigma$. In order to quantize the Hamiltonian constraint as a
well-defined operator on $\mathcal{H}_{kin}$, one has to express the
classical formula of $\mathcal{S}(N)$ in terms of holonomies $A(e)$
and other variables with clear quantum analogs. As a first attempt
\cite{thiemann1}, this can be realized by introducing a
triangulation $T(\epsilon)$, where the parameter $\epsilon$
describes how fine the triangulation is, and the triangulation will
fill out the spatial manifold $\Sigma$ when $\epsilon\rightarrow0$.
Given a tetrahedron $\Delta\in T(\epsilon)$, we use
$\{s_i(\Delta)\}_{i=1,2,3}$ to denote the three outgoing oriented
segments in $\Delta$ with a common beginning point
$v(\Delta)=s(s_i(\Delta))$, and use $a_{ij}(\Delta)$ to denote the
arc connecting the end points of $s_i(\Delta)$ and $s_j(\Delta)$.
Then several loops $\alpha_{ij}(\Delta)$ are formed by
$\alpha_{ij}(\Delta):=s_i(\Delta)\circ a_{ij}(\Delta)\circ
s_j(\Delta)^{-1}$. Thus we have the identities:
\begin{eqnarray}
\{\int_{s_i(\Delta)}\mathbf{A}_a\dot{s}^a_i(\Delta),V_{R_{v(\Delta)}}\}&=&-A(s_i(\Delta))^{-1}
\{A(s_i(\Delta)),V_{R_{v(\Delta)}}\}+o(\epsilon),\nonumber\\
\{\int_{s_i(\Delta)}\mathbf{A}_a\dot{s}^a_i(\Delta),K\}&=&-A(s_i(\Delta))^{-1}\{A(s_i(\Delta)),K\}
+o(\epsilon),\nonumber\\
\int_{P_{ij}}\mathbf{F}_{ab}(x)&=&\frac{1}{2}A(\alpha_{ij}(\Delta))^{-1}
-\frac{1}{2}A(\alpha_{ij}(\Delta))+o(\epsilon^2),\nonumber
\end{eqnarray}
where $P_{ij}$ is the plane with boundary $\alpha_{ij}$. Note that
the above identities are constructed by taking account of internal
gauge invariance of the final formula of Hamiltonian constraint
operator. So we have the regularized expression of $\mathcal{S}(N)$
by the Riemannian sum \cite{thiemann1}:
\begin{eqnarray}
\mathcal{S}^\epsilon_{E}(N)&=&\frac{2}{3\kappa^2\beta}\sum_{\Delta\in
T(\epsilon)}N(v(\Delta))\epsilon^{ijk}
\times\nonumber\\
&&\mathrm{Tr}\big(A(\alpha_{ij}(\Delta))^{-1}A(s_k(\Delta))^{-1}\{A(s_k(\Delta)),V_{R_{v(\Delta)}}\}\big),\nonumber\\
\mathcal{T}^\epsilon(N)&=&\frac{\sqrt{2}}{6\kappa^4\beta^3}\sum_{\Delta\in
T(\epsilon)}N(v(\Delta))\epsilon^{ijk}\times\nonumber\\
&&\mathrm{Tr}\big(A(s_i(\Delta))^{-1}\{A(s_i(\Delta)),K\}A(s_j(\Delta))^{-1}\{A(s_j(\Delta)),K\}\times\nonumber\\
&&A(s_k(\Delta))^{-1}\{A(s_k(\Delta)),V_{R_{v(\Delta)}}\}\big),\nonumber\\
\mathcal{S}^\epsilon(N)&=&\mathcal{S}^\epsilon_{E}(N)-2(1+\beta^2)\mathcal{T}^\epsilon(N),\label{hamilton}
\end{eqnarray}
such that
$\lim_{\epsilon\rightarrow0}\mathcal{S}^\epsilon(N)=\mathcal{S}(N)$.
It is clear that the above regulated formula of $\mathcal{S}(N)$ is
invariant under internal gauge transformations. Since all
constituents in the expression have clear quantum analogs, one can
quantize the regulated Hamiltonian constraint as an operator on
$\mathcal{H}_{kin}$ (or $\mathcal{H}^G$) by replacing them by the
corresponding operators and Poisson brackets by canonical
commutators, i.e.,
\begin{eqnarray}
&&A(e)\mapsto\hat{A}(e),\ \ \ V_R\mapsto\hat{V}_R,\ \ \ \{\ ,\
\}\mapsto\frac{[\ ,\ ]}{i\hbar},\nonumber\\
&&\mathrm{and}\ \ \ \ \ \
K\mapsto\hat{K}^\epsilon=\frac{\gamma^{-2}}{i\hbar}[\hat{\mathcal{S}}^\epsilon_E(1),\hat{V}_\Sigma].\nonumber
\end{eqnarray}
Removing the regulator by $\epsilon\rightarrow0$, it turns out
that one can obtain a well-defined limit operator on
$\mathcal{H}_{kin}$ (or $\mathcal{H}^G$) with respect to a natural
operator topology.

Now we begin to construct the Hamiltonian constraint operator in
analogy with the classical expression (\ref{classicalH}). All we
should do is define the corresponding regulated operators on
different $\mathcal{H}'_\alpha$ separately, then remove the
regulator $\epsilon$ so that the limit operator is defined on
$\mathcal{H}_{kin}$ (or $\mathcal{H}^G$) cylindrically consistently.
In the following, given a vertex and three edges intersecting at the
vertex in a graph $\gamma$ of $\psi_\gamma\in
Cyl_\gamma(\overline{\mathcal{A/G}})$, we construct one
triangulation of the neighborhood of the vertex adapted to the three
edges. Then we average with respect to the triples of edges meeting
at the given vertex. Precisely speaking, one can make the
triangulations $T(\epsilon)$ with the following properties
\cite{thiemann1}\cite{thiemann2}.
\begin{itemize}
\item The chosen triple of edges in the graph $\gamma$ is embedded in a $T(\epsilon)$ for all
$\epsilon$, so that the vertex $v$ of $\gamma$ where the three edges
meet coincides with a vertex $v(\Delta)$ in $T(\epsilon)$.

\item For every triple of segments ($e_1,\ e_2,\ e_3$) of $\gamma$
such that $v=s(e_1)=s(e_2)=s(e_3)$, there is a tetrahedra $\Delta\in
T(\epsilon)$ such that $v=v(\Delta)=s(s_i(\Delta))$, and
$s_i(\Delta)\subset e_i,\ \forall\ i=1,2,3$. We denote such a
tetrahedra as $\Delta^0_{e_1,e_2,e_3}$.

\item For each tetrahedra $\Delta^0_{e_1,e_2,e_3}$ one can
construct seven additional tetrahedron $\Delta^\wp_{e_1,e_2,e_3},\
\wp=1,...,7$, by backward analytic extensions of $s_i(\Delta)$ so
that $U_{e_1,e_2,e_3}:=\cup_{\wp=0}^7\Delta^\wp_{e_1,e_2,e_3}$ is a
neighborhood of $v$.

\item The triangulation must be fine enough so that the
neighborhoods $U(v):=\cup_{e_1,e_2,e_3}U_{e_1,e_2,e_3}(v)$ are
disjoint for different vertices $v$ and $v'$ of $\gamma$. Thus for
any open neighborhood $U_\gamma$ of the graph $\gamma$, there exists
a triangulation $T(\epsilon)$ such that $\cup_{v\in
V(\gamma)}U(v)\subseteq U_\gamma$.

\item The distance between a vertex $v(\Delta)$ and the
corresponding arcs $a_{ij}(\Delta)$ is described by the parameter
$\epsilon$. For any two different $\epsilon$ and $\epsilon'$, the
arcs $a_{ij}(\Delta^\epsilon)$ and $a_{ij}(\Delta^{\epsilon'})$ with
respect to one vertex $v(\Delta)$ are semi-analytically
diffeomorphic with each other.

\item With the triangulations $T(\epsilon)$, the integral over
$\Sigma$ is replaced by the Riemanian sum:
\begin{eqnarray}
\int_\Sigma\ \ \ &=&\int_{U_\alpha}\ \ \ \ +\int_{\Sigma-U_\alpha}\ ,\ \ \ \nonumber\\
\int_{U_\alpha}\ \ \ &=&\sum_{v\in V(\alpha)}\int_{U(v)}\ \ \ \ +\int_{U_\alpha-\cup_{v}U(v)}\ ,\ \ \ \nonumber\\
\int_{U(v)}\ \ \
&=&\frac{1}{E(v)}\sum_{e_1,e_2,e_3}[\int_{U_{e_1,e_2,e_3}(v)}\ \ \ \
+\int_{U(v)-U_{e_1,e_2,e_3},(v)}\ ],\nonumber
\end{eqnarray}
where $n(v)$ is the valence of the vertex $v=s(e_1)=s(e_2)=s(e_3)$,
and $E(v)\equiv \left(
\begin{array}{ll}
n(v)\\
3
\end{array} \right)$ denotes the binomial coefficient which comes from
the averaging with respect to the triples of edges meeting at given
vertex $v$. One then observes that
\begin{eqnarray}
\int_{U_{e_1,e_2,e_3}(v)}\ \ \ \
=8\int_{\Delta^0_{e_1,e_2,e_3}(v)}\ \ \,\nonumber
\end{eqnarray}
in the limit $\epsilon\rightarrow0$.

\item The triangulations for the regions
\begin{eqnarray}
&&U(v)-U_{e_1,e_2,e_3}(v),\nonumber\\
&&U_\alpha-\cup_{v\in V(\alpha)}U(v),\nonumber\\
&&\Sigma-U_\alpha, \label{*}
\end{eqnarray}
are arbitrary. These regions do not contribute to the construction
of the operator, since the commutator term
$[A(s_i(\Delta)),V_{R_{v(\Delta)}}]\psi_\alpha$ vanishes for all
tetrahedron $\Delta$ in the regions (\ref{*}).
\end{itemize}
Thus we find the regulated expression of Hamiltonian constraint
operator with respect to the triangulations $T(\epsilon)$ as
\cite{thiemann1}
\begin{eqnarray}
\hat{\mathcal{S}}^\epsilon_{E,\gamma}(N)&=&\frac{16}{3i\hbar\kappa^2\beta}\sum_{v\in
V(\gamma)}
\frac{N(v)}{{E}(v)}\sum_{v(\Delta)=v}\epsilon^{ijk}\times\nonumber\\
&&\mathrm{Tr}\big(\hat{A}(\alpha_{ij}(\Delta))^{-1}\hat{A}(s_k(\Delta))^{-1}[\hat{A}(s_k(\Delta)),
\hat{V}_{U^\epsilon_{v}}]\big),\nonumber\\
\hat{\mathcal{T}}^\epsilon_\gamma(N)&=&-\frac{4\sqrt{2}}{3i\hbar^3\kappa^4\beta^3}\sum_{v\in
V(\gamma)}\frac{N(v)}{E(v)}
\sum_{v(\Delta)=v}\epsilon^{ijk}\times\nonumber\\
&&\mathrm{Tr}\big(\hat{A}(s_i(\Delta))^{-1}[\hat{A}(s_i(\Delta)),\hat{K}^\epsilon]\hat{A}(s_j(\Delta))^{-1}
[\hat{A}(s_j(\Delta)),\hat{K}^\epsilon]\times\nonumber\\
&&\hat{A}(s_k(\Delta))^{-1}[\hat{A}(s_k(\Delta)),\hat{V}_{U^\epsilon_{v}}]\big),
\nonumber\\
\hat{\mathcal{S}}^\epsilon(N)\psi_\gamma&=&[\hat{\mathcal{S}}^\epsilon_{E,\gamma}(N)-2(1+\beta^2)
\hat{\mathcal{T}}_{\gamma}^\epsilon(N)]\psi_\gamma =\sum_{v\in
V(\gamma)}N(v)\hat{\mathcal{S}}^\epsilon_v\psi_\gamma,\nonumber
\end{eqnarray}
for any cylindrical function $\psi_\gamma\in
Cyl_\gamma(\overline{\mathcal{A/G}})$ is a finite linear combination
of spin-network states $T_s$ with $\gamma(s)=\gamma$.

By construction, the operation of $\hat{\mathcal{S}}^\epsilon(N)$ on
any $\psi_\gamma\in Cyl_\gamma(\overline{\mathcal{A/G}})$ is reduced
to a finite combination of that of $\hat{\mathcal{S}}^\epsilon_v$
with respect to different vertices of $\gamma$. Hence, for each
$\epsilon>0$, $\hat{\mathcal{S}}^\epsilon(N)$ is a well-defined
internal gauge invariant and diffeomorphism covariant operator on
$Cyl(\overline{\mathcal{A/G}})$.

The last step is to remove the regulator by taking the limit
$\epsilon\rightarrow0$. However, the action of the Hamiltonian
constraint operator on $\psi_\gamma$ adds arcs $a_{ij}(\Delta)$ with
a $\frac{1}{2}$-representation with respect to each $v(\Delta)$ of
$\gamma$\footnote{The Hamiltonian constraint operator depends indeed
on the choice of the representation $j$ on the arcs
$a_{ij}(\Delta)$, which is known as one of the regularization
ambiguities in the construction of quantum dynamics. For the
simplicity of the theory, one often choose the lowest label of
representation $j=\frac{1}{2}$.}, i.e. the action of
$\hat{\mathcal{S}}^\epsilon(N)$ on cylindrical functions is
graph-changing. Hence the operator does not converge with respect to
the weak operator topology in $\mathcal{H}_{kin}$ when
$\epsilon\rightarrow0$, since different
$\mathcal{H}'_{\alpha(\gamma)}$ with different graphs $\gamma$ are
mutually orthogonal. Thus one has to define a weaker operator
topology to make the operator limit meaningful. By physical
motivation and the naturally available Hilbert space
$\mathcal{H}_{Diff}$, the convergence of
$\hat{\mathcal{S}}^\epsilon(N)$ holds with respect to the so-called
Uniform Rovelli-Smolin Topology \cite{urst}, where one defines
$\hat{\mathcal{S}}^\epsilon(N)$ to converge if and only if
$\Psi_{Diff}[\hat{\mathcal{S}}^\epsilon(N)\phi]$ converge for all
$\Psi_{Diff}\in Cyl^\star_{Diff}$ and $\phi\in
Cyl(\overline{\mathcal{A/G}})$. Since the value of
$\Psi_{Diff}[\hat{\mathcal{S}}^\epsilon(N)\phi]$ is actually
independent of $\epsilon$ by the fifth property of the
triangulations, the sequence converges to a nontrivial result
$\Psi_{Diff}[\hat{\mathcal{S}}^{\epsilon_0}(N)\phi]$ with arbitrary
fixed $\epsilon_0>0$. Thus we have defined a diffeomorphism
covariant, densely defined, closed but non-symmetric operator,
$\hat{\mathcal{S}}(N)=\lim_{\epsilon\rightarrow0}\hat{\mathcal{S}}^\epsilon(N)=\hat{\mathcal{S}}^{\epsilon_0}(N)$,
on $\mathcal{H}_{kin}$ (or $\mathcal{H}^G$) representing the
Hamiltonian constraint. Moreover, a dual Hamiltonian constraint
operator $\hat{\mathcal{S}}'^\epsilon(N)$ is also defined on
$Cyl^\star$ depending on a specific value of $\epsilon$
\begin{eqnarray}
(\hat{\mathcal{S}}'^\epsilon(N)\Psi)[\phi]:=\Psi[\hat{\mathcal{S}}^\epsilon(N)\phi],\nonumber
\end{eqnarray}
for all $\Psi\in Cyl^\star$ and $\phi\in
Cyl(\overline{\mathcal{A/G}})$. For $\Psi_{Diff}\in
Cyl^\star_{Diff}\subset Cyl^\star$, one gets
\begin{eqnarray}
(\hat{\mathcal{S}}'(N)\Psi_{Diff})[\phi]=\Psi_{Diff}[\hat{\mathcal{S}}^\epsilon(N)\phi].\nonumber
\end{eqnarray}
which is independent of the value of $\epsilon$.

Several remarks on the Hamiltonian constraint operator are listed
in the following.
\begin{itemize}
\item {Finiteness of $\hat{\mathcal{S}}(N)$ on
$\mathcal{H}_{kin}$}

In ordinary quantum field theory, the continuous quantum field is
only recovered when one lets lattice spacing to approach zero, i.e.,
takes the continuous cut-off parameter to its continuous limit.
However, this will produce the well-known infinities in quantum
field theory and make the Hamiltonian operator ill-defined on the
Fock space. So it seems surprising that our operator
$\hat{\mathcal{S}}(N)$ is still well defined, when one takes the
limit $\epsilon\rightarrow0$ with respect to the Uniform
Rovelli-Smolin Topology so that the triangulation goes to the
continuum. The reason behind it is that the cut-off parameter is
essentially noneffective due to the diffeomorphism invariance of our
quantum field theory. This is why there is no UV divergence in the
background independent quantum gauge field theory with
diffeomorphism invariance. On the other hand, from a convenient
viewpoint, one may think the Hamiltonian constraint operator as an
operator dually defined on a dense domain in $\mathcal{H}_{Diff}$.
However, we will see that the dual Hamiltonian constraint operator
cannot leave $\mathcal{H}_{Diff}$ invariant.

\item {Implementation of Dual Quantum Constraint Algebra}

One important task is to check whether the commutator algebra
(quantum constraint algebra) among the corresponding quantum
operators of constraints both physically and mathematically
coincides with the classical constraint algebra by substituting
quantum constraint operators to classical constraint functionals and
commutators to Poisson brackets. Here the quantum anomaly has to be
avoided in the construction of constraint operators (see the
discussion for Eq.(\ref{anomaly})). First, the subalgebra of the
quantum diffeomorphism constraint algebra is free of anomaly by
construction:
\begin{eqnarray}
\hat{U}_\varphi \hat{U}_{\varphi'}\hat{U}^{-1}_\varphi
\hat{U}^{-1}_{\varphi'}=\hat{U}_{\varphi\circ\varphi'\circ\varphi^{-1}\circ\varphi'^{-1}},\nonumber
\end{eqnarray}
which coincides with the exponentiated version of the Poisson
bracket between two diffeomorphism constraints generating the
transformations $\varphi,\varphi'\in Diff(\Sigma)$.

Second, the quantum constraint algebra between the dual Hamiltonian
constraint operator $\mathcal{S}'(N)$ and the finite diffeomorphism
transformation $\hat{U}_\varphi$ on diffeomorphism-invariant states
coincides with the classical Poisson algebra between
$\mathcal{V}(\vec{N})$ and $\mathcal{S}(M)$. Given a cylindrical
function $\phi_\gamma$ associated with a graph $\gamma$ and the
triangulations $T(\epsilon)$ adapted to the graph $\alpha$, the
triangulations $T(\varphi\circ\epsilon)\equiv \varphi\circ
T(\epsilon)$ are compatible with the graph $\varphi\circ\gamma$.
Then we have by definition:
\begin{eqnarray}
&&\big(-([\hat{\mathcal{S}}(N),\hat{U}_\varphi])'\Psi_{Diff}\big)[\phi_\gamma]\nonumber\\
&=&([\hat{\mathcal{S}}'(N),\hat{U}'_\varphi]\Psi_{Diff})[\phi_\gamma]\nonumber\\
&=&\Psi_{Diff}[\hat{\mathcal{S}}^\epsilon(N)\phi_\gamma-\hat{\mathcal{S}}^\epsilon(N)
\phi_{\varphi\circ\gamma}]\nonumber \\
&=&\sum_{v\in
V(\gamma)}\{N(v)\Psi_{Diff}[\hat{\mathcal{S}}^\epsilon_v\phi_\gamma]-N(\varphi\circ
v)
\Psi_{Diff}[\hat{\mathcal{S}}^{\varphi\circ\epsilon}_{\varphi\circ v}\phi_{\varphi\circ\gamma}]\}\nonumber\\
&=&\sum_{v\in V(\gamma)}[N(v)-N(\varphi\circ v)]\Psi_{Diff}[\hat{\mathcal{S}}^\epsilon_v\phi_\gamma]\nonumber\\
&=&\big(\hat{\mathcal{S}}'(N-\varphi^*N)\Psi_{Diff}\big)[\phi_\gamma].\label{anomalyfree}
\end{eqnarray}
Thus there is no anomaly. However, Eq.(\ref{anomalyfree}) also
explains why the Hamiltonian constraint operator
$\hat{\mathcal{S}}(N)$ cannot leave $\mathcal{H}_{Diff}$ invariant.

Third, we compute the commutator between two Hamiltonian constraint
operators. Notice that
\begin{eqnarray}
&&[\hat{\mathcal{S}}(N),\hat{\mathcal{S}}(M)]\phi_\gamma\nonumber\\
&=&\sum_{v\in
V(\gamma)}[M(v)\hat{\mathcal{S}}(N)-N(v)\hat{\mathcal{S}}(M)]\hat{\mathcal{S}}^{\epsilon}_{v}\phi_\gamma\nonumber\\
&=&\sum_{v\in V(\gamma)}\sum_{v'\in
V(\gamma')}[M(v)N(v')-N(v)M(v')]\hat{\mathcal{S}}^{\epsilon'}_{v'}\hat{\mathcal{S}}^{\epsilon}_{v}\phi_\gamma,\nonumber
\end{eqnarray}
where $\gamma'$ is the graph changed from $\gamma$ by the action of
$\hat{\mathcal{S}}(N)$ or $\hat{\mathcal{S}}(M)$, which adds the
arcs $a_{ij}(\Delta)$ on $\gamma$, $T(\epsilon)$ is the
triangulation adapted to $\gamma$ and $T(\epsilon')$ adapted to
$\gamma'$. Since the newly added vertices by
$\hat{\mathcal{S}}^{\epsilon}_{v}$ is planar, they will never
contributes the final result. So one has
\begin{eqnarray}
&&[\hat{\mathcal{S}}(N),\hat{\mathcal{S}}(M)]\phi_\gamma\nonumber\\
&=&\sum_{v, v'\in V(\gamma), v\neq
v'}[M(v)N(v')-N(v)M(v')]\hat{\mathcal{S}}^{\epsilon'}_{v'}
\hat{\mathcal{S}}^{\epsilon}_{v}\phi_\gamma\nonumber\\
&=&\frac{1}{2}\sum_{v, v'\in V(\gamma), v\neq
v'}[M(v)N(v')-N(v)M(v')][\hat{\mathcal{S}}^{\epsilon'}_{v'}\hat{\mathcal{S}}^{\epsilon}_{v}-
\hat{\mathcal{S}}^{\epsilon'}_{v}\hat{\mathcal{S}}^{\epsilon}_{v'}]\phi_\gamma\nonumber\\
&=&\frac{1}{2}\sum_{v, v'\in V(\gamma), v\neq
v'}[M(v)N(v')-N(v)M(v')][(\hat{U}_{\varphi_{v',v}}-\hat{U}_{\varphi_{v,v'}})
\hat{\mathcal{S}}^{\epsilon}_{v'}\hat{\mathcal{S}}^{\epsilon}_{v}]\phi_\gamma,\nonumber\\
\label{hamiltonian constraint}
\end{eqnarray}
where we have used the facts that
$[\hat{\mathcal{S}}^{\epsilon}_{v},\hat{\mathcal{S}}^{\epsilon'}_{v'}]=0$
for $v\neq v'$and there exists a diffeomorphism $\varphi_{v,v'}$
such that
$\hat{\mathcal{S}}^{\epsilon'}_{v'}\hat{\mathcal{S}}^{\epsilon}_{v}=\hat{U}_{\varphi_{v',v}}
\hat{\mathcal{S}}^{\epsilon}_{v'}\hat{\mathcal{S}}^{\epsilon}_{v}$.
Obviously, we have in the Uniform Rovelli-Smolin Topology
\begin{eqnarray}
([\hat{\mathcal{S}}(N),\hat{\mathcal{S}}(M)])'\Psi_{Diff}=0\nonumber
\end{eqnarray}
for all $\Psi_{Diff}\in Cyl^\star_{Diff}$. As we have seen in
classical expression Eq.(\ref{s-algebra}), the Poisson bracket of
any two Hamiltonian constraints is given by a generator of the
diffeomrophism transformations. Therefore it is mathematically
consistent with the classical expression that two Hamiltonian
constraint operators commute on diffeomorphism invariant states, as
it is presented above. However, as it has been discussed in
\cite{GL}\cite{LM}, the domain of dual Hamiltonian constraint
operator can be extended to a slightly larger space (habitat) in
$Cyl^\star$, whose elements are not necessary diffeomorphism
invariant. And it turns out that the commutator between two
Hamiltonian constraint operators continues to vanish on the habitat,
which seems to be problematic. Fortunately, the quantum operator
corresponding to the right hand side of classical Poisson bracket
(\ref{s-algebra}) also annihilates every state in the habitat
\cite{GL}, so the quantum constraint algebra is consistent at this
level. But it is not clear that whether the quantum constraint
algebra, especially the commutator between two Hamiltonian
constraint is consistent with the classical one (\ref{s-algebra}) on
some larger space in $Cyl^\star$ containing more diffeomorphism
variant states\footnote{However, some scholars disagree with such an
argument involving the habitat and consider the habitat to be
unphysical and completely irrelevant (see, e.g.
Ref.\cite{insider}).}. On the other hand, more works on the
semi-classical analysis are also needed to test the classical limit
of Eq.(\ref{hamiltonian constraint}) and commutation relation
(\ref{s-algebra}). The way to do it is looking for some proper
semi-classical states for calculating the classical limit of the
operators. But due to the graph-changing property of the Hamiltonian
constraint operator, the semi-classical analysis for the Hamiltonian
constraint operator and the quantum constraint algebra is still an
open issue so far.

\item {General Regularization Scheme of the Hamiltonian Constraint}

In \cite{AL}, a general scheme of regulation is introduced for the
quantization of the Hamiltonian constraint, and includes Thiemann's
regularization we introduced above as a specific choice. Such a
general regularization can be summarized as follows: first, we
assign a partition of $\Sigma$ into cells $\Box$ of arbitrary shape.
In every cell of the partition we define edges $s_J$, $J=1,...,n_s$
and loops $\beta_i$, $i=1,...,n_\beta$, where $n_s$, $n_\beta$ may
be different for different cells. We use $\epsilon$ to represent the
scale of the cell $\Box$ Then fix an arbitrary chosen representation
$\rho$ of $SU(2)$. This structure is called a \textit{permissible
classical regulator} if the regulated Hamiltonian constraint
expression with respect to this partition has correct limit when
$\epsilon\rightarrow0$.

Second, we assign the diffeomorphism covariant property and let the
partition adapted to the choice of the graph. That is, given a
cylindrical function $\psi_\gamma\in
Cyl^3_\gamma(\overline{\mathcal{A/G}})$, we make the partition
sufficiently refined that every vertex $v\in V(\gamma)$ is contained
in exact one cell of the partition. And if $(\gamma,\ v)$ is
diffeomorphic to $(\gamma',\ v')$ then, for every $\epsilon$ and
$\epsilon'$, the quintuple $(\gamma,\ v,\ \Box,\ (s_J),\ (\beta_i))$
is diffeomorphic to the quintuple $(\gamma',\ v',\ \Box',\ (s'_J),\
(\beta'_i))$, where $\Box$ and $\Box'$ are the cells in the
partitions with respect to $\gamma$ and $\gamma'$ respectively,
containing $v$ and $v'$ respectively.

As a result, the Hamiltonian constraint operator in this general
regularization scheme is expressed as:
\begin{eqnarray}
\hat{\mathcal{S}}^\epsilon_{E,\gamma}(N)&=&\sum_{v\in
V(\gamma)}\frac{N(v)}{i\hbar\kappa^2\beta}
\sum_{i,J}C^{iJ}\mathrm{Tr}\big((\rho[A(\beta_i)]-\rho[A(\beta^{-1}_i)])\rho[A(s_J^{-1})][\rho[A(s_J)],\
\hat{V}_{U^\epsilon_v}]\big),\nonumber\\
\hat{\mathcal{T}}^\epsilon_\gamma(N)&=&\sum_{v\in V(\gamma)}\frac{i
N(v)}{\hbar^3\kappa^4\beta^3}\sum_{I,J,K}T^{IJK}\mathrm{Tr}\big(\rho[A(s_I^{-1})][\rho[A(s_I)],\
\hat{K}]\rho[A(s_J^{-1})][\rho[A(s_J)],\
\hat{K}]\nonumber\\ &\times&\rho[A(s_K^{-1})][\rho[A(s_K)],\ \hat{V}_{U^\epsilon_v}]\big),\nonumber\\
\hat{\mathcal{S}}^\epsilon(N)\psi_\gamma&=&[\hat{\mathcal{S}}^\epsilon_{E,\gamma}(N)-2(1+\beta^2)
\hat{\mathcal{T}}_{\gamma}^\epsilon(N)]\psi_\gamma,\nonumber
\end{eqnarray}
where $C^{iJ}$ and $T^{IJK}$ are fixed constants independent of the
value of $\epsilon$, the values of them are determined such that the
above expressions have correct classical limits. After removing the
regulator $\epsilon$ via diffeomorphism invariance the same as we
did above, we obtain a well-defined diffeomorphism covariant
operator on $\mathcal{H}_{kin}$ (or $\mathcal{H}^G$) in the sense of
the Uniform Rovelli-Smolin Topology, or dual-define the operator on
some suitable domain in $Cyl^\star$. Note that such a general scheme
of construction exhibits that there is a great deal of freedom in
choosing the regulators, so that there are considerable ambiguities
in our quantization for seeking a proper quantum dynamics for
gravity, which is also an open issue today.

\end{itemize}

\subsection{Master Constraint Programme}

Although the Hamiltonian constraint operator introduced above is
densely defined on $\mathcal{H}_{kin}$ and diffeomorphism covariant,
there are still several unsettled problems which are listed below.
\begin{itemize}

\item It is unclear whether the commutator between two Hamiltonian
constraint operators reproduces the classical Poisson bracket
between two Hamiltonian constraints. Hence it is unclear if the
quantum Hamiltonian constraint produces the correct quantum dynamics
with correct classical limit \cite{GL}\cite{LM}.

\item The dual Hamiltonian constraint operator does not leave the
Hilbert space $\mathcal{H}_{Diff}$ invariant. Thus the inner product
structure of $\mathcal{H}_{Diff}$ cannot be employed in the
construction of physical inner product.

\item Classically the collection of Hamiltonian constraints does not
form a Lie algebra. So one cannot employ group averaging strategy in
solving the Hamiltonian constraint quantum mechanically, since the
strategy depends crucially on group structure.
\end{itemize}
One may see that all above issues come from the properties of the
constraint algebra at classical level. However, if one could
construct an alternative classical constraint algebra, giving the
same constraint phase space, which is a Lie algebra (no structure
functions), where the subalgebra of diffeomorphism constraints forms
an ideal, then the programme of solving the constraints would be in
a much better position. Such a constraint Lie algebra was first
introduced by Thiemann in \cite{thiemann3}. The central idea is to
introduce the master constraint:
\begin{eqnarray}
\textbf{M}:=\frac{1}{2}\int_\Sigma
d^3x\frac{|\widetilde{C}(x)|^2}{\sqrt{|\det
q(x)|}},\label{mconstraint}
\end{eqnarray}
where $\widetilde{C}(x)$ is the scalar constraint in
Eq.(\ref{scalar}). One then gets the master constraint algebra:
\begin{eqnarray}
\{\mathcal{V}(\vec{N}),\ \mathcal{V}(\vec{N}')\}&=&\mathcal{V}([\vec{N},\vec{N}']),\nonumber\\
\{\mathcal{V}(\vec{N}),\ \textbf{M}\}&=&0,\nonumber\\
\{\textbf{M},\ \textbf{M}\}&=&0.\nonumber
\end{eqnarray}
The master constraint programme has been well tested in various
examples \cite{thiemann9}\cite{DT2}\cite{DT3}\\\cite{DT4}\cite{DT5}.
In the following, we extend the diffeomorphism transformations such
that the Hilbert space $\mathcal{H}_{Diff}$ is separable. This
separability of $\mathcal{H}_{Diff}$ and the positivity and the
diffeomorphism invariance of $\textbf{M}$ will be working together
properly and provide us with powerful functional analytic tools in
the programme to solve the constraint algebra quantum mechanically.
The regularized version of the master constraint can be expressed as
\begin{eqnarray}
\textbf{M}^{\epsilon}:=\frac{1}{2}\int_\Sigma d^3y \int_\Sigma
d^3x\chi_\epsilon(x-y)\frac{\widetilde{C}(y)}{\sqrt{V_{U_y^\epsilon}}}
\frac{\widetilde{C}(x)}{\sqrt{V_{U^\epsilon_{x}}}}.\nonumber
\end{eqnarray}
Introducing a partition $\mathcal{P}$ of the 3-manifold $\Sigma$
into cells $C$, we have an operator $\hat{H}^\epsilon_{C}$ acting on
any cylindrical function $f_\gamma\in
Cyl_\gamma(\overline{\mathcal{A/G}})$ in $\mathcal{H}^G$ as
\begin{equation}
\hat{H}^\epsilon_{C}\ f_\gamma=\sum_{v\in
V(\gamma)}\frac{\chi_C(v)}{E(v)}\sum_{v(\Delta)=v}\hat{h}^{\epsilon,\Delta}_{v}
f_\gamma,\label{so1}
\end{equation}
via a family of state-dependent triangulations $T(\epsilon)$ on
$\Sigma$ as we did in the last section, where $\chi_C(v)$ is the
characteristic function of the cell $C(v)$ containing a vertex $v$
of the graph $\gamma$, and the expression of
$\hat{h}^{\epsilon,\Delta}_{v}$ reads
\begin{eqnarray}
\hat{h}^{\epsilon,\Delta}_{v}&=&\frac{16}{3i\hbar\kappa^2\beta}\epsilon^{ijk}\mathrm{Tr}
\big(\hat{A}(\alpha_{ij}(\Delta))^{-1}
\hat{A}(s_k(\Delta))^{-1}[\hat{A}(s_k(\Delta)),
\sqrt{\hat{V}_{U^\epsilon_{v}}}]\big)\nonumber\\
&&+2(1+\beta^2)\frac{4\sqrt{2}}{3i\hbar^3\kappa^4\beta^3}\epsilon^{ijk}\mathrm{Tr}\big(\hat{A}(s_i(\Delta))^{-1}
[\hat{A}(s_i(\Delta)),\hat{K}^\epsilon]\nonumber\\
&&\hat{A}(s_j(\Delta))^{-1}
[\hat{A}(s_j(\Delta)),\sqrt{\hat{V}_{U^\epsilon_{v}}}]\hat{A}(s_k(\Delta))^{-1}[\hat{A}(s_k(\Delta)),
\hat{K}^\epsilon]\big).\label{hath}
\end{eqnarray}
Note that $\hat{h}^{\epsilon,\Delta}_{v}$ is similar to that
involved in the regulated Hamiltonian constraint operator in the
last section, while the only difference is that now the volume
operator is replaced by its quare-root in Eq.(\ref{hath}). Hence the
action of $\hat{H}^\epsilon_{C}$ on $f_\gamma$ adds arcs
$a_{ij}(\Delta)$ with 1/2-representation with respect to each
$v(\Delta)$ of $\gamma$. Thus, for each $\epsilon>0$,
$\hat{H}^\epsilon_{C}$ is a Yang-Mills gauge invariant and
diffeomorphism covariant operator defined on
$Cyl(\overline{\mathcal{A/G}})$. The family of such operators can
give a limit operator $\hat{H}_{C}$ densely defined on
$\mathcal{H}^G$ by the uniform Rovelli-Smollin topology. Then a
master constraint operator, $\hat{\mathbf{M}}$, acting on any
$\Psi_{Diff}\in Cyl^\star_{Diff}$ can be defined as \cite{HM2}
\begin{equation}
(\hat{\mathbf{M}}\Psi_{Diff})[f_\gamma]:=\lim_{\mathcal{P}\rightarrow
\Sigma;\epsilon,\epsilon'\rightarrow\mathrm{0}}\Psi_{Diff}[\sum_{C\in\mathcal{P}}\frac{1}{2}\hat{H}^\epsilon_{C}
(\hat{H}^{\epsilon'}_{C})^\dagger f_\gamma],\label{master}
\end{equation}
for any $f_\gamma$ is a finite linear combination of spin-network
function. Note that $\hat{H}^\epsilon_{C}
(\hat{H}^{\epsilon'}_{C})^\dagger f_\gamma$ is also a finite linear
combination of spin-network functions on an extended graph with the
same skeleton of $\gamma$, hence the value of
$(\hat{\textbf{M}}\Psi_{Diff})[f_\gamma]$ is finite for any given
$\Psi_{Diff}$. Thus $\hat{\textbf{M}}\Psi_{Diff}$ lies in the
algebraic dual of the space of cylindrical functions. Furthermore,
we can show that $\hat{\mathbf{M}}$ leaves the diffeomorphism
invariant distributions invariant. For any diffeomorphism
transformation $\varphi$ on $\Sigma$,
\begin{eqnarray}
(\hat{U}'_\varphi\hat{\textbf{M}}\Psi_{Diff})[f_\gamma]&=&\lim_{\mathcal{P}\rightarrow
\Sigma;\epsilon,\epsilon'\rightarrow\mathrm{0}}\Psi_{Diff}[\sum_{C\in\mathcal{P}}\frac{1}{2}\hat{H}^\epsilon_{C}
(\hat{H}^{\epsilon'}_{C})^\dagger \hat{U}_\varphi
f_\gamma]\nonumber\\
&=&\lim_{\mathcal{P}\rightarrow
\Sigma;\epsilon,\epsilon'\rightarrow\mathrm{0}}\Psi_{Diff}[\hat{U}_\varphi\sum_{C\in\mathcal{P}}
\frac{1}{2}\hat{H}^{\varphi^{-1}(\epsilon)}_{\varphi^{-1}(C)}
(\hat{H}^{\varphi^{-1}(\epsilon')}_{\varphi^{-1}(C)})^\dagger f_\gamma]\nonumber\\
&=&\lim_{\mathcal{P}\rightarrow
\Sigma;\epsilon,\epsilon'\rightarrow\mathrm{0}}\Psi_{Diff}[\sum_{C\in\mathcal{P}}\frac{1}{2}\hat{H}^\epsilon_{C}
(\hat{H}^{\epsilon'}_{C})^\dagger f_\gamma],
\end{eqnarray}
where in the last step, we used the fact that the diffeomorphism
transformation $\varphi$ leaves the partition invariant in the limit
$\mathcal{P}\rightarrow\sigma$ and relabel $\varphi(C)$ to be $C$.
So we have the result
\begin{eqnarray}
(\hat{U}'_\varphi\hat{\textbf{M}}\Psi_{Diff})[f_\gamma]=(\hat{\textbf{M}}\Psi_{Diff})[f_\gamma].\label{diff}
\end{eqnarray}
So given a diffeomorphism invariant spin-network state $T_{[s]}$,
the result state $\hat{\textbf{M}}T_{[s]}$ must be a diffeomorphism
invariant element in the algebraic dual of
$Cyl(\overline{\mathcal{A/G}})$, which means that
\begin{eqnarray}
\hat{\textbf{M}}T_{[s]}=\sum_{[s_1]}c_{[s_1]}T_{[s_1]},\nonumber
\end{eqnarray}
then
\begin{eqnarray}
\lim_{\mathcal{P}\rightarrow
\Sigma;\epsilon,\epsilon'\rightarrow\mathrm{0}}T_{[s]}[\sum_{C\in\mathcal{P}}\frac{1}{2}\hat{H}^\epsilon_{C}
(\hat{H}^{\epsilon'}_{C})^\dagger
T_{s_2}]=\sum_{[s_1]}c_{[s_1]}T_{[s_1]}[T_{s_2}],\nonumber
\end{eqnarray}
where the cylindrical function
$\sum_{C\in\mathcal{P}}\frac{1}{2}\hat{H}^{\epsilon'}_{C}
(\hat{H}^{\epsilon}_{C})^\dagger T_{s_2}$ is a finite linear
combination of spin-network functions on some graphs $\gamma\ '$
with the same skeleton of $\gamma(s_2)$ up to finite number of arcs.
Hence fixing the diffeomorphism equivalence class $[s]$, only for
spin-networks $s_2$ lies in finite number of diffeomorphism
equivalence class the left hand side of the last equation is
non-zero. So there are also only finite number of classes $[s_1]$ in
the right hand side such that $c_{[s_1]}$ is non-zero. As a result,
$\hat{\textbf{M}}T_{[s]}$ is a finite linear combination of
diffeomorphism invariant spin-network states and lies in the Hilbert
space of diffeomorphism invariant states $\mathcal{H}_{Diff}$ for
any $[s]$. And $\hat{\textbf{M}}$ is densely defined on
$\mathcal{H}_{Diff}$.

Given two diffeomorphism invariant spin-network functions
$T_{[s_1]}$ and $T_{[s_2]}$, one can give the matrix elements of
$\hat{\mathbf{M}}$ as \cite{HM2}\cite{HM}
\begin{eqnarray}
&&<T_{[s_1]}|\hat{\textbf{M}}|T_{[s_2]}>_{Diff}\nonumber\\
&=&\overline{(\hat{\textbf{M}}T_{[s_2]})[T_{s_1\in[s_1]}]}\nonumber\\
&=&\lim_{\mathcal{P}\rightarrow
\Sigma;\epsilon,\epsilon'\rightarrow\mathrm{0}}\sum_{C\in\mathcal{P}}\frac{1}{2}
\overline{T_{[s_2]}[\hat{H}^\epsilon_{C}(\hat{H}^{\epsilon'}_{C})^\dagger T_{s_1\in[s_1]}]}\nonumber\\
&=&\lim_{\mathcal{P}\rightarrow
\Sigma;\epsilon,\epsilon'\rightarrow\mathrm{0}}\sum_{C\in\mathcal{P}}\frac{1}{2}
\frac{1}{n_{\gamma(s_2)}}\sum_{\varphi\in
Diff(\Sigma)/Diff_{\gamma(s_2)}}\sum_{\varphi'\in
GS_{\gamma(s_2)}}\nonumber\\
&\times&\overline{<\hat{U}_{\varphi}\hat{U}_{\varphi'}T_{s_2\in[s_2]}
|\hat{H}^\epsilon_{C}(\hat{H}^{\epsilon'}_{C})^\dagger T_{s_1\in[s_1]}>_{Kin}}\nonumber\\
&=&\sum_{s}\lim_{\mathcal{P}\rightarrow
\Sigma;\epsilon,\epsilon'\rightarrow\mathrm{0}}\sum_{C\in\mathcal{P}}\frac{1}{2}
\frac{1}{n_{\gamma(s_2)}}\sum_{\varphi\in
Diff(\Sigma)/Diff_{\gamma(s_2)}}\sum_{\varphi'\in
GS_{\gamma(s_2)}}\nonumber\\
&\times&\overline{<\hat{U}_{\varphi}\hat{U}_{\varphi'}T_{s_2\in[s_2]}
|\hat{H}^\epsilon_{C}T_{s}>_{Kin}<T_{s}|(\hat{H}^{\epsilon'}_{C})^\dagger T_{s_1\in[s_1]}>_{Kin}}\nonumber\\
&=&\sum_{[s]}\sum_{v\in
V(\gamma(s\in[s]))}\frac{1}{2}\lim_{\epsilon,\epsilon'\rightarrow\mathrm{0}}\nonumber\\
&\times&\overline{T_{[s_2]}[\hat{H}^{\epsilon}_{v}
T_{s,c\in[s,c]}]\sum_{s,c\in[s,c]}
<T_{s}|(\hat{H}^{\epsilon'}_{v})^\dagger
T_{s_1\in[s_1]}>_{Kin}},\label{matrix}
\end{eqnarray}
where $Diff_{\gamma}$ is the set of diffeomorphisms leaving the
colored graph $\gamma$ invariant, $GS_{\gamma}$ denotes the graph
symmetry quotient group $Diff_{\gamma}/TDiff_{\gamma}$ where
$TDiff_{\gamma}$ is the diffeomorphism which is trivial on the graph
$\gamma$, and $n_\gamma$ is the number of elements in $GS_\gamma$.
Note that we have used the resolution of identity trick in the
fourth step. Since only a finite number of terms in the sum over
spin-networks $s$, cells $C\in\mathcal{P}$, and diffeomorphism
transformations $\varphi$ are non-zero respectively, we can
interchange the sums and the limit. In the fifth step, we take the
limit $C\rightarrow v$ and split the sum $\sum_{s}$ into
$\sum_{[s]}\sum_{s\in[s]}$, where $[s,c]$ denotes the diffeomorphism
equivalent class associated with $s$. Here we also use the fact
that, given $\gamma(s)$ and $\gamma(s')$ which are different up to a
diffeomorphism transformation, there is always a diffeomorphism
$\varphi$ transforming the graph associated with
$\hat{H}^{\epsilon}_{v} T_{s}\ (v\in\gamma(s))$ to that of
$\hat{H}^{\epsilon}_{v'} T_{s'}\ (v'\in\gamma(s'))$ with
$\varphi(v)=v'$, hence $T_{[s_2]}[\hat{H}^{\epsilon}_{v}
T_{s\in[s]}]$ is constant for different $s\in[s]$.

Since the term $\sum_{s\in[s]}
<T_{s}|(\hat{H}^{\epsilon'}_{v})^\dagger T_{s_1\in[s_1]}>_{Kin}$ is
independent of the parameter $\epsilon'$, one can see that by fixing
a arbitrary family of state-dependent triangulations $T(\epsilon')$,
\begin{eqnarray}
&&\sum_{s\in[s]}
<T_{s}|(\hat{H}^{\epsilon'}_{v})^\dagger T_{s_1\in[s_1]}>_{Kin}\nonumber\\
&=&\sum_{\varphi}<U_\varphi T_{s}|(\hat{H}^{\epsilon'}_{v})^\dagger T_{s_1\in[s_1]}>_{Kin}\nonumber\\
&=&\sum_{\varphi}<\hat{H}^{\epsilon'}_{v}U_\varphi T_{s}|T_{s_1\in[s_1]}>_{Kin}\nonumber\\
&=&\sum_{\varphi}<U_\varphi\hat{H}^{\varphi^{-1}(\epsilon')}_{\varphi^{-1}(v)}T_{s}|T_{s_1\in[s_1]}>_{Kin}\nonumber\\
&=&\overline{T_{[s_1]}[\hat{H}^{\varphi^{-1}(\epsilon')}_{v\in
V(\gamma(s))}T_{s}]},
\end{eqnarray}
where $\varphi$ are the diffeomorphism transformations spanning the
diffeomorphism equivalent class $[s]$. Note that the kinematical
inner product in above sum is non-vanishing if and only if
$\varphi(\gamma(s)))$ coincides with the graph obtained from certain
skeleton $\gamma(s_1)$ by the action of
$(\hat{H}^{\epsilon'}_{v})^\dagger$ and $v\in
V(\varphi(\gamma(s)))$, i.e., the scale $\varphi^{-1}(\epsilon')$ of
the diffeomorphism images of the tetrahedrons added by the action
coincides with the scale of certain tetrahedrons in $\gamma(s)$ and
$\varphi^{-1}(v)$ is a vertex in $\gamma(s)$. Then we can express
the matrix elements (\ref{matrix}) as:
\begin{eqnarray}
&&<T_{[s_1]}|\hat{\textbf{M}}|T_{[s_2]}>_{Diff}\nonumber\\
&=&\sum_{[s]}\sum_{v\in
V(\gamma(s\in[s]))}\frac{1}{2}\lim_{\epsilon,\epsilon'\rightarrow\mathrm{0}}
\overline{T_{[s_2]}[\hat{H}^{\epsilon}_{v}
T_{s\in[s]}]}T_{[s_1]}[\hat{H}^{\epsilon'}_{v}T_{s\in[s]}]\nonumber\\
&=&\sum_{[s]}\sum_{v\in
V(\gamma(s\in[s]))}\frac{1}{2}\overline{(\hat{H}'_v T_{[s_2]})[
T_{s\in[s]}]}(\hat{H}'_v T_{[s_1]}) [ T_{s\in[s]}].\label{master2}
\end{eqnarray}
From Eq.(\ref{master2}) and the fact that the master constraint
operator $\hat{\mathbf{M}}$ is densely defined on
$\mathcal{H}_{Diff}$, it is obvious that $\hat{\mathbf{M}}$ is a
positive and symmetric operator in ${\cal H}_{Diff}$. Therefore, the
quadratic form $Q_{\mathbf{M}}$ associated with $\hat{\mathbf{M}}$
is closable \cite{rs}. The closure of $Q_{\mathbf{M}}$ is the
quadratic form of a unique self-adjoint operator
$\hat{\overline{\mathbf{M}}}$, called the Friedrichs extension of
$\hat{\mathbf{M}}$. We relabel $\hat{\overline{\mathbf{M}}}$ to be
$\hat{\mathbf{M}}$ for simplicity. From the construction of
$\hat{\mathbf{M}}$, the qualitative description of the kernel of the
Hamiltonian constraint operator in Ref.\cite{thiemann5} can be
transcribed to describe the solutions to the equation:
$\hat{\mathbf{M}}\Psi_{Diff}=0$. In particular, the diffeomorphism
invariant cylindrical functions based on at most 2-valent graphs are
obviously normalizable solutions. In conclusion, there exists a
positive and self-adjoint operator $\hat{\mathbf{M}}$ on
$\mathcal{H}_{Diff}$ corresponding to the master constraint
(\ref{mconstraint}), and zero is in the point spectrum of
$\hat{\mathbf{M}}$.

Note that the quantum constraint algebra can be easily checked to be
anomaly free. i.e.,
\begin{eqnarray}
[\hat{\mathbf{M}},\hat{U}'_\varphi]=0,\ \ \
[\hat{\mathbf{M}},\hat{\mathbf{M}}]=0.\nonumber
\end{eqnarray}
which is consistent with the classical master constraint algebra in
this sense. As a result, the difficulty of the original Hamiltonian
constraint algebra can be avoided by introducing the master
constraint algebra, due to the Lie algebra structure of the latter.
Since zero is in the spectrum of $\hat{\mathbf{M}}$
\cite{thiemann15}, the further task is to obtain the physical
Hilbert space $\mathcal{H}_{phys}$ which is the kernel of the master
constraint operator with some suitable physical inner product, and
the issue of quantum anomaly is represented in terms of the size of
$\mathcal{H}_{phys}$ and the existence of semi-classical states.
Note that we will see in the next section that the master constraint
programme can be straightforwardly generalized to include matter
fields \cite{HM}. We list some open problems in the master
constraint programme for further research.
\begin{itemize}

\item Kernel of Master Constraint Operator

Since the master constraint operator $\hat{\textbf{M}}$ is
self-adjoint, it is a practical problem to define DID of
$\mathcal{H}_{Diff}$:
\begin{eqnarray}
\mathcal{H}_{Diff}&\sim&\int^\oplus
d\mu(\lambda)\mathcal{H}^\oplus_\lambda,\nonumber\\
<\Phi|\Psi>_{Diff}&=&\int_\mathbf{R}d\mu(\lambda)<\Phi|\Psi>_{\mathcal{H}^\oplus_{\lambda}},\nonumber
\end{eqnarray}
where $\mu(\lambda)$ is the spectral measure with respect to the
master constraint operator $\hat{\mathbf{M}}$. It is expected that
we can identify $\mathcal{H}^\oplus_{\lambda=0}$ with the physical
Hilbert space. However, such a prescription is ambiguous in the case
that zero is only in the continuous spectrum, loses physical
information in the case that zero is an embedded eigenvalue and
unambiguous only if zero is an isolated eigenvalue in which case
however the whole machinery of the DID is not needed at all because
$\mathcal{H}^\oplus_{\lambda=0}\subset\mathcal{H}_{Diff}$ and the
physical inner product coincide with the kinematical
(differomorphism invariant) one \cite{thiemann9}. There are some
improved prescriptions also presented in \cite{thiemann9} by
decomposing the measure with respect to the spectrum types before
direct integral decomposition, some ambiguities can be canceled by
some physical criterion, e.g., a complete subalgebra of bounded
Dirac observables should be represented irreducibly as self-adjoint
operators on the physical Hilbert space, and the resulting physical
Hilbert space should admits a sufficient number of semiclassical
states. Nonetheless, due to the complicated structure of the master
constraint operator, it is difficult anyhow to manage the spectrum
analysis and direct integral decomposition. On the other hand, for
the self-adjointness of the master constraint operator and the
Lie-algebra structure of the constraint algebra, a formal group
averaging strategy was introduced in \cite{thiemann3} as a more
concrete way to get the physical Hilbert space. It is realized by a
formal rigged map $\eta_{phys}$:
\begin{eqnarray}
\eta_{phys}:&&Cyl^\star_{Diff}\rightarrow \Phi_{phys}\nonumber\\
&&f\mapsto\eta_{phys}(f):=\int_{\mathbf{R}}\frac{dt}{2\pi}<e^{i\hat{\mathbf{M}}t}f|\
.>_{Diff},\nonumber
\end{eqnarray}
where $e^{i\hat{\mathbf{M}}t}$ is a one parameter continuous
unitary group on $\mathcal{H}_{Diff}$ by the self-adjointness of
$\hat{\mathbf{M}}$, and $\Phi_{phys}$ is a subset of the algebraic
dual of $Cyl^\star_{Diff}$. It is trivial to see that
$\eta_{phys}(f)$ is invariant under the (dual) transformation of
$e^{i\hat{\mathbf{M}}t}$. Thus a inner product can be formally
defined between two algebraic functionals $\eta_{phys}(f)$ and
$\eta_{phys}(f')$ in $\Phi_{phys}$ via:
\begin{eqnarray}
<\eta_{phys}(f)|\eta_{phys}(f')>_{phys}&:=&\eta_{phys}(f)[f'],\nonumber\\
&=&\int_{\mathbf{R}}\frac{dt}{2\pi}<e^{i\hat{\mathbf{M}}t}f|f'>_{Diff}\nonumber\\
&=&\int_{\mathbf{R}}\frac{dt}{2\pi}\int_{\mathbf{R}}d\mu(\lambda)e^{i\lambda
t}<f(\lambda)|f'(\lambda)>_{\mathcal{H}^\oplus_\lambda}\nonumber\\
&=&\int_{\mathbf{R}}d\mu(\lambda)\delta(\lambda)<f(\lambda)|f'(\lambda)>_{\mathcal{H}^\oplus_\lambda}\nonumber\\
&=&[\int_{\mathbf{R}}d\mu(\lambda)\delta(\lambda)]<f(0)|f'(0)>_{\mathcal{H}^\oplus_{\lambda=0}},\nonumber
\end{eqnarray}
where we have used the spectrum decomposition with respect to the
self-adjoint operator $\hat{\mathbf{M}}$, the operator
$e^{i\hat{\mathbf{M}}t}$ is represented by multiplication by a
number $e^{i\lambda t}$ on each $\mathcal{H}^\oplus_\lambda$, and
the vector valued function $f(\lambda)$ is the spectrum
decomposition representation of state $f\in\mathcal{H}_{Diff}$.
Although we can see from the above argument that the physical inner
product is proportional to the inner product in the fiber Hilbert
space $\mathcal{H}^\oplus_{\lambda=0}$, unfortunately, the factor
$\int_{\mathbf{R}}d\mu(\lambda)\delta(\lambda)$ is divergent when
$\mu$ has pure point part, e.g. zero is in the discrete spectrum of
$\hat{\mathbf{M}}$. That is one reason why we claim that the above
argument is formal.

On the other hand, the group averaging strategy and the formal
physical inner product we just defined has potential relationships
with path-integral formulation and spin foam models due to the
positivity of the master constraint operator $\hat{\mathbf{M}}$
\cite{thiemann3}, and hopefully, we may obtain the physical
transition amplitude from this physical inner product in the future.
However, the whole technique of group averaging for solving the
master constraint is still formal so far, and the rigorous
calculations for it has not done yet as far as we know.

\item Dirac Observables

Classically, one can prove that a function $\mathcal{O}\in
C^\infty(\mathcal{M})$ is a weak observable with respect to the
scalar constraint if and only if
\begin{eqnarray}
\{\mathcal{O},\{\mathcal{O},\textbf{M}\}\}|_{\overline{\mathcal{M}}}=0.\nonumber
\end{eqnarray}
We define $\mathcal{O}$ to be a strong observable with respect to
the scalar constraint if and only if
\begin{eqnarray}
\{\mathcal{O},\textbf{M}\}|_{\mathcal{M}}=0,\nonumber
\end{eqnarray}
and to be a ultra-strong observable if and only if
\begin{eqnarray}
\{\mathcal{O},\mathcal{S}(N)\}|_{\mathcal{M}}=0.\nonumber
\end{eqnarray}
In quantum version, an observable $\hat{\mathcal{O}}$ is a weak
Dirac observable if and only if $\hat{\mathcal{O}}$ leaves
$\mathcal{H}_{phys}$ invariant, while $\hat{\mathcal{O}}$ is now
called a strong Dirac observable if and only if $\hat{\mathcal{O}}$
commutes with the master constraint operator $\hat{\textbf{M}}$.
Given a bounded self-adjoint operator $\hat{\mathcal{O}}$ defined on
$\mathcal{H}_{Diff}$, for instance, a spectral projection of some
observables leaving $\mathcal{H}_{Diff}$ invariant, if the uniform
limit exists, the bounded self-adjoint operator defined by group
averaging
\begin{eqnarray}
\widehat{[\mathcal{O}]}:=\lim_{T\rightarrow\infty}\frac{1}{2T}\int_{-T}^{T}dt\
\hat{U}(t)^{-1} \hat{\mathcal{O}}\hat{U}(t)\nonumber
\end{eqnarray}
commutes with $\hat{\textbf{M}}$ and hence becomes a strong Dirac
observable on the physical Hilbert space.

\item Testing the Classical Limit of the Master Constraint Operator

One needs to construct spatial diffeomorphism invariant
semiclassical states to calculate the expectation value and
fluctuation of the master constraint operator. If the results
coincide with the classical values up to $\hbar$ corrections, one
can go ahead to finish our quantization programme with confidence.

\end{itemize}

\newpage

\section{Quantum Matter Field on a Quantum Background}

In ordinary quantum field theory, the quantum field is defined on a
smooth background spacetime. However, it is expected that the smooth
structure of a spacetime may break down at Planck scale, so the
present treatment of quantum field theory is valid only in a
semiclassical sense. Thus we would like to modify the formulation of
present quantum field theory to make it compatible with the quantum
theory of gravity(spacetime) which we already established in
previous sections so as to explore the behavior of the quantum
matter field under Planck scale and at extremely strong
gravitational fields, e.g. inside the black hole or at the early age
of the universe.

In the following, an alternative quantization of scalar field will
be introduced, the advantage of such a quantization scheme is that
the quantum scalar field doesn't depend on the background. We will
also see that the quantization technique for the previous
Hamiltonian constraint can be generalized to quantize the
Hamiltonian of matter fields coupled to gravity. Then it is shown
that an operator corresponding to the Hamiltonian of the scalar
field can be well defined on the coupled diffeomorphism invariant
Hilbert space. It is even positive and self-adjoint without any
divergence. Thus quantum gravity acts exactly as a natural regulator
for the quantum scalar field in the polymer representation.
Moreover, to study the whole dynamical system of the scalar field
coupled to gravity, a Hamiltonian constraint operator is defined in
the coupled kinematical Hilbert space. The contribution of the
scalar field to the Hamiltonian constraint can be promoted to a
positive self-adjoint operator. To avoid possible quantum anomalies
and find the physical Hilbert space, we will also introduce the
master constraint programme for the coupled system. A self-adjoint
master constraint operator is obtained in the diffeomorphism
invariant Hilbert space, which assures the feasibility of the
programme.

\subsection{Polymer-like Representation of a Scalar Field}

We begin with the total Hamiltonian of the gravity coupled with a
massless real scalar field which is a linear combination of
constraints:
\begin{eqnarray}
\mathcal{H}_{tot}=\Lambda^iG_i+N^aC_a+NC,\nonumber
\end{eqnarray}
where $\Lambda^i$, $N^a$ and $N$ are Lagrange multipliers, and the
three constraints in the Hamiltonian are expressed as
\cite{AR}\cite{han}:
\begin{eqnarray}
G_i&=&D_a\widetilde{P}^a_i\ :=\
\partial_a\widetilde{P}^a_i+\epsilon_{ij}^{\
\ k}A_a^i\widetilde{P}^a_k,\\
C_a&=&\widetilde{P}^b_iF_{ab}^i-A^i_aG_i+\widetilde{\pi}\partial_a\phi,\\
C&=&\frac{\kappa\beta^2}{2\sqrt{|\det
q|}}\widetilde{P}^a_i\widetilde{P}^b_j[\epsilon^{ij}_{\ \
k}F^k_{ab}-2(1+\beta^2)K^i_{[a}K^j_{b]}]\nonumber\\
&+&\frac{1}{\sqrt{|\det
q|}}[\frac{\kappa^2\beta^2\alpha_M}{2}\delta^{ij}\widetilde{P}^{a}_{i}\widetilde{P}^{b}_{j}(\partial_a
\phi)\partial_b\phi
+\frac{1}{2\alpha_M}\widetilde{\pi}^2],\label{kgconstraint}
\end{eqnarray}
here the real number $\alpha_M$ is the coupling constant, and
$\widetilde{\pi}$ denotes the momentum conjugate to $\phi$:
\begin{eqnarray}
\widetilde{\pi}:=\frac{\partial\cal
L}{\partial\dot{\phi}}=\frac{\alpha_M}{N}\sqrt{|\det
q|}(\dot{\phi}-N^a\partial_a\phi).\nonumber
\end{eqnarray}
Thus one has the elementary Poisson brackets
\begin{eqnarray}
\{A^i_a(x),\widetilde{P}^b_j(y)\}&=&\delta^a_b\delta^i_j\delta(x,y),\nonumber\\
\{\phi(x),\widetilde{\pi}(y)\}&=&\delta(x,y).\nonumber
\end{eqnarray}
Note that the second term of the Hamiltonian constraint
(\ref{kgconstraint}) is just the Hamiltonian of the real scalar
field.

Then we look for the background independent representation for the
real scalar field coupled to gravity, following the polymer
representation of the scalar field \cite{ALS}. The classical
configuration space, $\mathcal{U}$, consists of all real-valued
smooth functions $\phi$ on $\Sigma$. Given a set of a finite number
of points $X=\{x_1,...,x_N\}$ in $\Sigma$, a equivalence relation
can be defined by: given two scalar field $\phi_1,
\phi_2\in\mathcal{U}$, $\phi_1\sim\phi_2$ if and only if $\exp[i
\lambda_i\phi_1(x_i)]=\exp[i \lambda_j\phi_2(x_j)]$ for all $x_i\in
X$ and all real number $\lambda_j$. Hence we obtain a bijection
between $\mathcal{U}/\sim$ and $\overline{\mathbf{R}}_X$, which is
$N$ copies of the Bohr compactification of $\mathbf{R}$
\cite{sternberg}. Since one can define a projective family with
respect to the set of point (graph for scalar field), thus a
projective limit $\overline{\mathcal{U}}$, which is a compact
topological space, is obtained as the quantum configuration space of
scalar field. Next, we denote by $Cyl_X(\overline{\mathcal{U}})$ the
vector space generated by finite linear combinations of the
following functions of $\phi$:
\begin{eqnarray}
T_{X,\bf{\lambda}}(\phi):=\prod_{x_j\in X}\exp[i
\lambda_j\phi(x_j)],\nonumber
\end{eqnarray}
where $\bf{\lambda}\equiv (\lambda_1, \lambda_2, \cdot\cdot\cdot,
\lambda_N)$ are arbitrary non-zero real numbers assigned at each
point. It is obvious that $Cyl_X(\overline{\mathcal{U}})$ has the
structure of a $*$-algebra. The vector space
$Cyl(\overline{\mathcal{U}})$ of all cylindrical functions on
$\mathcal{U}$ is defined by the linear span of the linear span of
$T_0=1$ and $T_{X,\bf{\lambda}}$. Completing
$Cyl(\overline{\mathcal{U}})$ with respect to the sup norm, one
obtains a unital Abelian $C$*-algebra
$\overline{Cyl(\overline{\mathcal{U}})}$. Thus one can use the GNS
structure to construct its cyclic representations. A preferred
positive linear functional $\omega_0$ on
$\overline{Cyl(\overline{\mathcal{U}})}$ is defined by
\begin{eqnarray}
\omega_0(T_{X,\bf{\lambda}})=\left\{%
\begin{array}{ll}
    1 & \hbox{if $\lambda_j=0$ $\forall j$} \\
    0 & \hbox{otherwise}, \\
\end{array}%
\right.\ \nonumber
\end{eqnarray}
which defines a diffeomorphism-invariant faithful Borel measure
$\mu$ on $\overline{\mathcal{U}}$ as
\begin{eqnarray}\label{smeasure}
\int_\mathcal{U}d\mu(T_{X,\bf{\lambda}})=\left\{%
\begin{array}{ll}
    1 & \hbox{if $\lambda_j=0$ $\forall j$} \\
    0 & \hbox{otherwise}. \\
\end{array}%
 \right.\
\end{eqnarray}
Thus one obtains the Hilbert space, $\mathcal{H}^{KG}_{kin}$ which
is defined by $L^2(\overline{\mathcal{U}}, d\mu)$, of square
integrable functions on a compact topological measure space
$\overline{\mathcal{U}}$ with respect to $\mu$. The inner product
can be expressed explicitly as:
\begin{equation}
<T_c|T_{c'}>^{KG}_{kin}=\delta_{cc'},\nonumber
\end{equation}
where the label $c:=(X, \bf{\lambda})$ are called scalar-network.

As one might expect, the quantum configuration space
$\overline{\mathcal{U}}$ is just the Gel'fand spectrum of
$\overline{Cyl(\overline{\mathcal{U}})}$. More concretely, for a
single point set $X_0\equiv \{x_0\}$,
$Cyl_{X_0}(\overline{\mathcal{U}})$ is the space of all almost
periodic functions on a real line $\mathbf{R}$. The Gel'fand
spectrum of the corresponding $C$*-algebra
$\overline{Cyl_{X_0}(\overline{\mathcal{U}})}$ is the Bohr
completion $\overline{\mathbf{R}}_{x_0}$ of $\mathbf{R}$ \cite{ALS},
which is a compact topological space such that
$\overline{Cyl_{X_0}(\overline{\mathcal{U}})}$ is the $C$*-algebra
of all continuous functions on $\overline{\mathbf{R}}_{x_0}$. Since
$\mathbf{R}$ is densely embedded in $\overline{\mathbf{R}}_{x_0}$,
$\overline{\mathbf{R}}_{x_0}$ can be regarded as a completion of
$\mathbf{R}$.

It is clear from Eq.(\ref{smeasure}) that an orthonomal basis in
$\mathcal{H}^{KG}_{kin}$ is given by the scalar vacuum $T_0=1$ and
so-called scalar-network functions $T_c(\phi)$, where
$c=(X,\bf{\lambda})$ and $\bf{\lambda}\equiv (\lambda_1, \lambda_2,
\cdot\cdot\cdot, \lambda_N)$ are non-zero real numbers. So the total
kinematical Hilbert space $\mathcal{H}_{kin}$ is the direct product
of the kinematical Hilbert space $\mathcal{H}_{kin}^{GR}$ for
gravity and the kinematical Hilbert space for real scalar field,
i.e.,
$\mathcal{H}_{kin}:=\mathcal{H}_{kin}^{GR}\otimes\mathcal{H}_{kin}^{KG}$.
Then the spin-scalar-network state $T_{s,c}\equiv T_s(A)\otimes
T_c(\phi)\in Cyl_{\gamma(s)}(\overline{\mathcal{A/G}})\otimes
Cyl_{X(c)}(\overline{\mathcal{U}})\equiv Cyl_{\gamma(s,c)}$ is a
gravity-scalar cylindrical function on graph
$\gamma(s,c)\equiv\gamma(s)\cup X(c)$. Note that generally $X(c)$
may not coincide with the vertices of the graph $\gamma(s)$. It is
straightforward to see that all of these functions constitutes an
orthonormal basis in $\mathcal{H}_{kin}$ as
\begin{eqnarray}
<T_{s'}(A)\otimes T_{c'}(\phi)|T_s(A)\otimes
T_c(\phi)>_{kin}=\delta_{s's}\delta_{c'c}\ .\nonumber
\end{eqnarray}
Note that none of $\mathcal{H}_{kin}$, $\mathcal{H}^{GR}_{kin}$ and
$\mathcal{H}^{KG}_{kin}$ is a separable Hilbert space.

Given a pair $(x_0, \lambda_0)$, there is an elementary
configuration for the scalar field, the so-called point holonomy,
\begin{eqnarray}
U(x_0,\lambda_0):=\exp[i\lambda_0\phi(x_0)].\nonumber
\end{eqnarray}
It corresponds to a configuration operator $\hat{U}(x_0,\lambda_0)$,
which acts on any cylindrical function $\psi(\phi)\in
Cyl_{X(c)}(\overline{\mathcal{U}})$ by
\begin{equation}
\hat{U}(x_0,\lambda_0)\psi(\phi)=U(x_0,\lambda_0)\psi(\phi).\nonumber
\end{equation}
All these operators are unitary. But since the family of operators
$\hat{U}(x_0,\lambda)$ fails to be weakly continuous in $\lambda$,
there is no field operator $\hat{\phi}(x)$ on
$\mathcal{H}^{KG}_{kin}$. The momentum functional smeared on a
3-dimensional region $R\subset\Sigma$ is expressed by
\begin{eqnarray}
\pi(R):=\int_R d^3x \widetilde{\pi}(x).\nonumber
\end{eqnarray}
The Poisson bracket between the momentum functional and a point
holonomy can be easily calculated to be
\begin{eqnarray}
\{\pi(R), U(x,\lambda)\}=-i\lambda\chi_R(x)U(x, \lambda),\nonumber
\end{eqnarray}
where $\chi_R(x)$ is the characteristic function for the region $R$.
So the momentum operator is defined by the action on scalar network
functions $T_{c=(X,\bf{\lambda})}$ as
\begin{eqnarray}
\hat{\pi}(R)T_{c}(\phi):=i\hbar\{\pi(R),
T_c(\phi)\}=\hbar[\sum_{x_j\in
X}\lambda_j\chi(x_j)]T_c(\phi).\nonumber
\end{eqnarray}
Now we can impose the quantum constraints on $\mathcal{H}_{kin}$ and
consider the quantum dynamics. First, the Gauss constraint can be
solved independently of $\mathcal{H}_{kin}^{KG}$, since it only
involves the gravitational field. It is also expected that the
diffeomorphism constraint can be implemented by the group averaging
strategy in the similar way as in the case of pure gravity. Given a
spatial diffeomorphism transformation $\varphi$, a unitary
transformation $\hat{U}_\varphi$ was induced by $\varphi$ in the
Hilbert space $\mathcal{H}_{kin}$, which is expressed as
\begin{eqnarray}
\hat{U}_\varphi
T_{s=(\gamma(s),\mathbf{j},\mathbf{i}),c=(X(c),\mathbf{\lambda})}
=T_{\varphi\circ
s=(\varphi(\gamma(s)),\mathbf{j},\mathbf{i}),\varphi\circ
c=(\varphi(X(c)),\mathbf{\lambda})}.\nonumber
\end{eqnarray}
Then the differomorphism invariant spin-scalar-network functions are
defined by group averaging as
\begin{eqnarray}
T_{[s,c]}:=\frac{1}{n_{\gamma(s,c)}}\sum_{\varphi\in
Diff(\Sigma)/Diff_{\gamma(s,c)}}\sum_{\varphi'\in
GS_{\gamma(s,c)}}\hat{U}_{\varphi}\hat{U}_{\varphi'}T_{s,c},
\end{eqnarray}
where $Diff_{\gamma}$ is the set of diffeomorphisms leaving the
colored graph $\gamma$ invariant, $GS_{\gamma}$ denotes the graph
symmetry quotient group $Diff_{\gamma}/TDiff_{\gamma}$ where
$TDiff_{\gamma}$ is the set of the diffeomorphisms which is trivial
on the graph $\gamma$, and $n_\gamma$ is the number of elements in
$GS_\gamma$. Following the standard strategy in quantization of pure
gravity, an inner product can be defined on the vector space spanned
by the diffeomorphism invariant spin-scalar-network functions (and
the vacuum states for gravity, scalar and both respectively) such
that they form an orthonormal basis as:
\begin{eqnarray}
<T_{[s,c]}|T_{[s',c']}>_{Diff}:=T_{[s,c]}[T_{s',c'\in[s',c']}]=\delta_{[s,c],[s',c']}.
\end{eqnarray}
After the completion procedure, we obtain the expected Hilbert space
of diffeomorphism invariant states for the scalar field coupled to
gravity, which is denoted by $\mathcal{H}_{Diff}$.

\subsection{Diffeomorphism Invariant Hamiltonian of a Scalar Field}

In the following discussion, we consider the quantum scalar field on
a fluctuating background. A similar idea was considered in
Ref.\cite{thiemann8}, where a Hamiltonian operator with respect to a
U(1) group representation of the scalar field is defined on a
kinematical Hilbert space $\mathcal{H}_{kin'}$ of matter coupled to
gravity. Then an effective Hamiltonian operator of the scalar field
can be constructed as a quadratic form via
\begin{eqnarray}
&&<\psi_{matter},\
\hat{H}^{eff}_{matter}(m)\ \psi'_{matter}>^{KG}_{kin'}\nonumber\\
&:=&<\psi_{grav}(m)\otimes\psi_{matter},\ \hat{H}_{matter}\
\psi_{grav}(m)\otimes\psi'_{matter}>_{kin'},
\end{eqnarray}
where $\psi_{grav}(m)\in \mathcal{H}^{GR}_{kin}$ presents a
semiclassical state of gravity approximating some classical
spacetime background $m$ where the quantum scalar field lives. Thus
the effective Hamiltonian operator $\hat{H}^{eff}_{matter}(m)$ of
scalar field contains also the information of the fluctuating
background \emph{metric}. In the light of this idea, we will
construct a Hamiltonian operator $\hat{\mathcal{S}}_{KG}$ for scalar
field in the polymer-like representation. It turns out that this
Hamiltonian operator can be defined in the Hilbert space
$\mathcal{H}_{Diff}$ of diffeomorphism invariant states for scalar
field coupled to gravity without UV-divergence. So the quantum
dynamics of the scalar field is obtained in a diffeomorphism
invariant way, which is expected in the programme of loop quantum
gravity. Thus, here an effective Hamiltonian operator of the scalar
field could be extracted in $\mathcal{H}_{Diff}$ by defining
$<\Psi_{[m]}(A,\phi),\ \hat{\mathcal{S}}_{KG}\
\Psi_{[m]}(A,\phi)>_{Diff}$ to be its expectation value on
diffeomorphism invariant states $\Psi(\phi)$ of the scalar field,
where the diffeomorphism invariant semiclassical state
$\Psi_{[m]}(A)$ represents certain fluctuating geometry with spatial
diffeomorphism invariance, and the label $[m]$ denotes the classical
geometry approximated by $\Psi_{[m]}(A)$. Moreover, the quadratic
properties of the scalar field Hamiltonian will provide powerful
functional analytic tools in the quantization procedure, such that
the self-adjointness of the Hamiltonian operator can be proved by a
theorem in functional analysis.

Then the crucial point is to define an operator corresponding to the
Hamiltonian functional $\mathcal{S}_{KG}$ of the scalar field, which
can be decomposed into two parts
\begin{eqnarray}
\mathcal{S}_{KG}&=&\mathcal{S}_{KG,\phi}+\mathcal{S}_{KG,Kin},\nonumber
\end{eqnarray}
where
\begin{eqnarray}
\mathcal{S}_{KG,\phi}&=&\frac{\kappa^2\beta^2\alpha_M}{2}\int_\Sigma
d^3x\frac{1}{\sqrt{|\det
q|}}\delta^{ij}\widetilde{P}^{a}_{i}\widetilde{P}^{b}_{j}(\partial_a\phi)\partial_b\phi,\nonumber\\
\mathcal{S}_{KG,Kin}&=&\frac{1}{2\alpha_M}\int_\Sigma
d^3x\frac{1}{\sqrt{|\det q|}}\widetilde{\pi}^2\nonumber.
\end{eqnarray}
We will employ the following identities:
\begin{eqnarray}
\widetilde{P}^a_i=\frac{1}{2\kappa\beta}\widetilde{\eta}^{abc}\epsilon_{ijk}e^j_be^k_c\
\ \ \ \mathrm{and} \ \ \ \
e^i_a(x)=\frac{2}{\kappa\beta}\{A^i_a(x),V_{U_x}\},\nonumber
\end{eqnarray}
where $\widetilde{\eta}^{abc}$ denotes the Levi-Civita tensor
tensity and $V_{U_x}$ is the volume of an arbitrary neighborhood
$U_x$ containing the point $x$. By using the point-splitting
strategy, the regulated version of the Hamiltonian is obtained as:
\begin{eqnarray}
\mathcal{S}_{KG,\phi}
&=&\frac{\kappa^2\beta^2\alpha_M}{2}\int_\Sigma d^3y\int_\Sigma
d^3x\chi_\epsilon(x-y)\delta^{ij}\times\nonumber\\
&&\frac{1}{\sqrt{V_{U^\epsilon_x}}}\widetilde{P}^{a}_{i}(x)(\partial_a\phi(x))\frac{1}{\sqrt{V_{U^\epsilon_y}}}
\widetilde{P}^{b}_{j}(y)\partial_b\phi(y)\nonumber\\
&=&\frac{32\alpha_M}{\kappa^4\beta^4}\int_\Sigma d^3y\int_\Sigma
d^3x\chi_\epsilon(x-y)\delta^{ij}\times\nonumber\\
&&\widetilde{\eta}^{aec}(\partial_a\phi(x))\mathrm{Tr}\big(\tau_i\{\mathbf{A}_e(x),V_{U^\epsilon_x}^{3/4}\}
\{\mathbf{A}_c(x),V_{U^\epsilon_x}^{3/4}\}\big)\times\nonumber\\
&&\widetilde{\eta}^{bfd}(\partial_b\phi(y))\mathrm{Tr}\big(\tau_j\{\mathbf{A}_f(y),V_{U^\epsilon_y}^{3/4}\}
\{\mathbf{A}_d(y),V_{U^\epsilon_y}^{3/4}\}\big),\nonumber\\
\mathcal{S}_{KG,Kin} &=&\frac{1}{2\alpha_M}\int_\Sigma
d^3x\widetilde{\pi}(x)\int_\Sigma
d^3y\widetilde{\pi}(y)\times\nonumber\\
&&\int_\Sigma
d^3u\frac{\det(e_a^i(u))}{(V_{U^\epsilon_u})^{3/2}}\int_\Sigma
d^3w\frac{\det(e_a^i(w))}{(V_{U^\epsilon_w})^{3/2}}\chi_\epsilon(x-y)\chi_\epsilon(u-x)\chi_\epsilon(w-y)\nonumber\\
&=&\frac{1}{2\alpha_M}\frac{2^8}{9(\kappa\beta)^6}\int_\Sigma
d^3x\widetilde{\pi}(x)\int_\Sigma
d^3y\widetilde{\pi}(y)\times\nonumber\\
&&\int_\Sigma d^3u\
\widetilde{\eta}^{abc}\mathrm{Tr}\big(\{\mathbf{A}_a(u),\sqrt{V_{U^\epsilon_u}}\}
\{\mathbf{A}_b(u),\sqrt{V_{U^\epsilon_u}}\}\{\mathbf{A}_c(u),\sqrt{V_{U^\epsilon_u}}\}\big)\times\nonumber\\
&&\int_\Sigma d^3w\
\widetilde{\eta}^{def}\mathrm{Tr}\big(\{\mathbf{A}_d(w),\sqrt{V_{U^\epsilon_w}}\}
\{\mathbf{A}_e(w),\sqrt{V_{U^\epsilon_w}}\}\{\mathbf{A}_f(w),\sqrt{V_{U^\epsilon_w}}\}\big)\times\nonumber\\
&&\chi_\epsilon(x-y)\chi_\epsilon(u-x)\chi_\epsilon(w-y),\nonumber
\end{eqnarray}
where we denote by $\mathbf{A}_a\equiv A_a^i\tau_i$,
$\chi_\epsilon(x-y)$ the characteristic function of a box containing
$x$ with scale $\epsilon$ such that
$\lim_{\epsilon\rightarrow0}\chi_\epsilon(x-y)/{\epsilon^3}=\delta(x-y)$,
and $V_{U^\epsilon_x}$ is the volume of the box. In order to
quantize the Hamiltonian $\mathcal{S}_{KG}$ as a well-defined
operator in the polymer-like representation, we have to express the
classical formula of the Hamiltonian in terms of elementary
variables with clear quantum analogs by introducing a triangulation
$T(\epsilon)$ of $\Sigma$, where the parameter $\epsilon$ describes
how fine the triangulation is. The quantity regulated on the
triangulation is required to have correct limit when
$\epsilon\rightarrow0$. Given a tetrahedron $\Delta\in T(\epsilon)$,
we use $\{s_i(\Delta)\}_{i=1,2,3}$ to denote the three outgoing
oriented segments in $\Delta$ with a common beginning point
$v(\Delta)=s(s_i(\Delta))$ and use $a_{ij}(\Delta)$ to denote the
arcs connecting the end points of $s_i(\Delta)$ and $s_j(\Delta)$.
Then several loops $\alpha_{ij}(\Delta)$ are formed by
$\alpha_{ij}(\Delta):=s_i(\Delta)\circ a_{ij}(\Delta)\circ
s_j(\Delta)^{-1}$. Thus we have the identities:
\begin{eqnarray}
\{\int_{s(\Delta)}dt\
\mathbf{A}_a\dot{s}^a(t),V_{U^\epsilon_{s(s(\Delta))}}^{3/4}\}&=&-A(s(\Delta))^{-1}\{A(s(\Delta)),
V_{U^\epsilon_{s(s(\Delta))}}^{3/4}\}
+o(\epsilon),\nonumber
\end{eqnarray}
and
\begin{eqnarray}
\int_{s(\Delta)} dt\ \partial_a\phi\dot{s}^a(t)&=&\frac{1}{i\lambda}
U(s(s(\Delta)),\lambda)^{-1}[U(t(s(\Delta)),\lambda)-U(s(s(\Delta)),\lambda)]+o(\epsilon)
\nonumber
\end{eqnarray}
for nonzero $\lambda$, where $s(s(\Delta))$ and $t(s(\Delta))$
denote respectively the beginning and end points of segment
$s(\Delta)$ with scale $\epsilon$ associated with a tetrahedron
$\Delta$. Regulated on the triangulation, the scalar field
Hamiltonian reads
\begin{eqnarray}
\mathcal{S}^\epsilon_{KG,\phi}
&=&-\frac{4\alpha_M}{9\kappa^4\beta^4}\sum_{\Delta'\in
T(\epsilon)}\sum_{\Delta\in
T(\epsilon)}\chi_\epsilon(v(\Delta)-v(\Delta'))
\delta^{ij}\times\nonumber\\
&&\epsilon^{lmn}\frac{1}{\lambda}U(v(\Delta),\lambda)^{-1}[U(t(s_l(\Delta)),\lambda)-U(v(\Delta),\lambda)]
\times\nonumber\\
&&\mathrm{Tr}\big(\tau_iA(s_m(\Delta))^{-1}\{A(s_m(\Delta)),V_{U^\epsilon_{v(\Delta)}}^{3/4}\}A(s_n(\Delta))^{-1}
\{A(s_n(\Delta)),V_{U^\epsilon_{v(\Delta)}}^{3/4}\}\big)\times\nonumber\\
&&\epsilon^{kpq}\frac{1}{\lambda}U(v(\Delta'),\lambda)^{-1}[U(t(s_k(\Delta')),\lambda)-U(v(\Delta'),\lambda)]
\times\nonumber\\
&&\mathrm{Tr}\big(\tau_jA(s_p(\Delta'))^{-1}\{A(s_p(\Delta')),V_{U^\epsilon_{v(\Delta')}}^{3/4}\}A(s_q(\Delta'))^{-1}
\{A(s_q(\Delta')),V_{U^\epsilon_{v(\Delta')}}^{3/4}\}\big),\nonumber\\
\mathcal{S}^\epsilon_{KG,Kin}
&=&\frac{16}{81\alpha_M(\kappa\beta)^6}\sum_{\Delta\in
T(\epsilon)}\sum_{\Delta'\in
T(\epsilon)}{\pi}(\Delta){\pi}(\Delta')\times\nonumber\\
&&\sum_{\Delta''\in
T(\epsilon)}\epsilon^{imn}\mathrm{Tr}\big(A(s_i(\Delta''))^{-1}
\{A(s_i(\Delta'')),\sqrt{V_{U^\epsilon_{v(\Delta'')}}}\}\times\nonumber\\
&&A(s_m(\Delta''))^{-1}\{A(s_m(\Delta'')),\sqrt{V_{U^\epsilon_{v(\Delta'')}}}\}\times\nonumber\\
&&A(s_n(\Delta''))^{-1}\{A(s_n(\Delta'')),\sqrt{V_{U^\epsilon_{v(\Delta'')}}}\}\big)\times\nonumber\\
&&\sum_{\Delta'''\in
T(\epsilon)}\epsilon^{jkl}\mathrm{Tr}\big(A(s_j(\Delta'''))^{-1}
\{A(s_j(\Delta''')),\sqrt{V_{U^\epsilon_{v(\Delta''')}}}\}\times\nonumber\\
&&A(s_k(\Delta'''))^{-1}\{A(s_k(\Delta''')),\sqrt{V_{U^\epsilon_{v(\Delta''')}}}\}\times\nonumber\\
&&A(s_l(\Delta'''))^{-1}\{A(s_l(\Delta''')),\sqrt{V_{U^\epsilon_{v(\Delta''')}}}\}\big)\times\nonumber\\
&&\chi_\epsilon(v(\Delta)-v(\Delta'))\chi_\epsilon(v(\Delta'')-v(\Delta))\chi_\epsilon(v(\Delta''')-v(\Delta')).
\label{classicalH}
\end{eqnarray}
Note that the above regularization is explicitly dependent on the
parameter $\lambda$, which will lead to a kind of quantization
ambiguity of the real scalar field dynamics in polymer-like
representation. Introducing a partition $\mathcal{P}$ of the
3-manifold $\Sigma$ into cells $C$, we can smear the essential
"square roots" of $\mathcal{S}^\epsilon_{KG,\phi}$ and
$\mathcal{S}^\epsilon_{KG,Kin}$ in one cell $C$ respectively and
promote them as regulated operators in $\mathcal{H}_{kin}$ with
respect to triangulations $T(\epsilon)$ depending on
spin-scalar-network state $T_{s,c}$ as
\begin{eqnarray}
\hat{W}^{\epsilon,C}_{\phi,i}T_{s,c}&=&\sum_{v\in
V(\gamma(s,c))}\frac{\chi_C(v)}{E(v)}\sum_{v(\Delta)=v}\hat{h}^{\epsilon,\Delta}_{\phi,v,i}T_{s,c},\nonumber\\
\hat{W}^{\epsilon,C}_{Kin}T_{s,c}&=&\sum_{v\in
V(\gamma(s,c))}\frac{\chi_C(v)}{E(v)}\sum_{v(\Delta)=v}\hat{h}^{\epsilon,\Delta}_{Kin,v}T_{s,c},
\label{sqarerootH}
\end{eqnarray}
where $\chi_C(v)$ is the characteristic function of the cell $C$,
and
\begin{eqnarray}
\hat{h}^{\epsilon,\Delta}_{\phi,v,i}&:=&\frac{i}{\hbar^2}\epsilon^{lmn}\frac{1}{\lambda(v)}\hat{U}(v,\lambda(v))^{-1}
[\hat{U}(t(s_l(\Delta)),\lambda(v))
-\hat{U}(v,\lambda(v))]\times\nonumber\\
&&\mathrm{Tr}\big(\tau_i\hat{A}(s_m(\Delta))^{-1}[\hat{A}(s_m(\Delta)),\hat{V}_{U^\epsilon_{v}}^{3/4}]
\hat{A}(s_n(\Delta))^{-1}[\hat{A}(s_n(\Delta)),\hat{V}_{U^\epsilon_{v}}^{3/4}]\big),\nonumber\\
\hat{h}^{\epsilon,\Delta}_{Kin,v}&:=&\frac{1}{(i\hbar)^3}\hat{\pi}(v)\epsilon^{lmn}\mathrm{Tr}
\big(\hat{A}(s_l(\Delta))^{-1}[\hat{A}(s_l(\Delta)),
\sqrt{\hat{V}_{U^\epsilon_{v}}}]\times\nonumber\\
&&\hat{A}(s_m(\Delta))^{-1}[\hat{A}(s_m(\Delta)),\sqrt{\hat{V}_{U^\epsilon_{v}}}]\times\nonumber\\
&&\hat{A}(s_n(\Delta))^{-1}[\hat{A}(s_n(\Delta)),\sqrt{\hat{V}_{U^\epsilon_{v}}}]\big).\label{ambiguity}
\end{eqnarray}
Both operators in (\ref{sqarerootH}) and their adjoint operators are
densely defined on $\mathcal{H}_{kin}$. We now give several remarks
on their properties.
\begin{itemize}
\item {{Removal of regulator $\epsilon$}}

It is not difficult to see that the action of the operator
$\hat{W}^{\epsilon,C}_{\phi,i}$ on a spin-scalar-network function
$T_{s,c}$ is graph-changing. It adds finite number of vertices with
representation $\lambda(v)$ at $t(s_i(\Delta))$ with distance
$\epsilon$ from the vertex $v$. Recall that the action of the
gravitational Hamiltonian constraint operator on a spin network
function is also graph-changing. As a result, the family of
operators $\hat{W}^{\epsilon,C}_{\phi,i}$ also fails to be weakly
converged when $\epsilon\rightarrow0$. However, due to the
diffeomorphism covariant properties of the triangulation, the limit
operator can be well-defined via the uniform Rovelli-Smolin
topology, or equivalently, the operator can be dually defined on
diffeomorphism invariant states. But the dual operator cannot leave
$\mathcal{H}_{Diff}$ invariant.

\item {{Quantization ambiguity}}

As a main difference of the dynamics in polymer-like
representation from that in U(1) group representation
\cite{thiemann7}, a continuous label $\lambda$ appears explicitly
in the expression of (\ref{sqarerootH}). Hence there is an
one-parameter quantization ambiguity due to the real scalar field.
Recall that the construction of gravitational Hamiltonian
constraint operator also has a similar ambiguity due to the choice
of the representations $j$ of the edges added by its action. A
related quantization ambiguity also appears in the dynamics of
loop quantum cosmology \cite{boj6}.
\end{itemize}
Since our quantum field theory is expected to be diffeomorphism
invariant, we would like to define the Hamiltonian operator of
polymer scalar field in the diffeomorphism invariant Hilbert space
$\mathcal{H}_{Diff}$. For this purpose we fix the parameter
$\lambda$ to be a non-zero constant at every point. Then what we
will do is to employ the new quantization strategy developed in
Refs. \cite{thiemann3} and \cite{thiemann15}. We first construct a
quadratic form in the light of a new inner product defined in
Ref.\cite{thiemann15} on the algebraic dual $\mathcal{D}^\star$ of
the space of cylindrical functions which is spanned by
spin-scalar-networks $T_{s,c}$ (where the family of labels $s,c$
includes the vacuum states for gravity, scalar and both). Then we
prove that the quadratic form is closed. Note that, although the
calculation employing this inner product is formal, it can lead to a
well-defined expression of the desired quadratic form
Eq.(\ref{quadrticform4}). Since an arbitrary element of
$\mathcal{D}^\star$ is of the form $\Psi=\sum_{s,c}
c_{s,c}<T_{s,c}|\ \cdot>_{kin}$, one can formally define an inner
product $<\cdot\ |
 \cdot>_\star$ on
$\mathcal{D}^\star$ via
\begin{eqnarray}
<\Psi,\Psi'>_\star&:=&<\sum_{s,c} c_{s,c}<T_{s,c}|\
\cdot>_{kin}|\sum_{s',c'}
c'_{s',c'}<T_{s',c'}|\ \cdot>_{kin}>_\star\nonumber\\
&:=&\sum_{s,c;s',c'}c_{s,c}\overline{c'_{s',c'}}<T_{s,c}|T_{s',c'}>_{kin}
\frac{1}{\sqrt{\aleph([s,c])\aleph([s',c'])}}\nonumber\\
&=&\sum_{s,c}c_{s,c}\overline{c'_{s,c}}\frac{1}{\aleph([s,c])},\label{product}
\end{eqnarray}
where the Cantor aleph $\aleph$ denotes the cardinal of the set
$[s,c]$. Note that we exchange the coefficients on which the complex
conjugate was taken in Ref.\cite{thiemann15}, so that the inner
product $<\Psi_{Diff} |\Psi'_{Diff}>_\star$ reduces to
$<\Psi_{Diff}|\Psi'_{Diff}>_{Diff}$ for any
$\Psi_{Diff},\Psi'_{Diff}\in\mathcal{H}_{Diff}$. Completing the
quotient with respect to the null vectors by this inner product, one
gets a Hilbert space $\mathcal{H}_\star$. Our purpose is to
construct a quadratic form associated to some positive and symmetric
operator in analogy with the classical expression of
(\ref{classicalH}). So the quadratic form should first be given in a
positive and symmetric version. It is then natural to define two
quadratic forms on a dense subset of
$\mathcal{H}_{Diff}\subset\mathcal{H}_\star$ as:
\begin{eqnarray}
Q_{KG,\phi}(\Psi_{Diff},
\Psi'_{Diff})&:=&\lim_{\mathcal{P}\rightarrow\Sigma}\sum_{C\in\mathcal{P}}64\times
\frac{4\alpha_M}{9\kappa^4\beta^4}\delta^{ij}<\hat{W}'^{C}_{\phi,i}\Psi_{Diff}|\hat{W}'^{C}_{\phi,j}\Psi'_{Diff}>_\star,\nonumber\\
Q_{KG,Kin}(\Psi_{Diff},
\Psi'_{Diff})&:=&\lim_{\mathcal{P}\rightarrow\Sigma}\sum_{C\in\mathcal{P}}8^4\times
\frac{16}{81\alpha_M(\kappa\beta)^6}<\hat{W}'^{C}_{Kin}\Psi_{Diff}|\hat{W}'^{C}_{Kin}\Psi'_{Diff}>_\star,\nonumber\\
\label{quadrticform1}
\end{eqnarray}
for any two states $\Psi_{Diff}$ and $\Psi'_{Diff}$ which are finite
linear combinations of $T_{[s,c]}$, where the dual limit operator
$\hat{W}'^{C}$ of either family of $\hat{W}^{\epsilon,C}_{\phi,i}$
or $\hat{W}^{\epsilon,C}_{Kin}$ in (\ref{sqarerootH}) is naturally
defined on diffeomorphism invariant states as
\begin{eqnarray}
\hat{W}'^{C}\Psi_{Diff}[T_{s,c}]=\lim_{\epsilon\rightarrow0}\Psi_{Diff}[\hat{W}^{\epsilon,C}T_{s,c}].
\end{eqnarray}
To show that the quadratic forms are well defined, we write
\begin{eqnarray}
\hat{W}'^{C}_{\phi,i}\Psi_{Diff}=\sum_{s,c}w^{\Psi}_{\phi,i,s,c}(C)<T_{s,c}|\
\cdot>_\star&\Rightarrow&
w^{\Psi}_{\phi,i,s,c}(C)=(\hat{W}'^{C}_{\phi,i}\Psi_{Diff})[T_{s,c}],\nonumber\\
\hat{W}'^{C}_{Kin}\Psi_{Diff}=\sum_{s,c}w^{\Psi}_{Kin,s,c}(C)<T_{s,c}|\
\cdot>_\star&\Rightarrow&
w^{\Psi}_{Kin,s,c}(C)=(\hat{W}'^{C}_{Kin}\Psi_{Diff})[T_{s,c}].\nonumber
\end{eqnarray}
Then, by using the inner product (\ref{product}) the quadratic forms
in (\ref{quadrticform1}) become
\begin{eqnarray}
&&Q_{KG,\phi}(\Psi_{Diff},\Psi'_{Diff})\nonumber\\
&:=&\lim_{\mathcal{P}\rightarrow\Sigma}\sum_{C\in\mathcal{P}}64\times
\frac{4\alpha_M}{9\kappa^4\beta^4}\delta^{ij}\sum_{s,c}
w^{\Psi}_{\phi,i,s,c}(C)\overline{w^{\Psi'}_{\phi,j,s,c}(C)}\frac{1}{\aleph([s,c])}\nonumber\\
&=&\lim_{\mathcal{P}\rightarrow\Sigma}\sum_{C\in\mathcal{P}}64\times
\frac{4\alpha_M}{9\kappa^4\beta^4}\delta^{ij}\sum_{[s,c]}\frac{1}{\aleph([s,c])}\sum_{s,c\in[s,c]}
w^{\Psi}_{\phi,i,s,c}(C)\overline{w^{\Psi'}_{\phi,j,s,c}(C)},\nonumber\\
&&\overline{Q_{KG,Kin}(\Psi_{Diff},\Psi'_{Diff})}\nonumber\\
&:=&\lim_{\mathcal{P}\rightarrow\Sigma}\sum_{C\in\mathcal{P}}8^4\times
\frac{16}{81\alpha_M(\kappa\beta)^6}\sum_{s,c}
w^{\Psi}_{Kin,s,c}(C)\overline{w^{\Psi'}_{Kin,s,c}(C)}\frac{1}{\aleph([s,c])}\nonumber\\
&=&\lim_{\mathcal{P}\rightarrow\Sigma}\sum_{C\in\mathcal{P}}8^4\times
\frac{16}{81\alpha_M(\kappa\beta)^6}\sum_{[s,c]}\frac{1}{\aleph([s,c])}\sum_{s,c\in[s,c]}
w^{\Psi}_{Kin,s,c}(C)\overline{w^{\Psi'}_{Kin,s,c}(C)}.\nonumber\\
\label{quadrticform2}
\end{eqnarray}
Note that, since $\Psi_{Diff}$ is a finite linear combination of the
diffeomorphism invariant spin-scalar-network basis, taking account
of the operational property of $\hat{W}'^{C}$ there are only a
finite number of terms in the summation $\sum_{[s,c]}$ contributing
to (\ref{quadrticform2}). Hence we can interchange $\sum_{[s,c]}$
and $\lim_{\mathcal{P}\rightarrow\Sigma}\sum_{C\in\mathcal{P}}$ in
above calculation. Moreover, for a sufficiently fine partition such
that each cell contains at most one vertex, the sum over cells
therefore reduces to finite terms with respect to the vertices of
$\gamma(s,c)$. So we can interchange $\sum_{s,c\in[s,c]}$ and
$\lim_{\mathcal{P}\rightarrow\Sigma}\sum_{C\in\mathcal{P}}$ to
obtain:
\begin{eqnarray}
&&Q_{KG,\phi}(\Psi_{Diff},\Psi'_{Diff})\nonumber\\
&=&64\times\frac{4\alpha_M}{9\kappa^4\beta^4}\delta^{ij}
\sum_{[s,c]}\frac{1}{\aleph([s,c])}\sum_{s,c\in[s,c]}
\lim_{\mathcal{P}\rightarrow\Sigma}\sum_{C\in\mathcal{P}}
w^{\Psi}_{\phi,i,s,c}(C)\overline{w^{\Psi'}_{\phi,j,s,c}(C)}\nonumber\\
&=&64\times\frac{4\alpha_M}{9\kappa^4\beta^4}\delta^{ij}
\sum_{[s,c]}\frac{1}{\aleph([s,c])}\sum_{s,c\in[s,c]} \sum_{v\in
V(\gamma(s,c))}
(\hat{W}'^{v}_{\phi,i}\Psi_{Diff})[T_{s,c}]\overline{(\hat{W}'^{v}_{\phi,j}\Psi'_{Diff})[T_{s,c}]},\nonumber\\
&&Q_{KG,Kin}(\Psi_{Diff},\Psi'_{Diff})\nonumber\\
&=&8^4\times\frac{16}{81\alpha_M(\kappa\beta)^6}
\sum_{[s,c]}\frac{1}{\aleph([s,c])}\sum_{s,c\in[s,c]}
\lim_{\mathcal{P}\rightarrow\Sigma}\sum_{C\in\mathcal{P}}
w^{\Psi}_{Kin,s,c}(C)\overline{w^{\Psi'}_{Kin,s,c}(C)}\nonumber\\
&=&8^4\times\frac{16}{81\alpha_M(\kappa\beta)^6}
\sum_{[s,c]}\frac{1}{\aleph([s,c])}\sum_{s,c\in[s,c]} \sum_{v\in
V(\gamma(s,c))}(\hat{W}'^{v}_{Kin}\Psi_{Diff})[T_{s,c}]\overline{(\hat{W}'^{v}_{Kin}\Psi'_{Diff})[T_{s,c}]},\nonumber\\
\label{quadrticform3}
\end{eqnarray}
where the limit $\mathcal{P}\rightarrow\Sigma$ has been taken so
that $C\rightarrow v$. Since given $\gamma(s,c)$ and $\gamma(s',c')$
which are different up to a diffeomorphism transformation, there is
always a diffeomorphism $\varphi$ transforming the graph associated
with $\hat{W}^{\epsilon,v}T_{s,c}\ (v\in\gamma(s,c))$ to that of
$\hat{W}^{\epsilon,v'}T_{s',c'}\ (v'\in\gamma(s',c'))$ with
$\varphi(v)=v'$, $(\hat{W}'^{v}\Psi_{Diff})[T_{s,c\in[s,c]}]$ is
constant for different $(s,c)\in[s,c]$, i.e., all the
$\aleph([s,c])$ terms in the sum over $(s,c)\in[s,c]$ are identical.
Hence the final expressions of the two quadratic forms can be
written as:
\begin{eqnarray}
&&Q_{KG,\phi}(\Psi_{Diff},\Psi'_{Diff})\nonumber\\
&=&64\times\frac{4\alpha_M}{9\kappa^4\beta^4}\delta^{ij}
\sum_{[s,c]}\sum_{v\in V(\gamma(s,c))}
(\hat{W}'^{v}_{\phi,i}\Psi_{Diff})[T_{s,c\in[s,c]}]\overline{(\hat{W}'^{v}_{\phi,j}\Psi'_{Diff})[T_{s,c\in[s,c]}]},\nonumber\\
&&Q_{KG,Kin}(\Psi_{Diff},\Psi'_{Diff})\nonumber\\
&=&8^4\times\frac{16}{81\alpha_M(\kappa\beta)^6}
\sum_{[s,c]}\sum_{v\in V(\gamma(s,c))}
(\hat{W}'^{v}_{Kin}\Psi_{Diff})[T_{s,c\in[s,c]}]\overline{(\hat{W}'^{v}_{Kin}\Psi'_{Diff})[T_{s,c\in[s,c]}]}.\nonumber\\
\label{quadrticform4}
\end{eqnarray}
Note that both quadratic forms in (\ref{quadrticform4}) have finite
results and hence their form domains are dense in
$\mathcal{H}_{Diff}$. Moreover, both of them are obviously positive, and the following theorem will demonstrate their closedness. \\ \\
\textbf{Theorem 5.2.1}: \textit{Both $Q_{KG,\phi}$ and $Q_{KG,Kin}$
are densely defined, positive and closed quadratic forms on
$\mathcal{H}_{Diff}$, which are associated uniquely with two
positive self-adjoint operators  respectively on
$\mathcal{H}_{Diff}$ such that
\begin{eqnarray}
Q_{KG,\phi}(\Psi_{Diff},\Psi'_{Diff})&=&<\Psi_{Diff}|\hat{\mathcal{S}}_{KG,\phi}|\Psi'_{Diff}>_{Diff}\nonumber\\
Q_{KG,Kin}(\Psi_{Diff},\Psi'_{Diff})&=&<\Psi_{Diff}|\hat{\mathcal{S}}_{KG,Kin}|\Psi'_{Diff}>_{Diff}.\nonumber
\end{eqnarray}
Therefore the Hamiltonian operator
\begin{eqnarray}
\hat{\mathcal{S}}_{KG}:=\hat{\mathcal{S}}_{KG,\phi}+\hat{\mathcal{S}}_{KG,Kin}\label{HKG}
\end{eqnarray}
is positive and also have a unique self-adjoint extension.}\\ \\
\textbf{Proof}: We follow the strategy developed in
Refs.\cite{thiemann15} and \cite{HM2} to prove that both
$Q_{KG,\phi}$ and $Q_{KG,Kin}$ are closeable and uniquely induce two
positive self-adjoint operators $\hat{\mathcal{S}}_{KG,\phi}$ and
$\hat{\mathcal{S}}_{KG,Kin}$. One can formally define
$\hat{\mathcal{S}}_{KG,\phi}$ and $\hat{\mathcal{S}}_{KG,Kin}$
acting on diffeomorphism invariant spin-scalar network functions
via:
\begin{eqnarray}
\hat{\mathcal{S}}_{KG,\phi}\ T_{[s_1,c_1]}&:=&\sum_{[s_2,c_2]}Q_{KG,\phi}(T_{[s_2,c_2]},T_{[s_1,c_1]})T_{[s_2,c_2]},\label{define1}\\
\hat{\mathcal{S}}_{KG,Kin}\
T_{[s_1,c_1]}&:=&\sum_{[s_2,c_2]}Q_{KG,Kin}(T_{[s_2,c_2]},T_{[s_1,c_1]})T_{[s_2,c_2]}.\label{define2}
\end{eqnarray}
Then we need to show that both of the above operators are densely
defined on the Hilbert space $\mathcal{H}_{Diff}$. Given a
diffeomorphism invariant spin-scalar network function
$T_{[s_1,c_1]}$, there are only a finite number of terms
$T_{[s_1,c_1]} [\hat{W}^{\epsilon,v}T_{s,c\in[s,c]}]$ which are
nonzero in the sum over equivalent classes $[s,c]$ in
(\ref{quadrticform4}). On the other hand, given one
spin-scalar-network function $T_{s,c\in[s,c]}$, there are also only
a finite number of possible $T_{[s_2,c_2]}$ such that the terms
$\overline{T_{[s_2,c_2]}[\hat{W}^{\epsilon,v}T_{s,c\in[s,c]}]}$ are
nonzero. As a result, only a finite number of terms survive in both
sums over $[s_2,c_2]$ in Eqs. (\ref{define1}) and (\ref{define2}).
Hence both $\hat{\mathcal{S}}_{KG,\phi}$ and
$\hat{\mathcal{S}}_{KG,Kin}$ are well defined on spin-scalar-network
basis. Then it follows from Eqs. (\ref{quadrticform4}),
(\ref{define1}) and (\ref{define2}) that they are positive and
symmetric operators densely defined in $\mathcal{H}_{Diff}$, whose
quadratic forms coincide with $Q_{KG,\phi}$ and $Q_{KG,Kin}$ on
their form domains. Hence both $Q_{KG,\phi}$ and $Q_{KG,Kin}$ have
positive closures and uniquely induce self-adjoint (Friedrichs)
extensions of $\hat{\mathcal{S}}_{KG,\phi}$ and
$\hat{\mathcal{S}}_{KG,Kin}$ respectively \cite{rs}, which we denote
by $\hat{\mathcal{S}}_{KG,\phi}$ and $\hat{\mathcal{S}}_{KG,Kin}$ as
well. As a result, the Hamiltonian operator $\hat{\mathcal{S}}_{KG}$
defined by Eq.(\ref{HKG}) is also positive and symmetric. Hence it
has a unique
self-adjoint (Friedrichs) extension.\\
$\Box$

We notice that, from a different perspective, one can construct the
same Hamiltonian operator $\hat{\mathcal{H}}_{KG}$ without
introducing an inner product on $\mathcal{D}^\star$. The
construction is sketched as follows. Using the two well-defined
operators $\hat{W}^{\epsilon,C}_{\phi,i}$ and
$\hat{W}^{\epsilon,C}_{Kin}$ as in (\ref{sqarerootH}), as well as
their adjoint operators $(\hat{W}^{\epsilon,C}_{\phi,i})^\dagger$
and $(\hat{W}^{\epsilon,C}_{Kin})^\dagger$, one may define two
operators on $\mathcal{H}_{Diff}$ corresponding to the two terms in
(\ref{classicalH}) by
\begin{eqnarray}
(\hat{\mathcal{S}}_{KG,\phi}\Psi_{Diff})[T_{s,c}]&=&\lim_{\epsilon,\epsilon'\rightarrow0,\mathcal{P}\rightarrow\Sigma}
\Psi_{Diff}[\sum_{C\in\mathcal{P}}64\times
\frac{4\alpha_M}{9\kappa^4\beta^4}\delta^{ij}\hat{W}^{\epsilon,C}_{\phi,i}(\hat{W}^{\epsilon',C}_{\phi,j})^\dagger
T_{s,c}]\nonumber\\
(\hat{\mathcal{S}}_{KG,Kin}\Psi_{Diff})[T_{s,c}]&=&\lim_{\epsilon,\epsilon'\rightarrow0,\mathcal{P}\rightarrow\Sigma}
\Psi_{Diff}[\sum_{C\in\mathcal{P}}8^4\times
\frac{16}{81\alpha_M(\kappa\beta)^6}\hat{W}^{\epsilon,C}_{Kin}(\hat{W}^{\epsilon',C}_{Kin})^\dagger
T_{s,c}],\nonumber\\
\end{eqnarray}
for any spin-scalar-network $T_{s,c}$. In analogy with the
discussion about the master constraint operator and Ref.\cite{HM2},
it can be shown that both above operators leave $\mathcal{H}_{Diff}$
invariant and are densely defined on $\mathcal{H}_{Diff}$. Moreover,
the quadratic forms associated with them coincide with the quadratic
forms in (\ref{quadrticform4}). Thus the Hamiltonian operator
$\hat{\mathcal{S}}_{KG}:=\hat{\mathcal{S}}_{KG,\phi}+\hat{\mathcal{S}}_{KG,Kin}$
coincides with the one constructed in the quadratic form approach.

In summary, we have constructed a positive self-adjoint Hamiltonian
operator on $\mathcal{H}_{Diff}$ for the polymer-like scalar field,
depending on a chosen parameter $\lambda$. Thus there is an
1-parameter ambiguity in the construction. However, there is no UV
divergence in this quantum Hamiltonian without renormalization,
since quantum gravity plays the role of a natural regulator for the
polymer-like scalar field.

\subsection{Hamiltonian Constraint Equation for the Coupled System}

In this section we consider the whole dynamical system of scalar
field coupled to gravity. Recall that in perturbative quantum field
theory in curved spacetime, the definition of some basic physical
quantities, such as the expectation value of the energy-momentum, is
ambiguous and it is challenging difficult to calculate the
back-reaction of quantum fields on the background spacetime
\cite{wald1}. This is reflected by the fact that the semi-classical
Einstein equation,
\begin{equation}
R_{\alpha\beta}[g]-\frac{1}{2}R[g]g_{\alpha\beta}=\kappa
<\hat{T}_{\alpha\beta}[g]>,\label{ein}
\end{equation}
are known to be inconsistent and ambiguous
\cite{FM}\cite{thiemann2}. One could speculate that the difficulty
is related to the fact that the usual formulation of quantum field
theories are background dependent. Following this line of thought,
if the quantization programme is by construction non-perturbative
and background independent, it may be possible to solve the problems
fundamentally. In loop quantum gravity, there is no assumption of a
priori background metric at all. The quantum geometry and quantum
matter fields are coupled and fluctuating naturally with respect to
each other on a common manifold. On the other hand, there exists the
"time problem" in quantum theory of pure gravity, since all the
physical states have to satisfy certain version of quantum
Wheeler-DeWitt constraint equation. However, the situation could
improve when matter field is coupled to gravity
\cite{kuchar0}\cite{rovelli}. In the following construction, we
impose the quantum Hamiltonian constraint on $\mathcal{H}_{kin}$,
and thus define a quantum Wheeler-DeWitt constraint equation for the
scalar field coupled to gravity. Then one can gain an insight into
the problem of time from the coupled equation, and the back-reaction
of the quantum scalar field is included in the framework of loop
quantum gravity.

We now define an operator in $\mathcal{H}_{kin}$ corresponding to
the scalar field part $\mathcal{S}_{KG}(N)$ of the total Hamiltonian
constraint functional, which can be read out from Eqs.
(\ref{hamilton}) and (\ref{kgconstraint}) as
\begin{eqnarray}
\mathcal{S}_{KG}(N)&=&\mathcal{S}_{KG,\phi}(N)+\mathcal{S}_{KG,Kin}(N),\nonumber
\end{eqnarray}
where
\begin{eqnarray}
\mathcal{S}_{KG,\phi}(N)&=&\frac{\kappa^2\beta^2\alpha_M}{2}\int_\Sigma
d^3xN\frac{1}{\sqrt{|\det
q|}}\delta^{ij}\widetilde{P}^{a}_{i}\widetilde{P}^{b}_{j}(\partial_a\phi)\partial_b\phi,\nonumber\\
\mathcal{S}_{KG,Kin}(N)&=&\frac{1}{2\alpha_M}\int_\Sigma
d^3xN\frac{1}{\sqrt{|\det q|}}\widetilde{\pi}^2\nonumber.
\end{eqnarray}
In analogy with the regularization and quantization in the previous
section, the regulated version of quantum Hamiltonian constraint
$\hat{\mathcal{S}}^\epsilon_{KG}(N)$ of scalar field is expressed by
taking the limit $C\rightarrow v$:
\begin{eqnarray}
\hat{\mathcal{S}}^{\epsilon}_{KG}(N)T_{s,c}&:=&\sum_{v\in
V(\gamma(s,c))}N(v)
\big[64\times
\frac{4\alpha_M}{9\kappa^4\beta^4}\delta^{ij}(\hat{W}^{\epsilon,v}_{\phi,i})^\dagger\hat{W}^{\epsilon,v}_{\phi,j}\nonumber\\
&+&8^4\times
\frac{16}{81\alpha_M(\kappa\beta)^6}(\hat{W}^{\epsilon,v}_{Kin})^\dagger\hat{W}^{\epsilon,v}_{Kin}\big]T_{s,c},\label{scalarconstraint}
\end{eqnarray}
where for any $v\in V(\gamma(s,c))$, the operators
\begin{eqnarray}
\hat{W}^{\epsilon,v}_{\phi,i}T_{s,c}&=&\frac{1}{E(v)}\sum_{v(\Delta)=v}\hat{h}^{\epsilon,\Delta}_{\phi,v,i}T_{s,c},\nonumber\\
\hat{W}^{\epsilon,v}_{Kin}T_{s,c}&=&\frac{1}{E(v)}\sum_{v(\Delta)=v}\hat{h}^{\epsilon,\Delta}_{Kin,v}T_{s,c},\nonumber
\end{eqnarray}
and their adjoints are all densely defined in $\mathcal{H}_{kin}$.
Hence the family of Hamiltonian constraint operators
(\ref{scalarconstraint}) is also densely defined, and the regulator
$\epsilon$ can be removed via the Uniform Rovelli-Smollin topology,
or equivalently the limit operator dually acts on diffeomorphism
invariant states as
\begin{eqnarray}
(\hat{\mathcal{S}}'_{KG}(N)\Psi_{Diff})[f]=
\lim_{\epsilon\rightarrow0}\Psi_{Diff}[\hat{\mathcal{S}}^{\epsilon}_{KG}(N)f],
\end{eqnarray}
for any $f\in Cyl(\overline{\mathcal{A/G}})\otimes
Cyl(\overline{\mathcal{U}})$. Similar to the dual of
$\hat{\mathcal{S}}_{GR}(N)$, the operator
$\hat{\mathcal{S}}'_{KG}(N)$ fails to commute with the dual of
finite diffeomorphism transformation operators, unless the smearing
function $N(x)$ is a constant function over $\Sigma$. In fact, the
dual Hamiltonian constraint operator smeared by $N=1$ is just the
diffeomorphism invariant Hamiltonian we just defined in the last
subsection. From Eq.(\ref{scalarconstraint}), it is not difficult to
prove that for positive $N(x)$ the Hamiltonian constraint operator
$\hat{\mathcal{S}}_{KG}(N)$ of a scalar field is positive and
symmetric in $\mathcal{H}_{kin}$ and hence has a unique self-adjoint
extension \cite{HM}. Our construction of $\hat{\mathcal{S}}_{KG}(N)$
is similar to that of the Higgs field Hamiltonian constraint in
Ref.\cite{thiemann7}. However, like the case of
$\hat{\mathcal{S}}_{KG}$, there is a one-parameter ambiguity in our
construction of $\hat{\mathcal{S}}_{KG}(N)$ due to the real scalar
field, which is manifested as the continuous parameter $\lambda$ in
the expression of $\hat{h}^{\epsilon,\Delta}_{\phi,v,i}$ in
(\ref{ambiguity}). Note that now $\lambda$ is not required to be a
constant, i.e., its value can be changed from one point to another.
Thus the total Hamiltonian constraint operator of scalar field
coupled to gravity has been obtained as
\begin{eqnarray}
\hat{\mathcal{S}}(N)=\hat{\mathcal{S}}_{GR}(N)+\hat{\mathcal{S}}_{KG}(N).\label{Hconstraint}
\end{eqnarray}
Again, there is no UV divergence in this quantum Hamiltonian
constraint. Recall that, in standard quantum field theory the UV
divergence can only be cured by a renormalization procedure, in
which one has to multiply the Hamiltonian by a suitable power of the
regulating parameter $\epsilon$. However, now $\epsilon$ has
naturally disappeared from the expression of (\ref{Hconstraint}). So
renormalization is not needed for the polymer-like scalar field
coupled to gravity, since quantum gravity has played the role of a
natural regulator. This result heightens our confidence that the
issue of divergences in quantum field theory can be cured in the
framework of loop quantum gravity.

Now we have obtained the desired matter-coupled quantum Hamiltonian
constraint equation
\begin{eqnarray}
-\big(\hat{\mathcal{S}}_{KG}'(N)\Psi_{Diff}\big)[f]=\big(\hat{\mathcal{S}}_{GR}'(N)
\Psi_{Diff}\big)[f].\label{evo constr}
\end{eqnarray}
Comparing it with the well-known Sch\"{o}rdinger equation for a
particle,
\begin{eqnarray}
{i\hbar\frac{\partial}{\partial
t}}\psi(x,t)=H(\hat{x},\widehat{-i\hbar\frac{\partial}{\partial
x}})\psi(x,t),\nonumber
\end{eqnarray}
where $\psi(x,t)\in L^2(\mathbf{R},dx)$ and $t$ is a parameter
labeling time evolution, one may take the viewpoint that the matter
field constraint operator $\hat{\mathcal{S}}_{KG}'(N)$ plays the
role of $i\hbar\frac{\partial}{\partial t}$. Then $\phi$ appears as
the parameter labeling the evolution of the gravitational field
state. In the reverse viewpoint, the gravitational field would
become the parameter labeling the evolution of the quantum matter
field. Note that such an idea has been successfully applied in a
loop quantum cosmology model to help us to understand the quantum
nature of big bang in the deep Planck regime \cite{APS}\cite{APS1}.

\subsection{Master Constraint for the Coupled System}

Recall that in order to avoid possible quantum anomalies and find
the physical Hilbert space of quantum gravity, the master constraint
programme was first introduced in the last section. The central idea
is to construct an alternative classical constraint algebra, giving
the same constraint phase space, which is a Lie algebra (no
structure functions) and where the subalgebra of diffeomorphism
constraints forms an ideal. A self-adjoint master constraint
operator for loop quantum gravity is then proposed on
$\mathcal{H}_{Diff}$. The master constraint programme can be
generalized to matter fields coupled to gravity in a straightforward
way. We now take the massless real scalar field to demonstrate the
construction of a master constraint operator according to the same
strategy as we did in the last section. By this approach one not
only avoids possible quantum anomalies which might appear in the
conventional canonical quantization method, but also might give a
qualitative description of the physical Hilbert space for the
coupled system. We introduce the master constraint for the scalar
field coupled to gravity as
\begin{eqnarray}
\textbf{M}:=\frac{1}{2}\int_\Sigma d^3x\frac{|{C}(x)|^2}{\sqrt{|\det
q(x)|}},\label{mconstraint}
\end{eqnarray}
where ${C}(x)$ is the Hamiltonian constraint in
(\ref{kgconstraint}). After solving the Gaussian constraint, one
gets the master constraint algebra as a Lie algebra:
\begin{eqnarray}
\{\mathcal{V}(\vec{N}),\ \mathcal{V}(\vec{N}')\}&=&\mathcal{V}([\vec{N},\vec{N}']),\nonumber\\
\{\mathcal{V}(\vec{N}),\ \textbf{M}\}&=&0,\nonumber\\
\{\textbf{M},\ \textbf{M}\}&=&0,\label{malgebra}
\end{eqnarray}
where the subalgebra of diffeomorphism constraints forms an ideal.
So it is possible to define a corresponding master constraint
operator on $\mathcal{H}_{Diff}$. In the following, the positivity
and the diffeomorphism invariance of $\textbf{M}$ will be working
together properly and provide us with powerful functional analytic
tools in the quantization procedure.

The regulated version of the master constraint can be expressed via
a point-splitting strategy as:
\begin{eqnarray}
\textbf{M}^{\epsilon}:=\frac{1}{2}\int_\Sigma d^3y \int_\Sigma
d^3x\chi_\epsilon(x-y)\frac{C(y)}{\sqrt{V_{U_y^\epsilon}}}
\frac{{C}(x)}{\sqrt{V_{U^\epsilon_{x}}}}.
\end{eqnarray}
Introducing a partition $\mathcal{P}$ of the 3-manifold $\Sigma$
into cells $C$, we have an operator $\hat{H}^\epsilon_{C}$
acting on any spin-scalar-network state $T_{s,c}$ via a family of state-dependent
triangulation $T(\epsilon)$,
\begin{eqnarray}
\hat{H}^\epsilon_{C}T_{s,c}&=&\sum_{v\in
V(\gamma(s,c))}\frac{\chi_C(v)}{E(v)}\sum_{v(\Delta)=v}\hat{h}^{\epsilon,\Delta}_{GR,v}T_{s,c}
\nonumber\\
&+&\sum_{v\in V(\gamma(s,c))}\frac{\chi_C(v)}{E(v)}
\big[64\times
\frac{4\alpha_M}{9\kappa^4\beta^4}\delta^{ij}
(\hat{w}^{\epsilon,v}_{\phi,i})^\dagger\hat{w}^{\epsilon,v}_{\phi,j}\nonumber\\
&+&8^4\times
\frac{16}{81\alpha_M(\kappa\beta)^6}(\hat{w}^{\epsilon,v}_{Kin})^\dagger\hat{w}^{\epsilon,v}_{Kin}\big]T_{s,c},
\end{eqnarray}
where
\begin{eqnarray}
\hat{h}^{\epsilon,\Delta}_{GR,v}&=&\frac{16}{3i\hbar\kappa^2\beta}
\epsilon^{ijk}\mathrm{Tr}\big(\hat{A}(\alpha_{ij}(\Delta))^{-1}\hat{A}(s_k(\Delta))^{-1}[\hat{A}(s_k(\Delta)),
\sqrt{\hat{V}_{U^\epsilon_{v}}}]\big)\nonumber\\
&+&(1+\beta^2)\frac{8\sqrt{2}}{3i\hbar^3\kappa^4\beta^3}\epsilon^{ijk}
\mathrm{Tr}\big(\hat{A}(s_i(\Delta))^{-1}[\hat{A}(s_i(\Delta)),\hat{K}^\epsilon]\nonumber\\
&\times&\hat{A}(s_j(\Delta))^{-1}[\hat{A}(s_j(\Delta)),\hat{K}^\epsilon]
\hat{A}(s_k(\Delta))^{-1}[\hat{A}(s_k(\Delta)),\sqrt{\hat{V}_{U^\epsilon_{v}}}]\big),\nonumber\\
\hat{w}^{\epsilon,v}_{\phi,i}&=&\frac{i}{\hbar^2}\sum_{v(\Delta)=v}
\epsilon^{lmn}\frac{1}{\lambda}\hat{U}(v,\lambda)^{-1}[\hat{U}(t(s_l(\Delta)),\lambda)-\hat{U}(v,\lambda)]\nonumber\\
&\times&\mathrm{Tr}\big(\tau_i\hat{A}(s_m(\Delta))^{-1}[\hat{A}(s_m(\Delta)),\hat{V}_{U^\epsilon_{v}}^{5/8}]
\hat{A}(s_n(\Delta))^{-1}[\hat{A}(s_n(\Delta)),\hat{V}_{U^\epsilon_{v}}^{5/8}]\big),\nonumber\\
\hat{w}^{\epsilon,v}_{Kin}&=&\frac{1}{(i\hbar)^3}\sum_{v(\Delta)=v}
\hat{\pi}(v)\epsilon^{lmn}\nonumber\\
&\times&\mathrm{Tr}\big(\hat{A}(s_l(\Delta))^{-1}[\hat{A}(s_l(\Delta)),\hat{V}^{5/12}_{U^\epsilon_{v}}]
\hat{A}(s_m(\Delta))^{-1}[\hat{A}(s_m(\Delta)),\hat{V}^{5/12}_{U^\epsilon_{v}}]\nonumber\\
&\times&\hat{A}(s_n(\Delta))^{-1}[\hat{A}(s_n(\Delta)),\hat{V}^{5/12}_{U^\epsilon_{v}}]\big).
\end{eqnarray}
Hence the action of $\hat{H}^\epsilon_{C}$ on a cylindrical function
$f_\gamma$ adds analytical arcs $a_{ij}(\Delta)$ with
$\frac{1}{2}$-representation and
points at $t(s_i(\Delta))$ with representation constant $\lambda$ with
respect to each vertex $v(\Delta)$ of $\gamma$. Thus, for each
$\epsilon>0$, $\hat{H}^\epsilon_{C}$ is a $SU(2)$ gauge
invariant and diffeomorphism covariant operator defined on
$Cyl(\overline{\mathcal{A/G}})\otimes
Cyl(\overline{\mathcal{U}})$. The limit operator $\hat{H}_C$ is
densely defined on $\mathcal{H}_{Kin}$ by the uniform Rovelli-Smolin
topology. And the same result holds for the adjoint operator $(\hat{H}^\epsilon_{C})^\dagger$

Then a master constraint operator, $\hat{\mathbf{M}}$, on ${\cal
H}_{Diff}$ can be defined by:
\begin{equation}
(\hat{\textbf{M}}\Psi_{Diff})[T_{s,c}]:=\lim_{\mathcal{P}\rightarrow
\Sigma;\epsilon,\epsilon'\rightarrow\mathrm{0}}\Psi_{Diff}[\sum_{C\in\mathcal{P}}
\frac{1}{2}\hat{H}^\epsilon_{C} (\hat{H}_{C}^{\epsilon'})^\dagger
T_{s,c}].\label{master}
\end{equation}
Since $\hat{H}^\epsilon_{C} (\hat{H}^{\epsilon'}_{C})^\dagger
T_{s,c}$ is a finite linear combination of spin-scalar-network
functions on an graph with skeleton $\gamma$, the value of
$(\hat{\textbf{M}}\Psi_{Diff})[T_{s,c}]$ is finite for a given
$\Psi_{Diff}$ that is a finite linear combination of $T_{[s,c]}$. So
$\hat{\textbf{M}}\Psi_{Diff}$ is in the algebraic dual of the space
of cylindrical functions. Moreover, we can show that it is
diffeomorphism invariant. For any diffeomorphism transformation
$\varphi$,
\begin{eqnarray}
(\hat{U}'_\varphi\hat{\textbf{M}}\Psi_{Diff})[f_\gamma]&=&\lim_{\mathcal{P}\rightarrow
\Sigma;\epsilon,\epsilon'\rightarrow\mathrm{0}}\Psi_{Diff}[\sum_{C\in\mathcal{P}}\frac{1}{2}\hat{H}^\epsilon_{C}
(\hat{H}^{\epsilon'}_{C})^\dagger\hat{U}_\varphi
f_\gamma]\nonumber\\
&=&\lim_{\mathcal{P}\rightarrow
\Sigma;\epsilon,\epsilon'\rightarrow\mathrm{0}}\Psi_{Diff}[\hat{U}_\varphi\sum_{C\in\mathcal{P}}
\frac{1}{2}\hat{H}^{\varphi^{-1}(\epsilon)}_{\varphi^{-1}(C)}
(\hat{H}^{\varphi^{-1}(\epsilon')}_{\varphi^{-1}(C)})^\dagger f_\gamma]\nonumber\\
&=&\lim_{\mathcal{P}\rightarrow
\Sigma;\epsilon,\epsilon'\rightarrow\mathrm{0}}\Psi_{Diff}[\sum_{C\in\mathcal{P}}\frac{1}{2}\hat{H}^\epsilon_{C}
(\hat{H}^{\epsilon'}_{C})^\dagger f_\gamma],
\end{eqnarray}
for any cylindrical function $f_\gamma$, where in the last step, we
used the fact that the diffeomorphism transformation $\varphi$
leaves the partition invariant in the limit
$\mathcal{P}\rightarrow\Sigma$ and relabel $\varphi(C)$ to be $C$.
So we have the result
\begin{eqnarray}
(\hat{U}'_\varphi\hat{\textbf{M}}\Psi_{Diff})[f_\gamma]=(\hat{\textbf{M}}\Psi_{Diff})[f_\gamma].\label{diff}
\end{eqnarray}
So given a diffeomorphism invariant spin-scalar-network state $T_{[s,c]}$,
the result state $\hat{\textbf{M}}T_{[s,c]}$ must be a diffeomorphism
invariant element in the algebraic dual of
$Cyl(\overline{\mathcal{A/G}})\otimes Cyl(\overline{\mathcal{U}})$, which means that
\begin{eqnarray}
\hat{\textbf{M}}T_{[s,c]}=\sum_{[s_1,c_1]}c_{[s_1,c_1]}T_{[s_1,c_1]},\nonumber
\end{eqnarray}
then
\begin{eqnarray}
\lim_{\mathcal{P}\rightarrow
\Sigma;\epsilon,\epsilon'\rightarrow\mathrm{0}}T_{[s,c]}[\sum_{C\in\mathcal{P}}\frac{1}{2}\hat{H}^\epsilon_{C}
(\hat{H}^{\epsilon'}_{C})^\dagger
T_{s_2,c_2}]=\sum_{[s_1,c_1]}c_{[s_1,c_1]}T_{[s_1,c_1]}[T_{s_2,c_2}],\nonumber
\end{eqnarray}
where the cylindrical function
$\sum_{C\in\mathcal{P}}\frac{1}{2}\hat{H}^{\epsilon'}_{C}
(\hat{H}^{\epsilon}_{C})^\dagger T_{s_2,c_2}$ is a finite linear
combination of spin-scalar-network functions on some graphs $\gamma\
'$ with the same skeleton of $\gamma(s_2,c_2)$ up to finite number
of arcs and vertices. Hence fixing the diffeomorphism equivalence
class $[s,c]$, only for spin-scalar-networks $s_2,c_2$ lying in a
finite number of diffeomorphism equivalence class on the left hand
side of the last equation is non-zero. So there are also only finite
number of classes $[s_1,c_1]$ in the right hand side such that
$c_{[s_1,c_1]}$ is non-zero. As a result,
$\hat{\textbf{M}}T_{[s,c]}$ is a finite linear combination of
diffeomorphism invariant spin-network states so lies in the Hilbert
space of diffeomorphism invariant states $\mathcal{H}_{Diff}$ for
any $[s,c]$. And $\hat{\textbf{M}}$ is densely defined on
$\mathcal{H}_{Diff}$.

We now compute the matrix elements of $\hat{\mathbf{M}}$. Given two
diffeomorphism invariant spin-scalar-network functions
$T_{[s_1,c_1]}$ and $T_{[s_2,c_2]}$, the matrix element of
$\hat{\mathbf{M}}$ is calculated as
\begin{eqnarray}
&&<T_{[s_1,c_1]}|\hat{\textbf{M}}|T_{[s_2,c_2]}>_{Diff}\nonumber\\
&=&\overline{(\hat{\textbf{M}}T_{[s_2,c_2]})[T_{s_1,c_1\in[s_1,c_1]}]}\nonumber\\
&=&\lim_{\mathcal{P}\rightarrow
\Sigma;\epsilon,\epsilon'\rightarrow\mathrm{0}}\sum_{C\in\mathcal{P}}\frac{1}{2}
\overline{T_{[s_2,c_2]}[\hat{H}^\epsilon_{C}(\hat{H}^{\epsilon'}_{C})^\dagger T_{s_1,c_1\in[s_1,c_1]}]}\nonumber\\
&=&\lim_{\mathcal{P}\rightarrow
\Sigma;\epsilon,\epsilon'\rightarrow\mathrm{0}}\sum_{C\in\mathcal{P}}\frac{1}{2}
\frac{1}{n_{\gamma(s_2,c_2)}}\sum_{\varphi\in
Diff/Diff_{\gamma(s_2,c_2)}}\sum_{\varphi'\in
GS_{\gamma(s_2,c_2)}}\nonumber\\
&\times&\overline{<\hat{U}_{\varphi}\hat{U}_{\varphi'}T_{s_2,c_2\in[s_2,c_2]}
|\hat{H}^\epsilon_{C}(\hat{H}^{\epsilon'}_{C})^\dagger T_{s_1,c_2\in[s_1,c_1]}>_{Kin}}\nonumber\\
&=&\sum_{s,c}\lim_{\mathcal{P}\rightarrow
\Sigma;\epsilon,\epsilon'\rightarrow\mathrm{0}}\sum_{C\in\mathcal{P}}\frac{1}{2}
\frac{1}{n_{\gamma(s_2,c_2)}}\sum_{\varphi\in
Diff/Diff_{\gamma(s_2,c_2)}}\sum_{\varphi'\in
GS_{\gamma(s_2,c_2)}}\nonumber\\
&\times&\overline{<\hat{U}_{\varphi}\hat{U}_{\varphi'}T_{s_2,c_2\in[s_2,c_2]}
|\hat{H}^\epsilon_{C}T_{s,c}>_{Kin}<\Pi_{s,c}|(\hat{H}^{\epsilon'}_{C})^\dagger T_{s_1,c_1\in[s_1,c_1]}>_{Kin}}\nonumber\\
&=&\sum_{[s,c]}\sum_{v\in
V(\gamma(s,c\in[s,c]))}\frac{1}{2}\lim_{\epsilon,\epsilon'\rightarrow\mathrm{0}}\nonumber\\
&\times&\overline{T_{[s_2,c_2]}[\hat{H}^{\epsilon}_{v}
T_{s,c\in[s,c]}]\sum_{s,c\in[s,c]}
<T_{s,c}|(\hat{H}^{\epsilon'}_{v})^\dagger T_{s_1,c_1\in[s_1,c_1]}>_{Kin}},\label{matrix}
\end{eqnarray}
where $Diff_{\gamma}$ is the set of diffeomorphisms leaving the
colored graph $\gamma$ invariant, $GS_{\gamma}$ denotes the graph
symmetry quotient group $Diff_{\gamma}/TDiff_{\gamma}$ where
$TDiff_{\gamma}$ is the diffeomorphisms which is trivial on the
graph $\gamma$, and $n_\gamma$ is the number of elements in
$GS_\gamma$. Note that we have used the resolution of identity trick
in the fourth step. Since only a finite number of terms in the sum
over spin-scalar-networks $(s,c)$, cells $C\in\mathcal{P}$, and
diffeomorphism transformations $\varphi$ are non-zero respectively,
we can interchange the sums and the limit. In the fifth step, we
take the limit $C\rightarrow v$ and split the sum $\sum_{s,c}$ into
$\sum_{[s,c]}\sum_{s,c\in[s,c]}$, where $[s,c]$ denotes the
diffeomorphism equivalence class associated with $(s,c)$. Here we
also use the fact that, given $\gamma(s,c)$ and $\gamma(s',c')$
which are different up to a diffeomorphism transformation, there is
always a diffeomorphism $\varphi$ transforming the graph associated
with $\hat{H}^{\epsilon}_{v,\gamma(s,c)}T_{s,c}\ (v\in\gamma(s,c))$
to that of $\hat{H}^{\epsilon}_{v'\gamma(s',c')} T_{s',c'}\
(v'\in\gamma(s',c'))$ with $\varphi(v)=v'$, hence
$T_{[s_2,c_2]}[\hat{H}^{\epsilon}_{v,\gamma(s,c)} T_{s,c\in[s,c]}]$
is constant for different $(s,c)\in[s,c]$.

Since the term $\sum_{s,c\in[s,c]}
<T_{s,c}|(\hat{H}^{\epsilon'}_{v})^\dagger T_{s_1,c_1\in[s_1,c_1]}>_{Kin}$
is independent of the parameter $\epsilon'$, one can see that by
fixing a family of arbitrary state-dependent triangulations $T(\epsilon')$,
\begin{eqnarray}
&&\sum_{s,c\in[s,c]}
<T_{s,c}|(\hat{H}^{\epsilon'}_{v})^\dagger T_{s_1,c_1\in[s_1,c_1]}>_{Kin}\nonumber\\
&=&\sum_{\varphi}<U_\varphi T_{s,c}|(\hat{H}^{\epsilon'}_{v})^\dagger T_{s_1,c_1\in[s_1,c_1]}>_{Kin}\nonumber\\
&=&\sum_{\varphi}<\hat{H}^{\epsilon'}_{v}U_\varphi T_{s,c}|T_{s_1,c_1\in[s_1,c_1]}>_{Kin}\nonumber\\
&=&\sum_{\varphi}<U_\varphi\hat{H}^{\varphi^{-1}(\epsilon')}_{\varphi^{-1}(v)}T_{s,c}|T_{s_1,c_1\in[s_1,c_1]}>_{Kin}\nonumber\\
&=&\overline{T_{[s_1,c_1]}[\hat{H}^{\varphi^{-1}(\epsilon')}_{v\in
V(\gamma(s,c))}T_{s,c}]},
\end{eqnarray}
where $\varphi$ are the diffeomorphism transformations spanning the
diffeomorphism equivalence class $[s,c]$. Note that the kinematical
inner product in the above sum is non-vanishing if and only if
$\varphi(\gamma(s,c)))$ coincides with the extended graph obtained
from certain skeleton $\gamma(s_1,c_1)$ by the action of
$(\hat{H}^{\epsilon'}_{v})^\dagger$ and $v\in
V(\varphi(\gamma(s,c)))$, i.e., the scale $\varphi^{-1}(\epsilon')$
of the diffeomorphism images of the tetrahedrons added by the action
coincides with the scale of certain tetrahedrons in $\gamma(s,c)$
and $\varphi^{-1}(v)$ is a vertex in $\gamma(s,c)$. Then we can
express the matrix elements (\ref{matrix}) as:
\begin{eqnarray}
&&<T_{[s_1,c_1]}|\hat{\textbf{M}}|T_{[s_2,c_2]}>_{Diff}\nonumber\\
&=&\sum_{[s,c]}\sum_{v\in
V(\gamma(s,c\in[s,c]))}\frac{1}{2}\lim_{\epsilon,\epsilon'\rightarrow\mathrm{0}}
\overline{T_{[s_2,c_2]}[\hat{H}^{\epsilon}_{v}
T_{s,c\in[s,c]}]}T_{[s_1,c_1]}[\hat{H}^{\epsilon'}_{v}T_{s,c\in[s,c]}]\nonumber\\
&=&\sum_{[s,c]}\sum_{v\in
V(\gamma(s,c\in[s,c]))}\frac{1}{2}\overline{(\hat{H}'_v T_{[s_2,c_2]})[
T_{s,c\in[s,c]}]}(\hat{H}'_v T_{[s_1,c_1]}) [
T_{s,c\in[s,c]}].\label{master2}
\end{eqnarray}
From Eq.(\ref{master2}) and the result that the master constraint
operator $\hat{\mathbf{M}}$ is densely defined on
$\mathcal{H}_{Diff}$, it is obvious that $\hat{\mathbf{M}}$ is a
positive and symmetric operator on ${\cal H}_{Diff}$. Hence, it is
associated with a unique self-adjoint operator
$\hat{\overline{\mathbf{M}}}$, called the Friedrichs extension of
$\hat{\mathbf{M}}$. We relabel $\hat{\overline{\mathbf{M}}}$ to be
$\hat{\mathbf{M}}$ for simplicity. In conclusion, there exists a
positive and self-adjoint operator $\hat{\mathbf{M}}$ on
$\mathcal{H}_{Diff}$ corresponding to the master constraint
(\ref{mconstraint}). It is then possible to obtain the physical
Hilbert space of the coupled system by the direct integral
decomposition of $\mathcal{H}_{Diff}$ with respect to
$\hat{\mathbf{M}}$.

Note that the quantum constraint algebra can be easily checked to be
anomaly free. Eq.(\ref{diff}) assures that the master constraint
operator commutes with finite diffeomorphism transformations, i.e.,
\begin{eqnarray}
[\hat{\mathbf{M}},\hat{U}'_\varphi]=0.
\end{eqnarray}
Also it is obvious that the master constraint operator commutes with
itself,
\begin{eqnarray}
[\hat{\mathbf{M}},\hat{\mathbf{M}}]=0.
\end{eqnarray}
So the quantum constraint algebra is precisely consistent with the
classical constraint algebra (\ref{malgebra}) in this sense. As a
result, the difficulty of the original Hamiltonian constraint
algebra can be avoided by introducing the master constraint algebra,
due to the Lie algebra structure of the latter.

\newpage

\section{The Semiclassical Limit of Quantum Dynamics}

As shown in previous chapters, both the Hamiltonian constraint
operator $\hat{\mathcal{S}}(N)$ and the master constraint operator
$\hat{\textbf{M}}$ can be well defined in the framework of loop
quantum gravity. However, since the Hilbert spaces
$\mathcal{H}_{kin}$ and $\mathcal{H}_{Diff}$, the operators
$\hat{\mathcal{S}}(N)$ and $\hat{\textbf{M}}$ are constructed in
such ways that are drastically different from usual quantum field
theory, one has to check whether the constraint operators and the
corresponding algebras have correct semiclassical limits with
respect to suitable semiclassical states.

\subsection{The Construction of Coherent States}

In order to find the proper semiclassical states and check the
classical limit of the theory, the idea of a non-normalizable
coherent state defined by a generalized Laplace operator and its
heat kernel was introduced for the first time in \cite{ALM2}.
Recently, kinematical coherent states were constructed in two
different approaches. One leads to the so-called complexifier
coherent states proposed by Thiemann et al
\cite{thiemann10}\cite{thiemann11}\cite{thiemann12}\cite{thiemann13}.
The other was proposed by Varadarajan
\cite{var1}\cite{var2}\cite{var3} and further developed by Ashtekar
et al \cite{AL2}\cite{shadow}.

The complexifier approach is motivated by the coherent state
construction for compact Lie groups \cite{hall}. One begins with a
positive function $C$ (complexifier) on the classical phase space
and arrives at a "coherent state" $\psi_m$, which more possibly
belongs to the dual space $Cyl^\star$ rather than
$\mathcal{H}_{kin}$. However, one may consider the so-called
"cut-off state" of $\psi_m$ with respect to a finite graph as a
graph-dependent coherent state in $\mathcal{H}_{kin}$
\cite{thiemann2}. By construction, the coherent state $\psi_m$ is an
eigenstate of an annihilation operator coming also from the
complexifier $C$ and hence has the desired semiclassical properties
\cite{thiemann11}\cite{thiemann12}. We now sketch the basic idea of
its construction. Given the Hilbert space $\mathcal{H}$ for a
dynamical system with constraints and a subalgerba of observables
$\mathcal{S}$ in the space $\mathcal{L}(\mathcal{H})$ of linear
operators on $\mathcal{H}$, the semiclassical states with respect to
$\mathcal{S}$ are defined in Definition 3.1.5. Kinematical coherent
states $\{\Psi_m\}_{m\in\mathcal{M}}$ are semiclassical states which
in addition satisfy the annihilation operator property
\cite{thiemann10}\cite{thiemann2}, namely there exists a certain
non-self-adjoint operator $\hat{z}=\hat{a}+i\lambda\hat{b}$ with
$\hat{a},\ \hat{b}\in\mathcal{S}$ and a certain squeezing parameter
$\lambda$, such that
\begin{eqnarray}
\hat{z}\, \Psi_m=z(m)\Psi_m. \label{annihilation}
\end{eqnarray}
Note that Eq.(\ref{annihilation}) implies that the minimal
uncertainty relation is saturated for the pair of elements
$(\hat{a},\ \hat{b})$, i.e.,
\begin{eqnarray}
\Psi_m([\hat{a}-\Psi_m(\hat{a})]^2)=\Psi_m([\hat{b}-\Psi_m(\hat{b})]^2)=\frac{1}{2}|\Psi_m([\hat{a},
\hat{b}])|.
\end{eqnarray}
Note also that coherent states are usually required to satisfy the
additional peakedness property, namely for any $m\in\mathcal{M}$ the
overlap function $|<\Psi_m,\Psi_{m'}>|$ is concentrated in a phase
volume $\frac{1}{2}|\Psi_m([\hat{q}, \hat{p}])|$, where $\hat{q}$ is
a configuration operator and $\hat{p}$ a momentum operator. So the
central element in the construction is to define a suitable
"annihilation operator" $\hat{z}$ in analogy with the simplest case
of harmonic oscillator. A powerful tool named as "complexifier" is
introduced in Ref.\cite{thiemann10} to define a meaningful $\hat{z}$
operator which can give rise to kinematical coherent states for a general quantum system.\\ \\
\textbf{Definition 6.1.1}: \textit{Given a phase space
$\mathcal{M}=\mathrm{T}^*\mathcal{C}$ for some dynamical system with
configuration coordinates $q$ and momentum coordinates $p$, a
complexifier, $C$, is a positive smooth function on $\mathcal{M}$, such that\\
$(1)$ $C/\hbar$ is dimensionless;\\
$(2)$ $\lim_{||p||\rightarrow\infty}\frac{|C(m)|}{||p||}=\infty$ for
some suitable norm on the space of the momentum;\\
$(3)$ Certain complex coordinates $(z(m), \bar{z}(m))$ of
$\mathcal{M}$ can be constructed from $C$. }
\\ \\
Given a well-defined complexifier $C$ on phase space $\mathcal{M}$,
the programme for constructing coherent states associated with $C$
can be carried out as the following.
\begin{itemize}
\item {Complex polarization}

The condition (3) in Definition 7.3.1 implies that the complex
coordinate $z(m)$ of $\mathcal{M}$ can be constructed via
\begin{eqnarray}
z(m):=\sum_{n=0}^\infty\frac{i^n}{n!}\{q,C\}_{(n)}(m),\label{complex}
\end{eqnarray}
where the multiple Poisson bracket is inductively defined by
$\{q,C\}_{(0)}=q,\ \{q,C\}_{(n)}=\{\{q,C\}_{(n-1)},C\}$. One will
see that $z(m)$ can be regarded as the classical version of an
annihilation operator.

\item {Defining the annihilation operator}

After the quantization procedure, a Hilbert space
$\mathcal{H}=L^2(\mathcal{C}, d\mu)$ with a suitable measure $d\mu$
on a suitable configuration space $\mathcal{C}$ can be constructed.
It is reasonable to assume that $C$ can be defined as a positive
self-adjoint operator $\hat{C}$ on $\mathcal{H}$. Then a
corresponding operator $\hat{z}$ can be defined by transforming the
Poisson brackets in Eq.(\ref{complex}) into commutators, i.e.,
\begin{eqnarray}
\hat{z}:=\sum_{n=0}^\infty\frac{i^n}{n!}\frac{1}{(i\hbar)^n}[\hat{q},\hat{C}]_{(n)}=e^{-\hat{C}/\hbar}\hat{q}
e^{\hat{C}/\hbar},
\end{eqnarray}
which is called as an annihilation operator.

\item {Constructing coherent states}

Let $\delta_{q'}(q)$ be the $\delta$-distribution on $\mathcal{C}$
with respect to the measure $d\mu$. Since $\hat{C}$ is assumed to be
positive and self-adjoint, the conditions (1) and (2) in Definition
7.3.1 imply that $e^{-\hat{C}/\hbar}$ is a well-defined "smoothening
operator". So it is quite possible that the heat kernel evolution of
the $\delta$-distribution, $e^{-\hat{C}/\hbar}\delta_{q'}(q)$, is a
square integrable function in $\mathcal{H}$, which is even analytic.
Then one may analytically extend the variable $q'$ in
$e^{-\hat{C}/\hbar}\delta_{q'}(q)$ to complex values $z(m)$ and
obtain a class of states $\psi'_m$ as
\begin{eqnarray}
\psi'_m(q):=[e^{-\hat{C}/\hbar}\delta_{q'}(q)]_{q'\rightarrow
z(m)},\label{coherent}
\end{eqnarray}
such that one has
\begin{eqnarray}
\hat{z}\,
\psi'_m(q):=[e^{-\hat{C}/\hbar}\hat{q}\delta_{q'}(q)]_{q'\rightarrow
z(m)}=[q'e^{-\hat{C}/\hbar}\delta_{q'}(q)]_{q'\rightarrow
z(m)}=z(m)\, \psi'_m(q).
\end{eqnarray}
Hence $\psi'_m$ is automatically an eigenstate of the annihilation
operator $\hat{z}$. So it is natural to define coherent states
$\psi_m(q)$ by normalizing $\psi'_m(q)$.\\
\end{itemize}
One may check that all the coherent state properties usually
required are likely to be satisfied by the above complexifier
coherent states $\psi_m(q)$ \cite{thiemann2}. As a simple example,
in the case of one-dimensional harmonic oscillator with Hamiltonian
$H=\frac{1}{2}(\frac{p^2}{2m}+\frac{1}{2}m\omega^2q^2)$, one may
choose the complexifier $C=p^2/(2m\omega)$. It is straightforward to
check that the coherent state constructed by the above procedure
coincides with the usual harmonic oscillator coherent state up to a
phase \cite{thiemann2}. So the complexifier coherent state can be
considered as a suitable generalization of the concept of usual
harmonic oscillator coherent state.

The complexifer approach can be used to construct kinematical
coherent states in loop quantum gravity. Given a suitable
complexifier $C$, for each analytic path $e\subset\Sigma$ one can
define
\begin{eqnarray}
A^{\mathbf{C}}(e):=\sum_{n=0}^\infty\frac{i^n}{n!}\{A(e),C\}_{(n)},\label{cconnection}
\end{eqnarray}
where $A(e)\in SU(2)$ is assigned to $e$. As the complexifier $C$ is
assumed to give rise to a positive self-adjoint operator $\hat{C}$
on the kinematical Hilbert space $\mathcal{H}_{kin}$, one further
supposes that $\hat{C}/\hbar T_s=\tau\lambda_sT_s$, where $\tau$ is
a so-called classicality parameter, $\{T_s(A)\}_s$ form a basis in
$\mathcal{H}_{kin}$ and are analytic in
$A\in\overline{\mathcal{A}}$. Moreover the $\delta$-distribution on
the quantum configuration space $\overline{\mathcal{A}}$ can be
formally expressed as $\delta_{A'}(A)=\sum_s
T_s(A')\overline{T_s(A)}$. Thus by applying Eq.(\ref{coherent}) one
obtains coherent states
\begin{eqnarray}
\psi'_{A^\mathbf{C}}(A)=(e^{-\hat{C}/\hbar})\delta_{A'}(A)|_{A'\rightarrow
A^\mathbf{C}}=\sum_se^{-\tau\lambda_s}T_s(A^{\mathbf{C}})\overline{T_s(A)}.\label{cstate}
\end{eqnarray}
However, since there are an uncountably infinite number of terms in
the expression (\ref{cstate}), the norm of $\psi'_{A^\mathbf{C}}(A)$
would in general be divergent. So $\psi'_{A^\mathbf{C}}(A)$ is
generally not an element of $\mathcal{H}_{kin}$ but rather a
distribution on a dense subset of $\mathcal{H}_{kin}$. In order to
test the semiclassical limit of quantum geometric operators on
$\mathcal{H}_{kin}$, one may further consider the "cut-off state" of
$\psi'_{A^\mathbf{C}}(A)$ with respect to a finite graph $\gamma$ as
a graph-dependent coherent state in $\mathcal{H}_{kin}$
\cite{thiemann2}. So the key input in the construction is to choose
a suitable complexifer. There are vast possibilities of choice. For
example, a candidate complexifier $C$ is considered in
Ref.\cite{lecture} such that the corresponding operator acts on
cylindrical functions $f_{\gamma}$ by
\begin{eqnarray}
(\hat{C}/\hbar)f_{\gamma}=\frac{1}{2}(\sum_{e\in
E(\gamma)}l(e)\hat{J}_e^2)f_{\gamma},
\end{eqnarray}
where $\hat{J}_e^2$ is the Casimir operator defined by
Eq.(\ref{casimir}) associated to the edge $e$, the positive numbers
$l(e)$ satisfying $l(e\circ e')=l(e)+l(e')$ and $l(e^{-1})=l(e)$
serves as a classicalization parameter. Then it can be shown from
Eq.(\ref{cconnection}) that $A^{\mathbf{C}}(e)$ is an element of
$SL(2, \mathbf{C})$. So the classical interpretation of the
annihilation operators is simply the generalized complex $SU(2)$
connections. It has been shown in Refs. \cite{thiemann11} and
\cite{thiemann12} that the "cut-off state" of the corresponding
coherent state,
\begin{eqnarray}
\psi_{A^\mathbf{C},\gamma}(A)=\psi'_{A^\mathbf{C},\gamma}(A)/||\psi'_{A^\mathbf{C},\gamma}(A)||,
\end{eqnarray}
with
\begin{eqnarray}
\psi'_{A^\mathbf{C},\gamma}(A):=\sum_{s,\gamma(s)=\gamma}e^{-\frac{1}{2}\sum_{e\in
E(\gamma(s))}l(e)j_e (j_e+1)}T_s(A^{\mathbf{C}})\overline{T_s(A)}.
\end{eqnarray}
has desired semiclassical properties in testing the kinematical
operators (e.g. holonomy and flux). But unfortunately, these cut-off
coherent states cannot be directly used to test the semiclassical
limit of the Hamiltonian constraint operator $\hat{\mathcal{S}}(N)$,
since $\hat{\mathcal{S}}(N)$ is graph-changing so that its
expectation values with respect to these cut-off states are always
zero! So further work in this approach is expected in order to
overcome the difficulty. Anyway, the complexifier approach provides
a clean construction mechanism and manageable calculation method for
semiclassical analysis in loop quantum gravity.

We now turn to the second approach. As we have seen, loop quantum
gravity is based on quantum geometry, where the fundamental
excitations are one-dimensional polymer-like. On the other hand, low
energy physics is based on quantum field theories which are
constructed in a flat spacetime continuum. The fundamental
excitations of these fields are 3-dimensional, typically
representing wavy undulations on the background Minkowskian
geometry. The core strategy in this approach is then to relate the
polymer excitations of quantum geometry to Fock states used in low
energy physics and to locate Minkowski Fock states in the background
independent framework. Since the quantum Maxwell field can be
constructed in both Fock representation and polymer-like
representation, one first gains insights from the comparison between
the two representations, then generalizes the method to quantum
geometry. A "Laplacian operator" can be defined on
$\mathcal{H}_{kin}$ \cite{ALM2}\cite{AL2}, from which one may define
a candidate coherent state $\Phi_0$, also in $Cyl^\star$,
corresponding to the Minkowski spacetime. To calculate the
expectation values of kinematical operators, one considers the
so-called "shadow state" of $\Phi_0$, which is the restriction of
$\Phi_0$ to a given finite graph. However, the construction of
shadow states is subtly different from that of cut-off states.

We will only describe the simple case of the Maxwell field to
illustrate the ideas of the construction
\cite{var1}\cite{var2}\cite{AL}. Following the quantum geometry
strategy discussed in Sec.4, the quantum configuration space
$\overline{\mathbf{A}}$ for the polymer representation of the $U(1)$
gauge theory can be similarly constructed. A generalized connection
$\mathbf{A}\in\overline{\mathbf{A}}$ assigns each oriented analytic
edge in $\Sigma$ an element of $U(1)$. The space
$\overline{\mathbf{A}}$ carries a diffeomorphism and gauge invariant
measure $\mu_0$ induced by the Haar measure on $U(1)$, which gives
rise to the Hilbert space,
$\mathcal{H}_0:=L^2(\overline{\mathbf{A}}, d\mu_0)$, of polymer
states. The basic operators are holonomy operators
$\hat{\mathbf{A}}(e)$ labeled by one-dimensional edges $e$, which
act on cylindrical functions by multiplication, and smeared electric
field operators $\hat{E}(g)$ for suitable test one-forms $g$ on
$\Sigma$, which are self-adjoint. Note that, since the gauge group
$U(1)$ is Abelian, it is more convenient to smear the electric
fields in 3 dimensions \cite{AL}. The eigenstates of $\hat{E}(g)$,
so-called flux network states $\mathcal{N}_{\alpha, \vec{n}}$,
provide an orthonormal basis in $\mathcal{H}_0$, which are defined
for any finite graph $\alpha$ with $N$ edges as:
\begin{equation}
\mathcal{N}_{\alpha,\vec{n}}(\mathbf{A}):=[{\mathbf{A}}(e_1)]^{n_1}
[\mathbf{A}(e_2)]^{n_2}\cdot\cdot\cdot[\mathbf{A}(e_N)]^{n_N},
\end{equation}
where $\vec{n}\equiv(n_1,\cdot\cdot\cdot,n_N)$ assigns an integer
$n_I$ to each edge $e_I$. The action of $\hat{E}(g)$ on the flux
network states reads
\begin{equation}
\hat{E}(g)\, \mathcal{N}_{\alpha, \vec{n}}=-\hbar(\sum_I
n_I\int_{e_I}g)\mathcal{N}_{\alpha, \vec{n}}.
\end{equation}
In this polymer-like representation, cylindrical functions are the
finite linear combinations of flux network states and span a dense
subspace of $\mathcal{H}_0$. Denote by $\mathbf{Cyl}$ the set of
cylindrical functions and by $\mathbf{Cyl}^\star$ its algebraic
dual. One then has a triplet
$\mathbf{Cyl}\subset\mathcal{H}_0\subset \mathbf{Cyl}^\star$ in
analogy with the case of loop quantum gravity.

The Schr\"{o}dinger or Fock representation of the Maxwell field, on
the other hand, depends on the Minkowski background metric. Here the
Hilbert space is given by $\mathcal{H}_F=L^2(\mathcal{S}', d\mu_F)$,
where $\mathcal{S}'$ is the appropriate space of tempered
distributions on $\Sigma$ and $\mu_F$ is the Gaussian measure. The
basic operators are connections $\hat{\mathbf{A}}(f)$ smeared in 3
dimensions with suitable vector densities $f$ and smeared electric
fields $\hat{E}(g)$. But $\hat{\mathbf{A}}(e)$ fail to be well
defined. To resolve this tension between the two representations,
one proceeds as follows. Let $\vec{x}$ be the Cartesian coordinates
of a point in $\Sigma=\mathbf{R}^3$. Introduce a test function by
using the Euclidean background metric on $\mathbf{R}^3$,
\begin{eqnarray}
f_r(\vec{x})=\frac{1}{(2\pi)^{3/2}r^3} \exp(-|\vec{x}|^2/2r^2),
\end{eqnarray}
which approximates the Dirac delta function for small $r$. The
Gaussian smeared form factor for an edge $e$ is defined as
\begin{equation}
X^a_{(e,r)}(\vec{x}):=\int_e ds\,f_r(\vec{e}(s)-\vec{x})\dot{e}^a.
\end{equation}
Then one can define a smeared holonomy for $e$ by
\begin{equation}
\mathbf{A}_{(r)}(e):=\exp[-i\int_{\mathbf{R}^3}X^a_{(e,r)}(\vec{x})A_a(\vec{x})],
\end{equation}
where $A_a(\vec{x})$ is the $U(1)$ connection one-form of the
Maxwell field on $\Sigma$. Similarly one can define Gaussian smeared
electric fields by
\begin{equation}
E_{(r)}(g):=\int_{\mathbf{R}^3}g_a(\vec{x})\int_{\mathbf{R}^3}f_r(\vec{y}-\vec{x})E^a(\vec{y}).
\end{equation}
In this way one obtains two Poission bracket algebras. One is formed
by smeared holonomies and electric fields with
\begin{eqnarray}
\{\mathbf{A}_{(r)}(e), \mathbf{A}_{(r)}(e')\}=0=\{E(g), E(g')\}\\
\nonumber \{\mathbf{A}_{(r)}(e),
E(g)\}=-i(\int_{\mathbf{R}^3}X^a_{(e,r)}g_a)\, \mathbf{A}_{(r)}(e).
\end{eqnarray}
The other is formed by unsmeared holonomies and Gaussian smeared
electric fields with
\begin{eqnarray}
\{\mathbf{A}(e), \mathbf{A}(e')\}=0=\{E_{(r)}(g), E_{(r)}(g')\}\\
\nonumber \{\mathbf{A}(e),
E_{(r)}(g)\}=-i(\int_{\mathbf{R}^3}X^a_{(e,r)}g_a)\, \mathbf{A}(e).
\end{eqnarray}
Obviously, there is an isomorphism between them,
\begin{eqnarray}
I_r: \, (\mathbf{A}_{(r)}(e), E(g)) \mapsto (\mathbf{A}(e),
E_{(r)}(g)).
\end{eqnarray}
Using the isomorphism $I_r$, one can pass back and forth between the
polymer and the Fock representations. Specifically, the image of the
Fock vacuum can be shown to be the following element of
$\mathbf{Cyl}^\star$ \cite{var1}\cite{var2},
\begin{equation}
(V|\, = \, \sum_{\alpha,{\vec n}}\, \exp [ -\frac{\hbar}{2}
\sum_{IJ} G_{IJ}n_I n_J ]\, (\mathcal{N}_{\alpha, \vec{n}}|
,\label{Fvacuum}
\end{equation}
where $(\mathcal{N}_{\alpha, \vec{n}}|\in\mathbf{Cyl}^\star$ maps
the flux network function $|\mathcal{N}_{\alpha, \vec{n}}\rangle$ to
one and every other flux network functions to zero. While the states
$(\mathcal{N}_{\alpha, \vec{n}}|$ do not have any knowledge of the
underlying Minkowskian geometry, this information is coded in the
matrix $G_{IJ}$ associated with the edges of the graph $\alpha$,
given by \cite{AL}
\begin{equation}
G_{IJ}\ =  \int_{e_I}dt \dot{e}^a_I(t) \int_{e_J}dt' \dot{e_J}^b
(t')\, \int d^3x\, \delta_{ab}(\vec{x})\,
[f_r(\vec{x}-\vec{e}_I(t))\,
|\Delta|^{-\frac{1}{2}}\,f(\vec{x},\vec{e}_J(t'))],
\end{equation}
where $\delta_{ab}$ is the flat Euclidean metric and $\Delta$ its
Laplacian. Therefore, one can single out the Fock vacuum state
directly in the polymer representation by invoking Poincar\'e
invariance without any reference to the Fock space. Similarly, one
can directly locate in $\mathbf{Cyl}^\star$ all coherent states as
the eigenstates of the exponentiated annihilation operators. Since
$\mathbf{Cyl}^\star$ does not have an inner product, one uses the
notion of shadow states to do semiclassical analysis in the polymer
representation. From Eq.(\ref{Fvacuum}), the action of the Fock
vacuum $(V|$ on $\mathcal{N}_{\alpha, \vec{n}}$ reads
\begin{eqnarray}
(V|\mathcal{N}_{\alpha, \vec{n}}\rangle\
=\int_{\overline{\mathbf{A}}_\alpha}d\mu^0_\alpha\,
\overline{V}_\alpha \mathcal{N}_{\alpha, \vec{n}},
\end{eqnarray}
where the state $V_\alpha$ is in the Hilbert space
$\mathcal{H}_\alpha$ for the graph $\alpha$ and given by
\begin{equation}
V_\alpha (\mathbf{A})\, = \, \sum_{\vec n}\, \exp [ -\frac{\hbar}{2}
\sum_{IJ} G_{IJ}n_I n_J ]\, \mathcal{N}_{\alpha,
\vec{n}}(\mathbf{A}) .
\end{equation}
Thus for any cylindrical functions $\varphi_\alpha$ associated with
$\alpha$,
\begin{equation}
(V|\varphi_\alpha\rangle\ =\langle V_\alpha |\varphi_\alpha\rangle,
\end{equation}
where the inner product in the right hand side is taken in
$\mathcal{H}_\alpha$. Hence $V_\alpha(\mathbf{A})$ are referred to
as "shadows" of $(V|$ on the graphs $\alpha$. The set of all shadows
captures the full information in $(V|$. By analyzing shadows on
sufficiently refined graphs, one can introduce criteria to test if a
given element of $\mathbf{Cyl}^\star$ represents a semi-classical
state \cite{AL}. It turns out that the state $(V|$ does satisfy this
criterion and hence can be regarded as semi-classical in the polymer
representation.

The mathematical and conceptual tools gained from simple models like
the Maxwell fields are currently being used to construct
semiclassical states of quantum geometry. A candidate kinematical
coherent state corresponding to the Minkowski spacetime has been
proposed by Ashtekar and Lewandowki in the light of a "Laplacian
operator" \cite{AL2}\cite{AL}. However, the detailed structure of
this coherent state is yet to be analyzed and there is no a priori
guarantee that it is indeed a semiclassical state.

One may find comparisons of the two approaches from both sides
\cite{complexifier}\cite{AL}. It turns out that Varadarajan's
Laplacian coherent state for the polymer Maxwell field can also be
derived from Thiemann's complexifier method. However, one cannot
find a complexifier to get the coherent state proposed by Ashtekar
et al. for loop quantum gravity. Both approaches have their own
virtues and need further developments. The complexifier approach
provides a clear construction mechanism and manageable calculation
method, while the Laplacian operator approach is related closely
with the well-known Fock vacuum state. One may also expect that a
judicious combination of the two approaches may lead to significant
progress in the semiclassical analysis of loop quantum gravity.

\subsection{Algebraic Quantum Gravity Approach}

As we have shown in the last subsection, although Thiemann's
complexifier coherent state has a clear calculable mechanism and
correct semi-classical properties in testing kinematical operators,
it fails to be a qualified semi-classical state for the quantum
dynamics since the semiclassical limit of the Hamiltonian constraint
operator $\hat{\mathcal{S}}(N)$ or master constraint operator
$\hat{\mathbf{M}}$ is clearly not correct, both
$\hat{\mathcal{S}}(N)$ and $\hat{\mathbf{M}}$ are graph-changing so
that their expectation values with respect to these cut-off coherent
states are always zero. So a possible way to avoid such a problem is
to define a non-graph-changing version of Hamiltonian constraint
operator or similarly, a master constraint operator. However, such a
modification is hard to make in the framework of loop quantum
gravity since the action of Hamiltonian constraint operator always
adds several arcs on certain graphs. But if the framework of loop
quantum gravity is suitably modified then it turns out that a
version of non-graph-changing Hamiltonian constraint operator can be
proposed and the semi-classical analysis for the quantum dynamics
can be carried out with the complexifier coherent states defined
previously. Such a modification is recently made by Thiemann in
\cite{AQG1}\cite{AQG2}\cite{AQG3} and is called algebraic quantum
gravity (AQG) approach. We describe it briefly in what follows.

Algebraic quantum gravity is a new approach to canonical quantum
gravity suggested by loop quantum gravity. But in contrast to loop
quantum gravity, the quantum kinematics of algebraic quantum gravity
is determined by an abstract $*$-algebra generated by a countable
set of elementary operators labeled by a single algebraic graph with
countably infinite number of edges, while in loop quantum gravity
the elementary operators are labeled by a collection of embedded
graphs with finite number of edges. Thus one can expect that in
algebraic quantum gravity, we lose the information of the
topological and differential structure of the manifold in all the
quantization procedure before we do semi-classical analysis. Hence
the quantum theory will be of course independent of the topology and
differential structure of the manifold but based only on an
algebraic graph, which only contains the information of the number
of vertices and their oriented
valence.\\ \\
\textbf{Definition 6.2.1}: \textit{An oriented algebraic graph is an
abstract graph specified by its adjacency matrix $\alpha$, which is
an $N\times N$ matrix. One of its entries $\alpha_{IJ}$ stand for
the number of edges that start at vertex $I$ and end at vertex $J$.
The valence of the vertex $I$ is given by
$v_I=\sum_J(\alpha_{IJ}+\alpha_{JI})$. We also use $V(\alpha)$ and
$E(\alpha)$ to denote the sets of vertices and edges respectively.}
\\ \\
In our quantization procedure, we fix a specific cubic algebraic
graph with a countably infinite number of edges $N=\aleph$ and the
valence of each vertex $v_I=2\times \dim(\Sigma)$. Such a specific
choice, although it detracts from the generality of the theory, is
practically sufficient for our use in the semiclassical analysis.

Given the algebraic graph $\alpha$, we define a quantum $*$-algebra
by associating with each edge $e$ an element $A(e)$ of a compact,
connected, semisimple Lie group $G$ and an element $E_j(e)$ take
value in its Lie algebra $\mathfrak{g}$. These elements are subject
to the commutation relations
\begin{eqnarray}
&&[\hat{A}(e),\hat{A}(e')]=0, \nonumber\\
&&[\hat{E}_j(e),\hat{A}(e')]=i\hbar Q^2\delta_{e,e'}\tau_j/2\hat{A}(e),\nonumber\\
&&[\hat{E}_j(e),\hat{A}(e')]=-i\hbar
Q^2\delta_{e,e'}f_{jkl}\hat{E}_l(e'),\nonumber
\end{eqnarray}
and $*$-relations
\begin{eqnarray}
\hat{A}(e)^*=[\hat{A}(e)^{-1}]^T,\ \ \ \ \
\hat{E}_j(e)^*=\hat{E}_j(e),\nonumber
\end{eqnarray}
where $Q$ stands for the coupling constant, $\tau_j$ is the
generators in the Lie algebra $\mathfrak{g}$ and $f_{jkl}$ is the
structure constant of $\mathfrak{g}$. We denote the abstract quantum
$*$-algebra generated by above elements and relations by
$\mathfrak{A}$.

A natural representation of $\mathfrak{A}$ is the infinite tensor
product Hilbert space $\mathcal{H}^{\otimes}=\otimes_e\mathcal{H}_e$
where $\mathcal{H}_e=L^2(G,d\mu_H)$\cite{thiemann13}, whose element
is denoted by $\otimes_f\equiv\otimes_ef_e$. Two elements
$\otimes_f$ and $\otimes_{f'}$ in $\mathcal{H}^{\otimes}$ are said
to be strongly equivalent if $\sum_e|<f_e,f'_e>_{\mathcal{H}_e}-1|$
converges. We denote by $[f]$ the strongly equivalence class
containing $\otimes_f$. It turns out that two elements in
$\mathcal{H}^{\otimes}$ are orthogonal if they lie in different
strongly equivalence classes. Hence the infinite tensor Hilbert
space $\mathcal{H}^{\otimes}$ can be decomposed as a direct sum of
the Hilbert subspaces (sectors) $\mathcal{H}^\otimes_{[f]}$ which
are the closure of strongly equivalence classes $[f]$. Furthermore,
although each sector $\mathcal{H}^\otimes_{[f]}$ is separable and
has a natural Fock space structure, the whole Hilbert space
$\mathcal{H}^{\otimes}$ is non-separable since there are uncountably
infinite number of strongly equivalence classes in it. Our basic
elements in the quantum algebra are represented on
$\mathcal{H}^{\otimes}$ in an obvious way
\begin{eqnarray}
\hat{A}(e)\otimes_f&:=&[A(e)f_e]\otimes[\otimes_{e'\neq
e}f_{e'}],\nonumber\\
\hat{E}_j(e)\otimes_f&:=&[i\hbar Q^2X^e_jf_e]\otimes[\otimes_{e'\neq
e}f_{e'}].\nonumber
\end{eqnarray}
As one might have expected, all these operators are densely defined
and $E_j(e)$ is essentially self-adjoint. Given a vertex $v\in
V(\alpha)$, the volume operator can be constructed by using the
operators we just defined
\begin{eqnarray}
\hat{V}_v:=\ell_p^3\sqrt{|\frac{1}{48}\sum_{e_1\cap e_2\cap
e_3=v}\epsilon_v(e_1, e_2,
e_3)\epsilon^{ijk}\hat{E}_i(e_1)\hat{E}_j(e_2)\hat{E}_k(e_3)|},\nonumber
\end{eqnarray}
where the values of $\epsilon_v(e_1, e_2, e_3)$ should be assigned
once for all for each vertex. When we embed the algebraic graph into
some manifold, the embedding should be consistent with the assigned
values of $\epsilon_v(e_1, e_2, e_3)$.

Then we discuss the quantum dynamics. By the regularization methods
frequently used in the last two sections, the half densitized
constraints can be quantized to be composite operators as we list
below.
\begin{itemize}
\item Gauss constraint
\begin{eqnarray}
\hat{G}_j(v):=\hat{Q}_v^{(1/2)}\sum_{e\ at\ v}\hat{E}_j(e);\nonumber
\end{eqnarray}

\item Spatial diffeomorphism constraint
\begin{eqnarray}
\hat{D}_j(v)&:=&\frac{1}{E(v)}\sum_{e_1\cap e_2\cap
e_3=v}\frac{\epsilon_v(e_1, e_2,
e_3)}{|L(v,e_1,e_2)|}\nonumber\\
&\times&\sum_{\beta\in
L(v,e_1,e_2)}\mathrm{Tr}\big(\tau_j[\hat{A}(\beta)-\hat{A}(\beta)^{-1}]\hat{A}(e_3)[\hat{A}(e_3)^{-1},\sqrt{\hat{V}_v}]\big);\nonumber
\end{eqnarray}

\item Euclidean Hamiltonian constraint (up to an overall factor)
\begin{eqnarray}
\hat{H}^{(r)}_E(v)&:=&\frac{1}{E(v)}\sum_{e_1\cap e_2\cap
e_3=v}\frac{\epsilon_v(e_1, e_2,
e_3)}{|L(v,e_1,e_2)|}\nonumber\\
&\times&\sum_{\beta\in
L(v,e_1,e_2)}\mathrm{Tr}\big([\hat{A}(\beta)-\hat{A}(\beta)^{-1}]\hat{A}(e_3)[\hat{A}(e_3)^{-1},\hat{V}_v^{(r)}]\big);\nonumber
\end{eqnarray}

\item Lorentzian Hamiltonian constraint (up to an overall factor)
\begin{eqnarray}
\hat{T}(v)&:=&\frac{1}{E(v)}\sum_{e_1\cap e_2\cap
e_3=v}\epsilon_v(e_1, e_2,
e_3)\nonumber\\
&\times&\mathrm{Tr}\big(\hat{A}(e_1)[\hat{A}(e_1)^{-1},[\hat{H}_E^{(1)},\hat{V}]]
\hat{A}(e_2)[\hat{A}(e_2)^{-1},[\hat{A}(e_3)^{-1},[\hat{H}_E^{(1)},\hat{V}]]\nonumber\\
&\times&\hat{A}(e_3)[\hat{A}(e_3)^{-1},\sqrt{\hat{V}_v}]\big),\nonumber\\
\hat{H}(v)&=&\hat{H}^{(1/2)}_E(v)+\hat{T}(v);
\end{eqnarray}

\end{itemize}
where $\hat{V}:=\sum_v \hat{V}_v$,
$\hat{H}_E^{(1)}:=\sum_v\hat{H}_E^{(1)}(v)$ and
\begin{eqnarray}
\hat{Q}_v^{(r)}&:=&\frac{1}{E(v)}\sum_{e_1\cap e_2\cap
e_3=v}\epsilon_v(e_1, e_2,
e_3)\nonumber\\
&\times&\mathrm{Tr}\big(\hat{A}(e_1)[\hat{A}(e_1)^{-1},\hat{V}_v^{(r)}]\hat{A}(e_2)[\hat{A}(e_2)^{-1},\hat{V}_v^{(r)}]
\hat{A}(e_3)[\hat{A}(e_3)^{-1},\hat{V}_v^{(r)}]\big).\nonumber
\end{eqnarray}
$L(v,e_1,e_2)$ denotes the set of minimal loops starting at $v$
along $e_1$ and ending at $v$ along $e_2^{-1}$. And a loop $\beta\in
L(v,e_1,e_2)$ is said to be minimal provided that there is no other
loop within $\alpha$ satisfying the same restrictions with fewer
edges traversed. Note that since we only have a single cubic
algebraic graph, the diffeomorphism constraint can only be
implemented by defining the operators corresponding to
diffeomorphism generators because a finite diffeomorphism
transformation is not meaningful in our algebraic treatment unless
the algebraic graph is embedded in a manifold. As a result, the
(extended) master constraint can be expressed as a quadratic
combination:
\begin{eqnarray}
\hat{\mathbf{M}}:=\sum_{v\in V(\alpha)}[\hat{G}_j(v)^\dagger
\hat{G}_j(v)+\hat{D}_j(v)^\dagger \hat{D}_j(v)+\hat{H}(v)^\dagger
\hat{H}(v)].\nonumber
\end{eqnarray}
It is trivial to see that all the above operators are
non-graph-changing and embedding independent because we have only
worked on a single algebraic graph so far. However, when we test the
semiclassical limit of these operators, especially the master
constraint operators, we should specify an embedding map $X$ which
map a algebraic graph to be an embedded one. With this specific
embedding, we can see the correspondence between the classical
algebra of elementary observables and the quantum $*$-algebra. We
define the holonomy and suitably modified flux by
\begin{eqnarray}
A(e)&:=&A(X(e)):=\mathcal{P}\exp(\int_{X(e)}A),\nonumber\\
E_j(e)&:=&-2\mathrm{Tr}\big[\tau_j\int_{S_e}\epsilon_{abc}\mathrm{d}x^a\wedge
\mathrm{d}x^bA(\rho_e(x))E^c(x)A(\rho_e(x))^{-1}\big],\nonumber
\end{eqnarray}
where $S_e$ is a face which intersects the edge $X(e)$ only at an
interior point $p_e$ of both $S_e$ and $X(e)$. We choose a system of
paths $\{\rho_e(x)\}_{x}$ for all $x\in S_e$, such that $\rho_e(x)$
starts at $s(X(e))$ along $X(e)$ until $p_e$ and then runs within
$S_e$ until $x$. As one might expect, the quantum $*$-algebra we
defined previously is just consistent with the classical Poisson
algebra generated by these holonomis and fluxes:
\begin{eqnarray}
&&\{{A}(e),{A}(e')\}=0, \nonumber\\
&&\{{E}_j(e),{A}(e')\}= Q^2\delta_{e,e'}\tau_j/2{A}(e),\nonumber\\
&&\{{E}_j(e),{A}(e')\}=-Q^2\delta_{e,e'}f_{jkl}{E}_l(e').\nonumber
\end{eqnarray}

Then we consider the coherent states. By employing the Laplacian
complexifier on each edge
\begin{eqnarray}
C_e:=-\frac{1}{2Q^2a^2_e}E_j(e)E_j(e),\nonumber
\end{eqnarray}
the coherent state is obtained as it was in the last section:
\begin{eqnarray}
\Psi^{t_e}_{e;(A,E)}(A)\equiv\Psi^{t_e}_{e;g(A,E)}(A(e))=\sum_\pi
\dim(\pi)e^{-t\lambda_\pi}\chi_\pi\big(g(A,E)A(e)\big),\nonumber
\end{eqnarray}
where $\lambda_\pi$ denotes the eigenvalue of the Laplacian on $G$
and $t=\ell_p^2/a_e^2$ represents the classicalization parameter.
The coherent state peaks at the complexified classical phase space
point
\begin{eqnarray}
g(A,E):=\sum_{n=0}^\infty\frac{(-i)^n}{n!}\{C_e, A(e)\}_n=\exp(i
E(e)/a^2_e)A(e),\nonumber
\end{eqnarray}
note that the parameter $a_e$ is specified such that $E(e)/a^2_e$ is
dimensionless. Hence the coherent state on the whole graph is
represented by an infinite tensor product state:
\begin{eqnarray}
\Psi^t_{A,E}(A):=\bigotimes_{e\in
E(\alpha)}\frac{\Psi^{t_e}_{e;(A,E)}(A)}{||\Psi^{t_e}_{e;(A,E)}(A)||}.\nonumber
\end{eqnarray}
The peakness, fluctuation and other semiclassical properties of
these states have been checked in \cite{thiemann11}\cite{thiemann12}
in which the most important part is that
\begin{eqnarray}
<\Psi_{A,E}|\hat{A}(e)|\Psi_{A,E}>=A(e)\ \ \ \ \
<\Psi_{A,E}|\hat{E}(e)|\Psi_{A,E}>=E(e)\nonumber
\end{eqnarray}
up to terms which vanish faster than any power of $t_e$ as
$t_e\rightarrow0$. And the fluctuations are small.

With the semiclassical state we just constructed, the expectation
value of the above (extended) master constraint operator can be
calculated and its semiclassical limit can be tested. In the
following, we summarize the result of the calculation. In
\cite{AQG2}, a semiclassical calculation for the master constraint
operator is carried out based on a cubic algebraic graph. The
calculation makes use of a simplifying assumption: we substitute the
gauge group for gravity $SU(2)$ by $U(1)^3$. And the result of the
calculation shows that in $U(1)^3$ case the (extended) master
constraint operator has correct semiclassical limit
\begin{eqnarray}
\lim_{t\rightarrow0}<\Psi^t_{m}|\hat{\mathbf{M}}|\Psi^t_{m}>
=\mathbf{M}^{\mathrm{cubic}}[m]\rightarrow\mathbf{M}[m]\ \
(\epsilon\rightarrow0)\nonumber
\end{eqnarray}
where $m$ represents a phase space point and $\epsilon$ is the
lattice parameter such that the lattice become continuum as
$\epsilon\rightarrow0$. In addition, it is shown that the
next-to-leading order terms which contribute to the fluctuation of
$\hat{\mathbf{M}}$ are finite.

Moreover, the calculation in \cite{AQG3} shows that the result of
the exact non-Abelian calculation matches precisely the results of
the Abelian approximation, provided that we replace the classical
$U(1)^3$ terms $\{h^j_e,p^e_j\}_{j=1,2,3}$ by
$\{\mathrm{Tr}(\tau_jA(e)),\mathrm{Tr}(\tau_jE(e))\}_{j=1,2,3}$,
which means that the theory of algebraic quantum gravity admits a
semiclassical limit whose infinitesimal gauge symmetry agrees with
that of general relativity.

\newpage
\section{Conclusion and Discussion}

As it was shown in the previous sections, loop quantum gravity
offers a conceptually clear and mathematically rigorous approach to
quantize general relativity. In this approach, we are seeking new
physics deeply below the Planck scale. In the kinematical framework,
a quantum Riemannian geometry is established and some geometrical
operators, e.g. area, volume, are well-defined, and their spectrum
are shown to be discrete, which means that the structure of the
space may be discrete below the Planck scale. Such a new phenomena
sheds light on quantum field theory, lattice gauge theory and their
renormalization. Moreover, the program in the quantum dynamics of
loop quantum gravity represents significant progress in the research
area of quantum gravity. Before loop quantum gravity, the quantum
Wheeler-DeWitt equation was only a formal equation and from concrete
calculations. However, in the framework of loop quantum gravity, we
already have a well-defined quantum Hamiltonian constraint operator
which has an explicit action on kinematical states, so that the
quantum Wheeler-DeWitt equation is well-defined in loop quantum
gravity. On the other hand, the matter field can also be quantized
in this framework and we show that the matter Hamiltonian is free of
UV-divergence and don't need a renormalization process. Furthermore,
with the coupled matter field, a matter coupled Hamiltonian
constraint operator is obtained so that the problem of time may be
solved and, such an idea is being translated into a new
understanding of the early universe in the context of loop quantum
cosmology.

Although great progress has been made, as an unfinished framework,
loop quantum gravity still has many issues to be solved in the
future research. To conclude this thesis, we list some of those in
the following:
\begin{itemize}
\item First of all, we don't have the complete solutions for either
Hamiltonian constraint equation or master constraint equation. Thus
one cannot explicitly construct the physical Hilbert space for loop
quantum gravity. So the quantum dynamics of gravity is essentially
unknown so far.

\item To make contact with experimental results, one should know the
observables in the quantum theory which have to be invariant under
gauge transformation. However, some of the Dirac observables that
have been constructed involve an infinite number of derivatives and
extremely hard to manage
\cite{thiemann14}\cite{dittrich}\cite{dittrich2}.

\item The semiclassical limit of loop quantum gravity is
unknown so far, although a great deal of progress has been made in
the context of algebraic quantum gravity. And in algebraic quantum
gravity, further research work is needed to show the fluctuation of
master constraint operator should be small.

\item As it was shown at the end of section 4.3, the regularization
process for the Hamiltonian constraint operator is ambiguous and
there is a list of free parameters. Thus it is also a research
project to remove as many ambiguities as possible. And some work has
been recently done in this direction \cite{perez2}.

\item The Immirzi parameter is another free parameter in the framework
of loop quantum gravity, which comes in with the classical
formulation. In the classical theory, different values of the
Immirzi parameter label equivalent classical theories since they are
connected by canonical transformations. However, in quantum theory,
it is problematic because the representation with different Immirzi
parameter are not unitarily equivalent.

\item The construction of loop quantum gravity crucially depends on
the compactness of its gauge group $SU(2)$, which comes from an
internal partial gauge fixing. And it is argued that the internal
Lorentz symmetry is broken in a non-natural way \cite{samuel}. So it
seems to be better to switch back to the complex Ashtekar variables
which are free of the internal gauge fixing and preserve the
internal Lorentz symmetry and, will also greatly simplify the
Hamiltonian constraint. However, the price is that we should work on
non-compact gauge group $SL(2,\mathbf{C})$, and there is no
satisfactory quantization programme for the
$SL(2,\mathbf{C})$-gravity so far, although some work has been done
in this direction \cite{FLncs}\cite{okolow1}\cite{okolow2}.

\item As it was shown in section 6.2, algebraic quantum gravity provides a clear way to
make many calculations accessible and the semiclassical analysis can
be carried out in this framework. And it seems that the construction
of algebraic quantum gravity admits the non-compact internal gauge
group (or non-compact reduced configuration space) since there is
only one graph in quantization process. However, in this framework,
the restriction of the possible representation is so loose that
there even exists a possible representation in which the spectrum of
geometrical operators are continuous.

\item The transition amplitude calculation for loop quantum gravity
is accessible in the so-called spin foam model and depends on a
unclear conjecture of GFT/spinfoam duality. On the other hand, it is
still not clear how to build a path-integral formulation and connect
with spin foam models from the canonical approach, and such a
connection may give the needed support for GFT/spinfoam duality.

\item We have constructed the dynamics of matter quantum
field theory on a quantum background in section 5. However, it is
not clear how we can make the connection with the ordinary quantum
field theory on curved spacetime. Moreover, it is still an problem
how to find the non-perturbative correspondence of Hadamard states
in perturbative quantum field theory in curved spacetime, although
some hints of a connection may come out in spin foam calculations
\cite{rovelli10}\cite{Gravpropagator}.

\end{itemize}

\newpage

\section*{Vita}
Muxin Han was born in Beijing, People's Republic of China. He
studied as an undergraduate student in the Department of Physics at
Beijing Normal University from 2001 to 2005, and obtained his
Bachelor of Science degree at Beijing Normal University in 2005.
Muxin came to the United States and began his graduate studies at
Louisiana State University in August of 2005. His major is physics.
\end{document}